\documentclass[traditabstract]{aa}
\usepackage{amsmath}
\usepackage{txfonts}
\usepackage{ifpdf}

\ifpdf
\usepackage[pdftex]{graphicx}
\usepackage{epstopdf}
\usepackage[pdftex,colorlinks=false,pdfstartview=Fit,urlcolor=pdfwww,
  bookmarks=true,bookmarksnumbered=true]{hyperref}

\else
\usepackage[dvips]{graphicx}
\usepackage[dvips]{hyperref}

\fi
\usepackage{url}
\usepackage{framed}
\usepackage{natbib,ifthen}

\bibpunct{(}{)}{;}{a}{}{,} 

\newcommand{\sqbeam}{F_l}
\newcommand{\fnurl}[1]{\footnote{\url{#1}}}

\newcommand{\artdeco}{\texttt{artDeco}}
\newcommand{\Artdeco}{\texttt{ArtDeco}}
\newcommand{\lmax}{\ensuremath{l_{\text{max}}}}
\newcommand{\kmax}{\ensuremath{k_{\text{max}}}}
\newcommand{\psipol}{\ensuremath{\psi_{\text{pol}}}}

\newcommand{\blue}{\emph{blue}}
\newcommand{\red}{\emph{red}}
\newcommand{\green}{\emph{green}}
\newcommand{\magenta}{\emph{purple}}
\newcommand{\black}{\emph{black}}

\newcommand{\solid}{\emph{solid}}
\newcommand{\dashed}{\emph{dashed}}

\newcommand{\ftop}{\emph{top}}
\newcommand{\fmiddle}{\emph{middle}}
\newcommand{\fbottom}{\emph{bottom}}
\newcommand{\fleft}{\emph{left}}
\newcommand{\fright}{\emph{right}}

\begin{document}

\title{ArtDeco: A beam-deconvolution code for absolute CMB measurements}

\author
{E.~Keih\"anen\inst{1} \and
 M.~Reinecke\inst{2}}

\institute{
University of Helsinki, Department of Physics,
P.O.~Box 64, FIN-00014, Helsinki, Finland \\
\email{elina.keihanen@helsinki.fi}
\and
Max-Planck-Institut f\"ur Astrophysik, Karl-Schwarzschild-Str.~1, 85741 Garching, Germany\\
\email{martin@mpa-garching.mpg.de}
}

\abstract{
We present a method for beam-deconvolving cosmic microwave background (CMB) anisotropy measurements.
The code takes as input the time-ordered data
along with the corresponding detector pointings and known beam shapes,
and produces as output the harmonic $a_{Tlm}$, $a_{Elm}$, and $a_{Blm}$
coefficients of the observed sky. From these one can derive temperature and
Q and U polarisation maps.
The method is applicable to absolute CMB measurements with wide sky coverage, and is independent of the scanning strategy.
We tested the code with extensive simulations,
mimicking the resolution and data volume of Planck 30GHz and 70GHz channels, but with exaggerated beam asymmetry.
We applied it to multipoles up to $l=1700$ and examined the results in both pixel space and harmonic space.
We also tested the method in presence of white noise.
}


\maketitle

\section{Introduction}

Removing systematic effects plays an important role in the data analysis of modern
high-sensitivity cosmic microwave background (CMB) experiments,
such as the WMAP and Planck missions \citep{jarosik-etal-2011,tauber-etal-2010}.
In this work we concentrate on one source of systematic effects,
the beam asymmetry.
We present an efficient beam deconvolution code, called \artdeco.
It can be applied to any absolute CMB experiment with sufficient sky coverage.

The required input consists of the time-ordered data (TOD),
the corresponding detector pointings, and known beam shapes.
The primary output of the code consists of the harmonic $a_{Tlm}$, $a_{Elm}$, and $a_{Blm}$
coefficients of the sky. From these one can derive temperature and
Q and U polarisation maps.

This is the usual deconvolution map-making problem \citep[see, e.g.,][]
{armitage-wandelt-2004,armitage-wandelt-2009,harrison-etal-2011}.
Our method differs from the earlier works by an efficient reformulation of the deconvolution
problem through heavy use of Wigner functions.
The formulation is general, and makes no assumptions on the scanning pattern.
Similar ideas have been studied by G.~Pr\'ezeau, independently of us (private communication).

In the CMB literature the concept of map-making often refers to a procedure that involves
removal of correlated noise. 
Methods for doing this have been treated in several papers
\citep{keihanen-etal-2005,keihanen-etal-2010,poutanen-etal-2006,ashdown-etal-2007b,
ashdown-etal-2007a,kurki-suonio-etal-2009,sutton-etal-2009}.
The methods discussed in these works construct pixelised temperature and Q, U polarisation maps,
but do not attempt to correct for an asymmetric beam shape.

Our method steps in at the point where the correlated noise has already been cleaned from the data.
The input TOD is assumed to include signal and uncorrelated noise only.
We are assuming here that the correlated noise component can be removed from the data to a sufficient
degree prior to the beam deconvolution step. The noise removal step itself is affected by beam asymmetries
because signal differences caused by beam mismatch can be falsely interpreted as noise. Fortunately, in destriping methods
it is possible to strongly reduce this effect by masking out the regions with the strongest foreground contamination that generate
most of the effect. Also, destriping methods produce as natural output the noise-cleaned TOD,
which is the required input in deconvolution map-making. 
Accordingly, destriping methods are the natural counterparts of our deconvolution method.

Since the aim of this paper is to present a new deconvolution method, rather than assess the performance
of any particular instrument, we tested the code with a somewhat idealised simulation.
In particular, we ignored other systematic effects such as frequency response.
On the other hand, we chose strongly asymmetric beam shapes to show the beam effects more clearly.

\section{Beam deconvolution}

\subsection{Constructing the linear system}

In this section we present the deconvolution problem using a formalism
based on Wigner functions.
We write the solution in a form that allows us to solve the problem numerically in an efficient way.
Some details are left for Appendix \ref{appendix:definitions}.

We start by writing the signal $t_j$ received by a detector at a given time $\tau_j$ as
\begin{equation}
   t_j = \sum_{slmk} a_{slm}b^\ast_{slk} 
         D^{l\ast}_{mk}(\omega_j) +n_j\text{.}
\end{equation}
The $a_{slm}$ ($s=0,\pm2$) are harmonic coefficients that represent temperature and polarisation
of the sky signal, $b_{slk}$ are corresponding beam coefficients, $n_{j}$ represents noise,
and $D^l_{mk}(\omega_j)$ is a Wigner function that defines a rotation of the beam
from a fiducial orientation at the north pole to its actual position and orientation.
For exact definitions, see Appendix \ref{appendix:definitions}.

We have adopted a compact notation where $\omega_{j}$
denotes a combination 
of three angles $\{\phi_{j},\theta_{j},\psi_{j}\}$, which define the detector's orientation.
Angles $\theta_{j}$ and $\phi_{j}$ define a point on the celestial sphere,
and $\psi_{j}$ defines the beam orientation.
The angles can be interpreted as Euler rotation angles.

We perform a simple linear least-squares fit of $a_{slm}$ to the data
up to some multipole \lmax.

In general, a linear least-squares fit leads to an equation of the form
\begin{equation}
    x = (A^\dagger C_n^{-1}A +C_s^{-1})^{-1}A^\dagger C_n^{-1} y \text{,} \label{linear_fit}
\end{equation}
where $x$ is the vector of unknowns, $y$ is the data vector,
$C_n$ is noise covariance, and $C_s$ represents \emph{a priori} information on the covariance 
of $x$ \citep{wiener-1949}.
Matrix $A$ relates the unknowns to the data through
\begin{equation}
  y = A x +n.
\end{equation}

In the present case matrix $A$ is
\begin{equation}
   A^j_{slm} = \sum_k b^\ast_{slk}D^{l\ast}_{mk}(\omega_j)
\end{equation}
and $x$ represents $a_{slm}$.

We made two simplifying assumptions, which are not of fundamental nature, however.
Firstly, we assume the noise is white with constant variance, and assign equal weights to all samples $j$.
Secondly, we use no \emph{a priori} information on the sky signal,
i.e.\ the harmonic coefficients are assumed to be uncorrelated and can host a signal
of infinite variance.
Under these assumptions
the covariance matrices drop out of the equation ($C_{n}$=const, $C_{s}^{-1}=0$).
The least-squares solution for $a_{slm}$ is obtained by solving the linear system of equations
\begin{equation}
   \sum_{s'l'm'}\sum_j A^{j\ast}_{slm}A^{j}_{s'l'm'}a_{s'l'm'} 
    = \sum_j A^{j\ast}_{slm} t_j .  \label{alm_equation}
\end{equation}

In the following we process Eq.~(\ref{alm_equation}) further to bring it into a form where it can be solved in an efficient way.
We treat the right- and left-hand sides separately.

\subsubsection{Right-hand side}

The right-hand side yields
\begin{equation}
   \sum_j A^{j\ast}_{slm} t_j  = 
     \sum_{kj} b_{slk} D^l_{mk}(\omega_j) t_j
      = \sum_k b_{slk} T^l_{mk}.
\end{equation}
Here we define
\begin{equation}
   T^l_{mk} = \sum_j D^{l}_{mk}(\omega_j) t_j .
\end{equation}
The triplets $\omega_{j}=\{\phi_{j},\theta_{j},\psi_{j}\}$
define the detector orientation and are provided as input to the code.
To numerically evaluate the transform, we replace the sum over TOD samples by
a sum over a three-dimensional grid as follows.
We divide the space spanned by all possible triplets $\omega=\{\phi,\theta,\psi\}$
into bins of equal volume and denote by 
$t(\omega)$ the sum of values $t_j$ 
falling into one bin.
In ${\theta,\phi}$ we use HEALPix\fnurl{http://sourceforge.net/projects/healpix}
pixelisation \citep{gorski-etal-2005}, and in $\psi$ we divide the space uniformly.
Most elements of $t(\omega)$ are zero, since a particular HEALPix pixel
is typically observed only in a few orientations of the beam.

We call quantity $t(\omega)$ the \emph{3D signal map},
and quantity
\begin{equation}
   T^l_{mk} =
    \sum_\omega D^{l}_{mk}(\omega) t(\omega)
     \label{twigner_trans}
\end{equation}
the \emph{Wigner transform of the signal map}.

Here $\sum_\omega$ denotes a sum over all bins.
Since most bins are empty, the sum needs to be carried out only over a subset of all $\omega$ values.
Expansion \eqref{twigner_trans}
can be evaluated numerically quite efficiently
using FFTs and fast algorithms for the generation of Wigner functions.

Quantities $t(\omega)$ and $T^l_{mk}$ carry no memory of the order in which the sky was scanned.
They just record which pixels of the sky were observed in which orientation of the beam.

\subsubsection{Left-hand side}

Consider then the matrix on the left-hand side of Eq.~\eqref{alm_equation},
\begin{equation}
\sum_j A^{j\ast}_{slm}A^{j}_{s'l'm'} =
    \sum_{jkk'} b_{slk}
    D^l_{mk}(\omega_j) D^{l'\ast}_{m'k'}(\omega_j) 
    b^\ast_{s'l'k'}. \label{lhs1}
\end{equation}

The multiplier matrix is too large to be inverted.
Instead, we use the conjugate gradient (CG) iteration method to solve
Eq.~\eqref{alm_equation}.

We pixelise the three-dimensional pointing space in the same way we did with the signal,
and denote by $n(\omega)$ the number of hits into bin $\omega$.
In line with the definition of the 3D signal map,
we call $n(\omega)$ the \emph{3D hit map}.

We replace the sum over samples $j$ by a sum over bins and write
\begin{equation}
 \sum_{j} D^{l\ast}_{mk}(\omega_j) 
          D^{l'}_{m'k'}(\omega_j)
 = \sum_\omega D^{l\ast}_{mk}(\omega) 
               D^{l'}_{m'k'}(\omega) 
               n(\omega) .
       \label{dproduct_sum1}
\end{equation}
As the next step, we write the Wigner functions as
\begin{equation}
  D^l_{mk}(\theta,\phi,\psi) = 
    e^{-im\phi} d^l_{mk}(\theta) e^{-ik\psi} ,
\end{equation}
where $d^l_{mk}$ are the reduced Wigner functions.
Inserting this into \eqref{dproduct_sum1}, we obtain
\begin{align}
 & \sum_{j} D^{l\ast}_{mk}(\omega_j)
          D^{l'}_{m'k'}(\omega_j) \nonumber \\
 =& \sum_{\theta}
     d^l_{mk}(\theta)d^{l'}_{m'k'}(\theta) 
     \sum_{\phi,\psi} e^{i(m-m')\phi}e^{i(k-k')\psi}n(\theta,\phi,\psi) \nonumber \\
 =& \sum_{\theta}
     d^l_{mk}(\theta)d^{l'}_{m'k'}(\theta)  N^{\ast}_{m-m',k-k'}(\theta) ,
     \label{dproduct_sum2}
\end{align}
where we have defined
\begin{equation}
     N_{mk}(\theta) = 
     \sum_{\phi,\psi} e^{-im\phi}n(\theta,\phi,\psi)e^{-ik\psi} .
       \label{Nmap}
\end{equation}
The sums over $\theta,\phi,\psi$ are carried out over 3D bins.
Substituting this back into \eqref{lhs1} and rearranging,
we bring the left-hand side of the deconvolution equation into the form
\begin{align}
   & \sum_{s'l'm'}\sum_j A^{j\ast}_{slm}A^{j}_{s'l'm'}a_{s'l'm'}
   \label{lhs2} \\
   =& \sum_{k} b_{slk}
    \sum_{\theta} d^l_{mk}(\theta)
     \sum_{m'k'} N^{\ast}_{m-m',k-k'}(\theta) 
    \sum_{l's'} d^{l'}_{m'k'}(\theta) b^\ast_{s'l'k'} a_{s'l'm'} . 
    \nonumber
\end{align}
This can be evaluated much more quickly than the original formula~\eqref{lhs1},
since the summation over the long TOD vector has been replaced by an operation
on the more compact $N$ object. Also, the convolution operations in indices $k$ and $m$ can be carried
out efficiently with FFTs.

The terms are written in the order in which they are applied in the code.

\subsection{Formulation through 3j coefficients and preconditioner}

Eq.~\eqref{lhs1} can be written in a yet more compact form.
 
The product of two Wigner functions can be expressed in terms of $3j$ symbols (Appendix \ref{wigner_expansion})
as
\begin{align}
   D^{l\ast}_{mk}(\omega) D^{l'}_{m'k'}(\omega) =&
    (-1)^{m'+k'}\sum_{l_2} (2l_2+1) D^{l_2}_{m'-m,k'-k}(\omega) \\
    & \times
        \left( \begin{array}{ccc}
            l & l' & l_2 \\ m & -m' & (m'\!-\!m)
         \end{array} \right)
          \left( \begin{array}{ccc}
             l & l' & l_2 \\ k & -k' & (k'\!-\!k)
         \end{array} \right).       \nonumber
\end{align}


Substituting this into \eqref{dproduct_sum1} and that into \eqref{lhs1},
we arrive at the formula
\begin{align}
& \sum_j A^{j\ast}_{slm}A^{j}_{s'l'm'} \nonumber\\
=&  \sum_{kk'} (-1)^{m'+k'}b_{slk} b^\ast_{s'l'k'}
   \sum_{l_2} (2l_2+1)
          W^{l_2}_{m'-m,k'-k}  \label{lhs3} \\
      & \times
        \left( \begin{array}{ccc}
            l & l' & l_2 \\ m & -m' & (m'\!-\!m)
         \end{array} \right)
          \left( \begin{array}{ccc}
             l & l' & l_2 \\ k & -k' & (k'\!-\!k)\end{array} \right)
         .  \nonumber
\end{align}
Indices $l$ and $l'$ run from 0 to \lmax.
The sum over $l_{2}$ covers the values for which $|l-l'| \le l_2 \le l+l'$.
We have defined the \emph{Wigner transform of the hit map} as
\begin{equation}
   W^{l}_{mk} = \sum_\omega 
    D^{l}_{mk}(\omega) n(\omega).
    \label{nwigner_trans}
\end{equation}
We use the formula \eqref{lhs3} to explicitly construct the diagonal of the matrix,
which serves as a preconditioner to speed up the CG iteration.

The formulation presented here is general, and makes no assumptions on
the scanning pattern or on the sky coverage.

\section{Implementation}

\subsection{User interface}
The \artdeco\ code takes two lists of input data objects containing the
beam coefficients $b_{slk}$ and the so-called 3D maps for the respective
detector for all detectors that are to be processed simultaneously.
The beam coefficients can be read from FITS files in the format that is also
being used by the HEALPix package for storing spherical harmonic coefficients.

The 3D maps contain the quantities $n(\omega)$ and $t(\omega)$ for a given
detector; they are created from time streams of detector pointings and sky signal, respectively,
by another stand-alone module called \texttt{todtobin}. This module achieves the
angular discretisation of the detector pointings by sorting them into pixels
of a HEALPix map with a user-defined $N_\text{side}$ parameter, which are in
turn subdivided into $n_\psi$ bins for the discretisation in $\psi$.
The choice of $N_\text{side}$ and $n_\psi$ has direct consequences on the
run-time of the deconvolver and the accuracy of its results: if high values
are chosen, the 3D map will become very large, and the deconvolver will
require more memory and CPU time. On the other hand, too low values will result
in noticeable discretisation errors and inaccurate output $a_{Xlm}$. As a guideline,
$N_\text{side}$ should be chosen in such a way that the smallest features of
all beams are well resolved at the angular resolution
\begin{equation}
\delta\alpha\approx 3500\arcmin/N_\text{side}
\end{equation}
associated with this $N_\text{side}$.
In a similar fashion, $n_\psi$ should be large enough to resolve azimuthal
features of the beams (assuming the beams are centred on a pole). For
completely axisymmetric beams, $n_\psi=1$ is sufficient. Generally,
$n_\psi=10k_\text{max}$ (where $k_\text{max}$ is the highest
azimuthal multipole used to describe the beams) is a good choice.

The computational complexity of \texttt{todtobin} is approximately
$\mathcal{O}(n_\text{samp} \log n_\text{samp})$, where $n_\text{samp}$ is the
total number of TOD samples, since its runtime is dominated by the sorting of
the input TOD according to HEALPix pixel number and $\psi$ bin. Consequently,
$N_\text{side}$ and $n_\psi$ only have a significant effect on the size
of the output 3D maps. Typical execution times for \texttt{todtobin} with
several billion samples (as used for the experiments shown in this paper) are of
the order of a few minutes.

\Artdeco\ takes several user-supplied parameters that influence its behaviour.
Most important among these are \lmax, which specifies the maximum multipole
up to which deconvolution is carried out, and $\kmax\le\lmax$, which limits
the maximum azimuthal multipole used for the beams. The explicit introduction
of \kmax\ is advantageous since most beams can be expressed accurately
even when using a $\kmax\ll\lmax$ (even $\kmax=0$ if the beam has rotational
symmetry and is unpolarised), and the reduction in \kmax\ reduces CPU and memory
requirements dramatically.

\Artdeco's output consists of a set of $a_{lm}$ in HEALPix-compatible format.
Depending on whether the user requested polarisation,
the file contains only $a_{Tlm}$ or additional $a_{Elm}$ and $a_{Blm}$
coefficients.

\subsection{Algorithm overview}
\label{algorithm}

The deconvolution algorithm has been implemented in C++, making use of
the MPI library for parallelisation.
The quantities that depend on $\theta$, such as $d$ and $N$, are distributed
to MPI tasks according to the $\theta$ angle. This allows a fairly good load
balance. Summation over $\theta$,
when needed, is done efficiently through the collective \texttt{MPI\_Reduce()} routine.
The $a_{slm}$ coefficients are distributed according to index $m$. The beam 
coefficients $b_{slk}$ are fully present on all tasks.

The procedure of evaluating Eq.~\eqref{lhs2} can be summarised as follows.
First we multiply the input $a$ by the beam coefficients $b$.
Then we perform a loop over index $m'$, inside which we broadcast the $a_{s'l'm'}$
coefficients for the current $m'$ to all processes, multiply by $b$, construct
the reduced Wigner matrices
for the local values of $\theta$, and accumulate over $l'$.
The multiplication by matrix $N$ is a convolution operation in indices $m,k$
and can be performed efficiently using 2D FFTs.
Each MPI task performs the multiplication by $N$ independently for the local $\theta$.
After that, we make another loop over index $m$, construct again the Wigner
matrices, sum over $\theta$, collect the result according to $m$ to the owner
process, and multiply by the beam coefficients.

The solution is assumed to have converged sufficiently as soon as the
residual of the current $a_{slm}$ estimate is less than $10^{-12}$
times the residual of a vector containing all zeroes.

To generate the $3j$ symbols needed to construct the preconditioner,
we use a C implementation of the recursive
algorithm described by \citet{schulten-gordon-1975}.

For a more detailed technical description of the algorithm, see Appendix
\ref{technical_tweaks}. CPU and memory requirements of various runs are
presented and discussed in Appendix \ref{benchmarks}.

\section{Simulation setup}
\label{sim_setup}

The deconvolution algorithm was tested using mock data produced by the Planck
simulation package \citep[Level-S,][]{reinecke-etal-2006}.

\subsection{Mission parameters}
We assumed a scanning strategy that caused the telescope's spin
axis to perform two cycloidal excursions from the ecliptic plane per year with
an amplitude of $7.5\degr$; in azimuthal direction, the spin axis always
pointed towards the Sun.

Time-ordered data were generated for a time span of 10\,000 hours,
corresponding to slightly more than two full sky surveys; their length is
of the order of $10^9$ samples.

\subsection{Detector geometry}
For the first set of experiments, data were simulated for two horns, i.e.\ two
pairs of detectors, which were modelled roughly following Planck's 30GHz
channel. In particular, the detectors had an average FWHM of 32\farcm 2,
the polarisation directions in each detector pair were rotated
by about (but not exactly) $90\degr$ against each other, and the polarisation
directions of the two horns were in turn shifted by $45\degr$.
Both horns followed the same scan path on the sky, although with a little
time delay. 

The beam shapes were assumed to be elliptical, with ellipticities of 1.7 and
1.8 for each detector in a pair, respectively. These values are considerably
higher than in a typical experiment and were chosen to demonstrate the
performance of the method even under unfavourable conditions.

In order to show the algorithm's performance at higher resolutions, another
data set was produced using two horns sensitive at 70GHz and with an
average FWHM of 13\arcmin.

Table \ref{table:beamparam} gives an overview over the detector properties.

\begin{table}
\caption{Overview of the detector parameters}
\begin{center}
\begin{tabular}{cccccc}
\hline\hline
Name\vphantom{\Large I} & Frequency & ellipticity & FWHM & \psipol & $\sigma_w$($\mu$K) \\
\hline
\hline
\vphantom{\Large I}$30_{0,0}$ & 30GHz & 1.7 & 32.2352' & $-22.3\degr$ & 514 \\
$30_{0,1}$ & 30GHz & 1.8 & 32.1377' & $\phantom{-}67.6\degr$ & 514\\
$30_{1,0}$ & 30GHz & 1.7 & 32.2352' & $\phantom{-}22.3\degr$ & 514 \\
$30_{1,1}$ & 30GHz & 1.8 & 32.1377' & $          -67.6\degr$ & 514 \\
$70_{0,0}$ & 70GHz & 1.7 & 13.0328' & $\phantom{-}22.2\degr$ & --\\
$70_{0,1}$ & 70GHz & 1.8 & 12.9916' & $          -67.5\degr$ & --\\
$70_{1,0}$ & 70GHz & 1.7 & 13.0328' & $          -22.2\degr$ & --\\
$70_{1,1}$ & 70GHz & 1.8 & 12.9916' & $\phantom{-}67.5\degr$ & --\\
\end{tabular}
\end{center}
\tablefoot{The first subscript denotes the horn, the second the number
of the detector inside the horn. $\sigma_w$ is the standard deviation of the
Gaussian random numbers that are added to each TOD sample in simulations
with white noise. Except for the ellipticities (which were exaggerated
to stress-test our algorithm), the values are
roughly comparable to those given in Tables 1 and 2 in
\cite{mandolesi-etal-2010}. The absolute value of the individual \psipol\ is
not relevant in the context of this paper, only their pairwise differences are.}
\label{table:beamparam}
\end{table}

\subsection{Input data}
To generate a theoretical power spectrum for the CMB emission, we used
the CAMB\fnurl{http://camb.info} code with the cosmological parameters
$\Omega_b=0.045$, $\Omega_{\text{cdm}}=0.255$ and $\Omega_\lambda=0.7$.
Using the HEALPix tool \texttt{syn\_alm\_cxx}, we created a random,
unconstrained realisation of polarised $a_{lm}$ from this power spectrum.

Emission maps of foreground components were generated using the Planck
pre-launch Sky model\fnurl{http://www.apc.univ-paris7.fr/~delabrou/PSM/psm.html} 
\citep[PSM,][]{delabrouille-etal-2012}, version 1.6.6. We chose to
include the contributions from galactic infrared emission
(galactic synchrotron radiation, free-free emission, as well as thermal and spinning dust).

The signal from strong point sources was taken into account as well;
to achieve this, a PSM-generated catalogue of point sources was directly
expanded into spherical harmonic coefficients, using a sufficiently high cutoff value
\lmax\ that the beam response was negligible at this
scale.

Polarised emission was disabled for all contributions except for the CMB, which
makes it very easy to detect leakage of the strong temperature signal in the
galactic plane into the polarisation signal.

We neglected bandpass effects and assumed an identical delta frequency
response for all detectors.

\begin{figure}
\centering
\includegraphics[width=0.9\columnwidth]{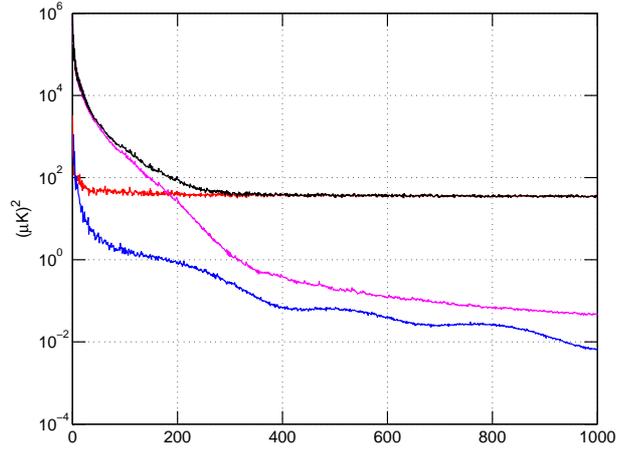}
\caption{Spectra of the 30GHz input sky components: CMB (\blue), diffuse foregrounds
(\magenta), point sources (\red), and total (\black).}
\label{fig:fig_cl_input}
\end{figure}

The power spectra of the signal components at 30 GHz are plotted in Fig.~\ref{fig:fig_cl_input}.

\subsection{Noise}
In some of our simulations, detector noise was added to the idealised sky
signal. Since the deconvolution method requires correlated noise to be
removed from the time-ordered data beforehand, we only added Gaussian white
noise with an appropriately chosen $\sigma$ to the simulated samples (see
Table \ref{table:beamparam}).

\subsection{3D map generation}
The mock TOD were processed with the \texttt{todtobin} tool to
generate the 3D input maps required by the deconvolver.
We set $N_\text{side}=1024$ for the 30GHz case, and
$N_\text{side}=2048$ for 70GHz; $n_\psi$ was always set to 256.

\section{Results}

\subsection{Constructing a sky map}

The primary output of the deconvolution code consists of the $a_{Xlm}$ (X=T,E,B) coefficients
of the sky, up to \lmax.
From these one can construct the usual temperature and Q,U polarisation maps
through harmonic expansion of the coefficients.
We did this with the \texttt{alm2map\_cxx} tool of the HEALPix package.

For comparison, we also produced binned maps directly from the TOD.
A binned map is constructed by assigning each TOD sample completely to the pixel
containing the centre of the beam. The binning operation can be formally written as
\begin{equation}
  m=(P^TC_n^{-1}P)^{-1}P^TC_n^{-1} y.
\end{equation}
Here $P$ is a pointing matrix, $C_n$ is white noise covariance, and $m$ is a (I,Q,U) map triplet.

Binned maps are distorted by beam effects if the beam is asymmetric.
In particular, the polarisation component
of a binned map is contaminated by temperature leakage through beam shape mismatch.
We aim at demonstrating that deconvolution can reduce these undesired effects.

Since our simulation includes point sources, the input $a_{Xlm}$ coefficients are not band-limited.
However, we can only solve the coefficients up to some limited \lmax.
This leads to ringing artefacts, if a map is constructed directly from the raw $a_{Xlm}$ coefficients.
Ringing is particularly prominent around point sources and other sharp features. 

Ringing is caused by the sharp cut-off of the spectrum at \lmax, and it occurs even if the
deconvolved coefficients are fully correct up to \lmax.
For this reason it is necessary to smooth the coefficients with a symmetric Gaussian beam
prior to constructing a map.
The required amount of smoothing depends on the spectrum of the underlying sky,
and on \lmax. The highest \lmax\ that one can solve in turn depends on the beam width.
In other words, it is not possible to recover structures significantly smaller than the beam itself.
This gives the following rule of thumb: the required smoothing width is given by the smallest
features of the original beam.

\begin{figure}
\centering
\includegraphics[width=0.48\columnwidth]{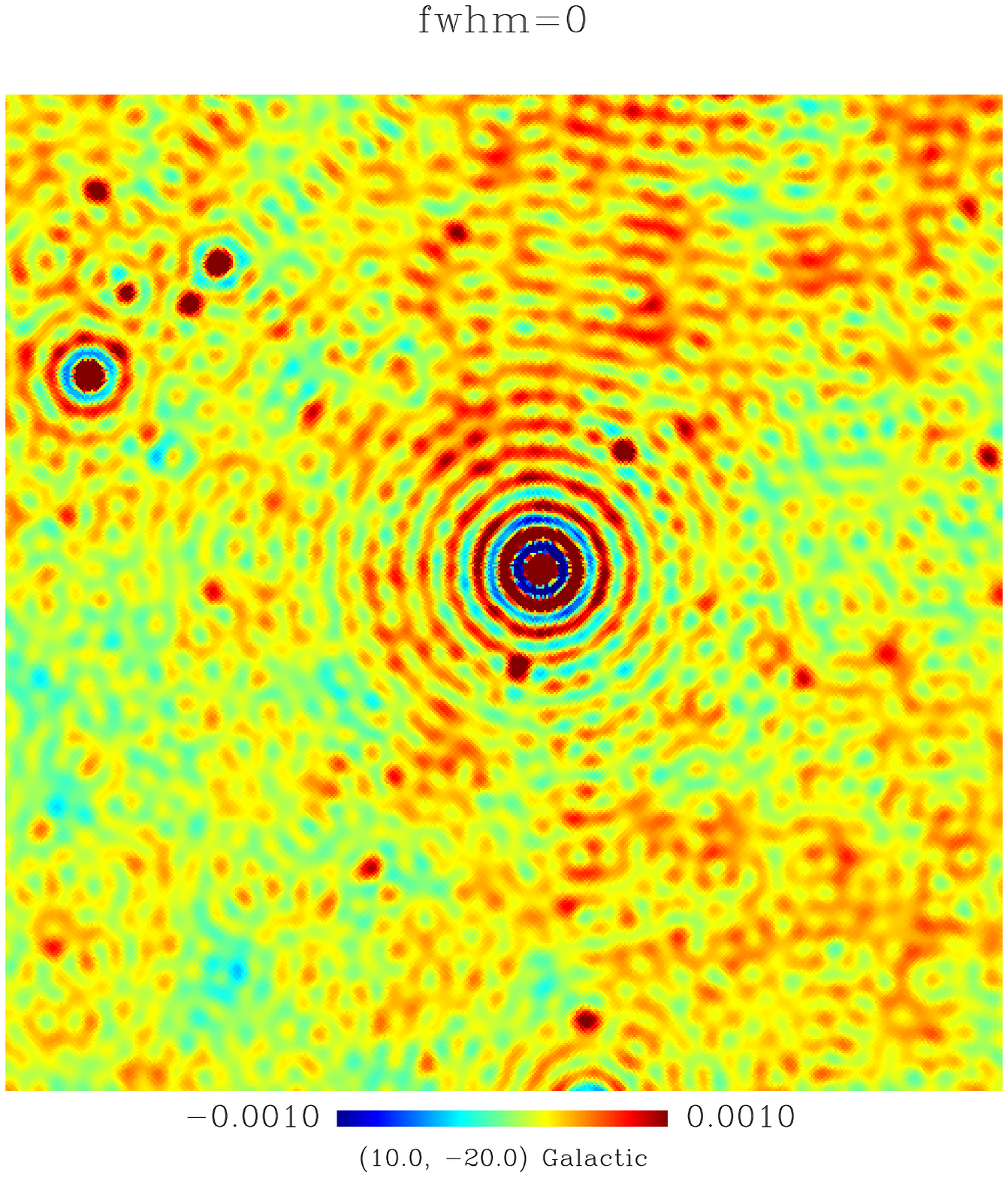}
\includegraphics[width=0.48\columnwidth]{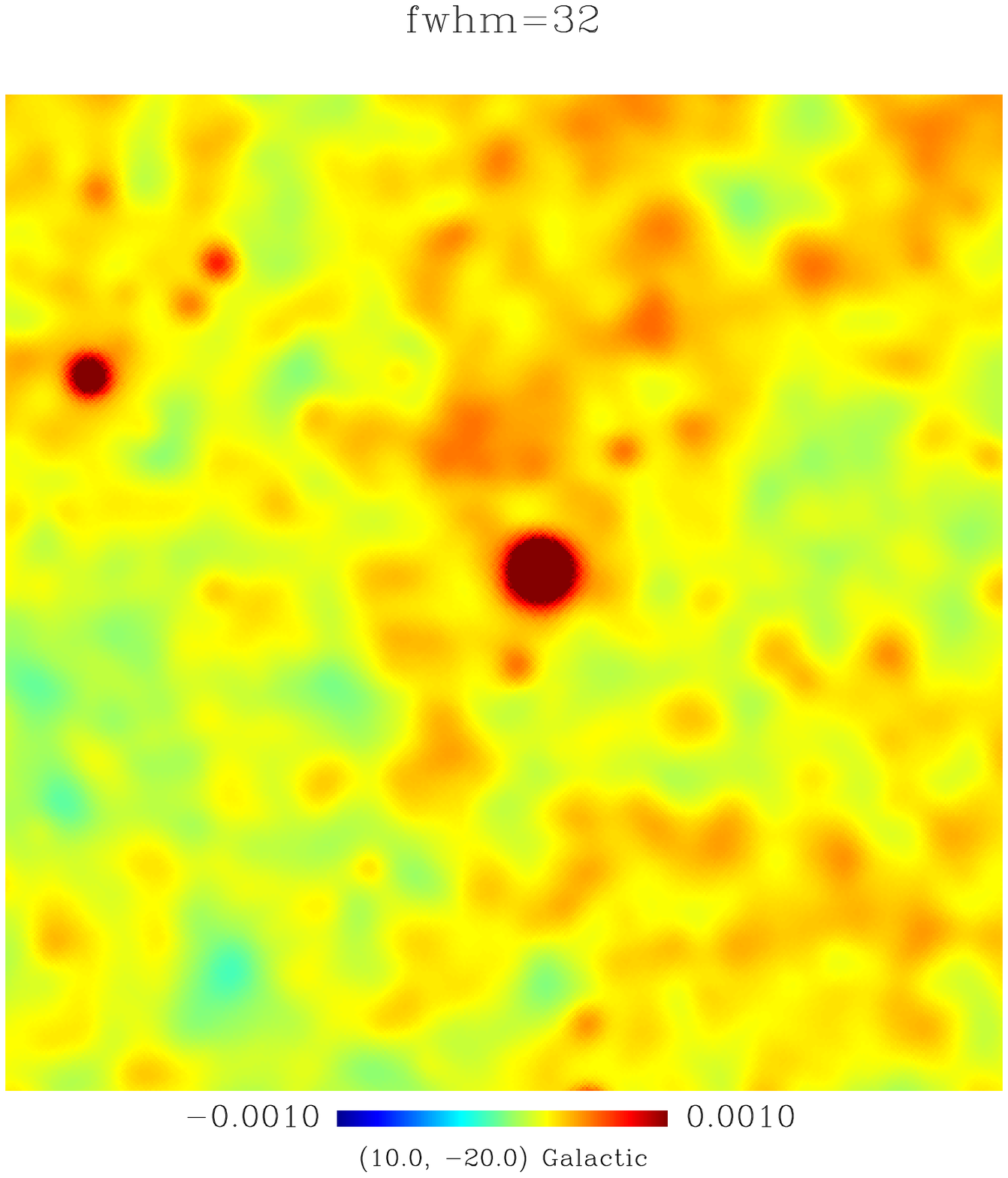}
\caption{Ringing due to insufficient smoothing.
A map constructed from the input $a_{Tlm}$ coefficients with \lmax=800 
with no smoothing (\fleft) or smoothed to FWHM=32\arcmin\ (\fright).
The cut-off of the spectrum at \lmax\ leads to strong ringing effects around a point source.
}
\label{fig:ringing}
\end{figure}

In Fig.~\ref{fig:ringing} we demonstrate  the ringing around a point source.
We show an image of a point source, constructed from the input $a_{Xlm}$, but including only
coefficients up to \lmax=800. Ringing artefacts are evident.
On the right we show the same region of the sky after the map is smoothed with a FWHM=32\arcmin\ Gaussian
symmetric beam. Smoothing removes the ringing, but naturally also smoothes out
the smallest structures in the map.

\subsection{Signal-only simulations with elliptic beam}
\label{run30}

We analyse first the noise-free 30GHz simulation in detail.
A detailed description of the simulation was given in Section \ref{sim_setup}.
We ran the code with various combinations of parameters \lmax, \kmax,
and analysed the results both in pixel space and in harmonic space.
We included data from all four detectors and considered both temperature and polarisation.

\subsubsection{Temperature map}

\begin{figure}
\centering
\includegraphics[height=0.9\columnwidth,angle=90]{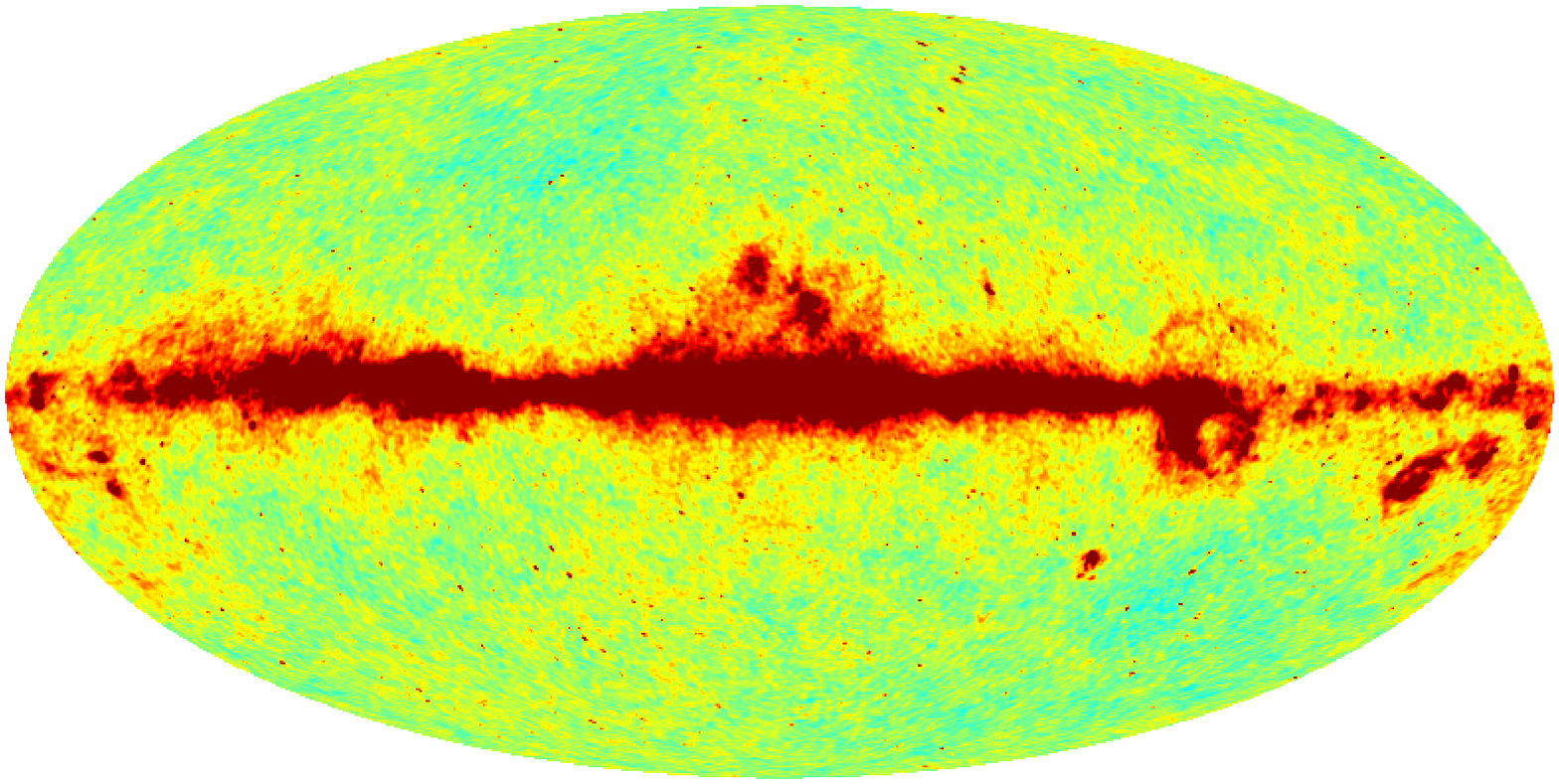}
\includegraphics[height=0.9\columnwidth,angle=90]{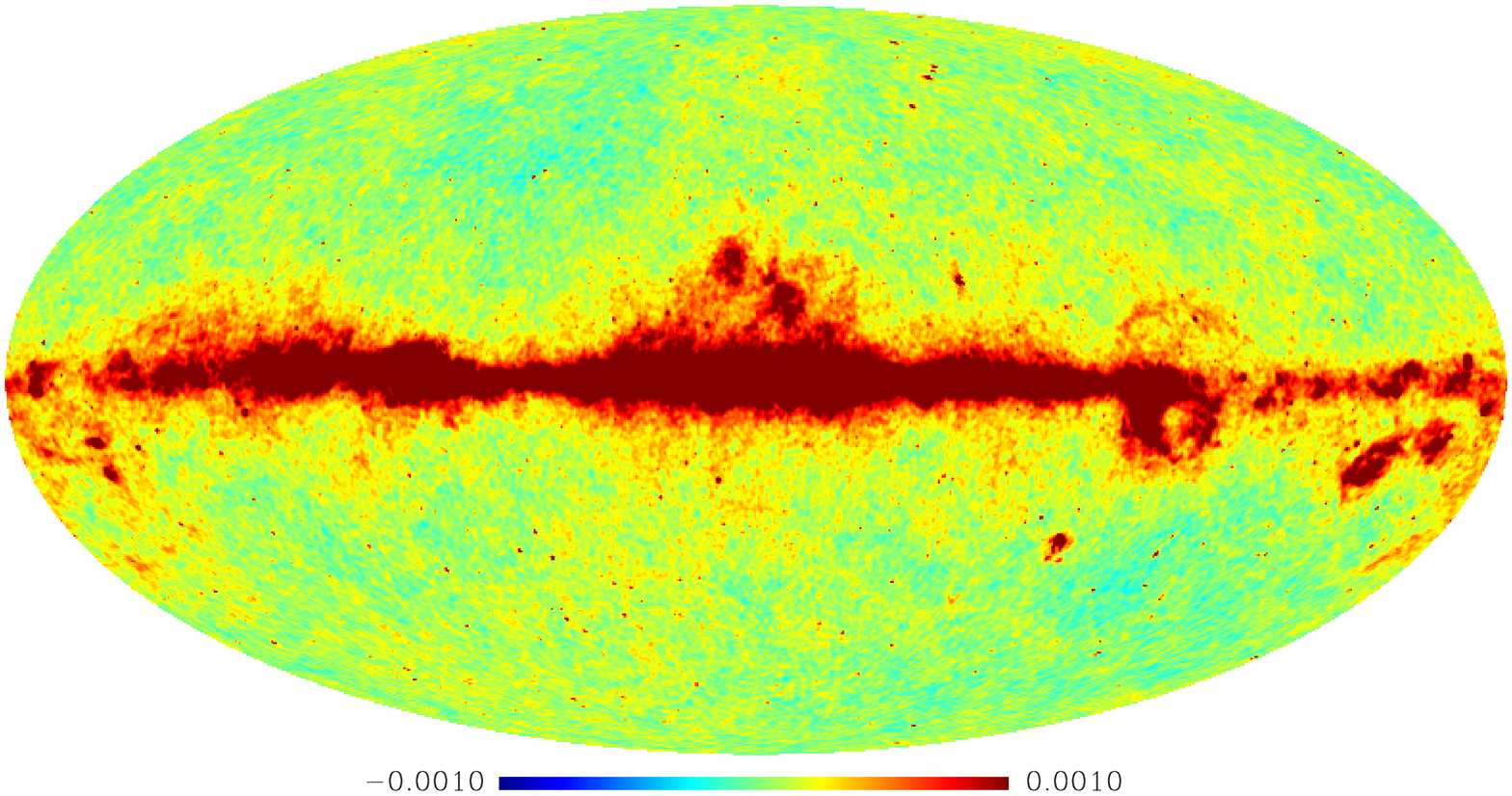}
\caption{30GHz temperature map without noise: binned (\ftop), and deconvolved and FWHM=32\arcmin\ smoothed (\fbottom).}
\label{fig:moll_signal_T}
\end{figure}

In Fig.~\ref{fig:moll_signal_T}
we show a full-sky Mollweide projection of the binned temperature map
together with that of the deconvolved map.
We used deconvolution parameters \lmax=800, \kmax=6.

We chose the value FWHM=32\arcmin\ as the baseline smoothing width for the 30GHz maps.
This corresponds to the mean width of the input beam.
In some cases it was necessary to apply even stronger smoothing.

\begin{figure}
\centering
\includegraphics[width=0.49\columnwidth]{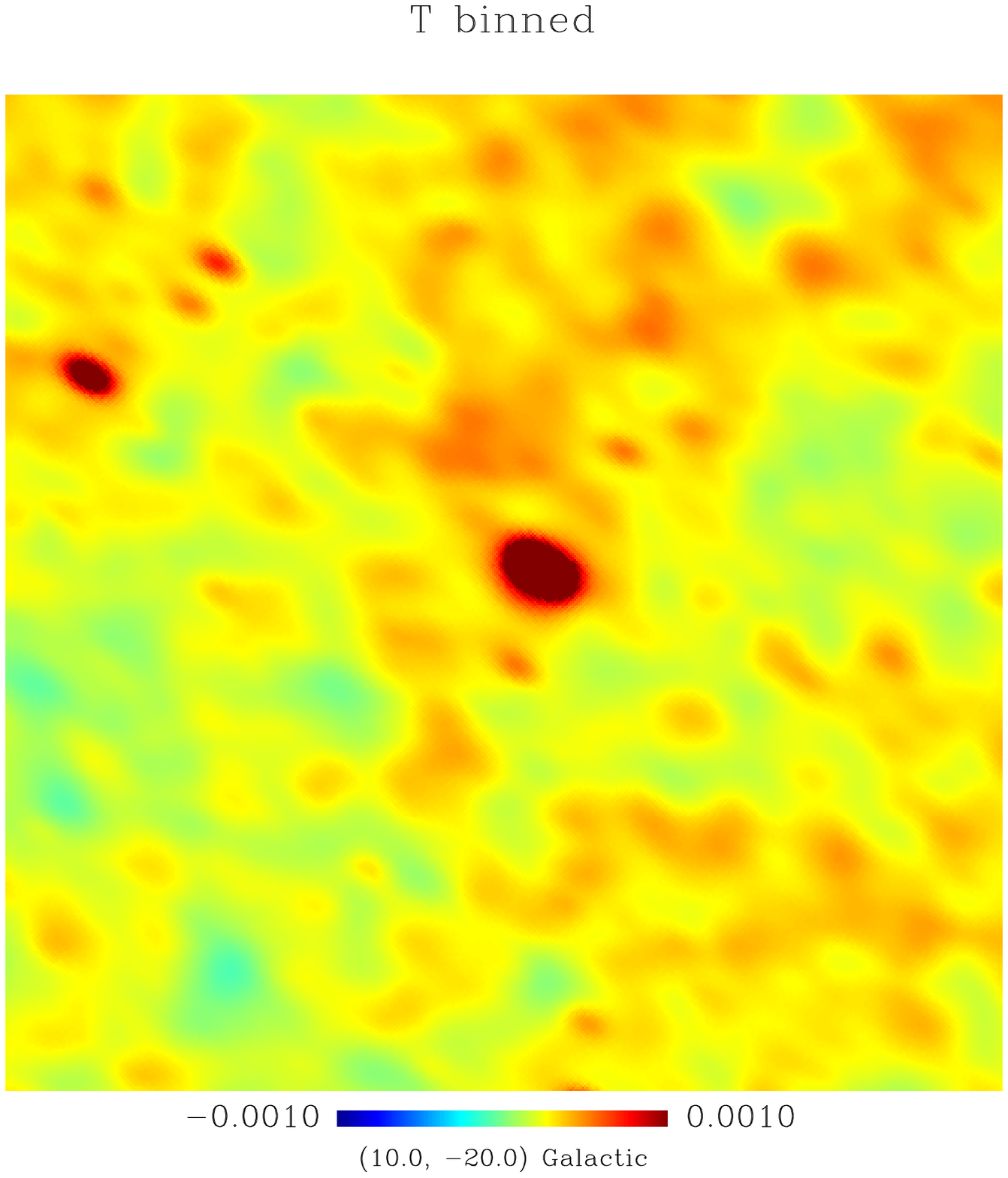}
\includegraphics[width=0.49\columnwidth]{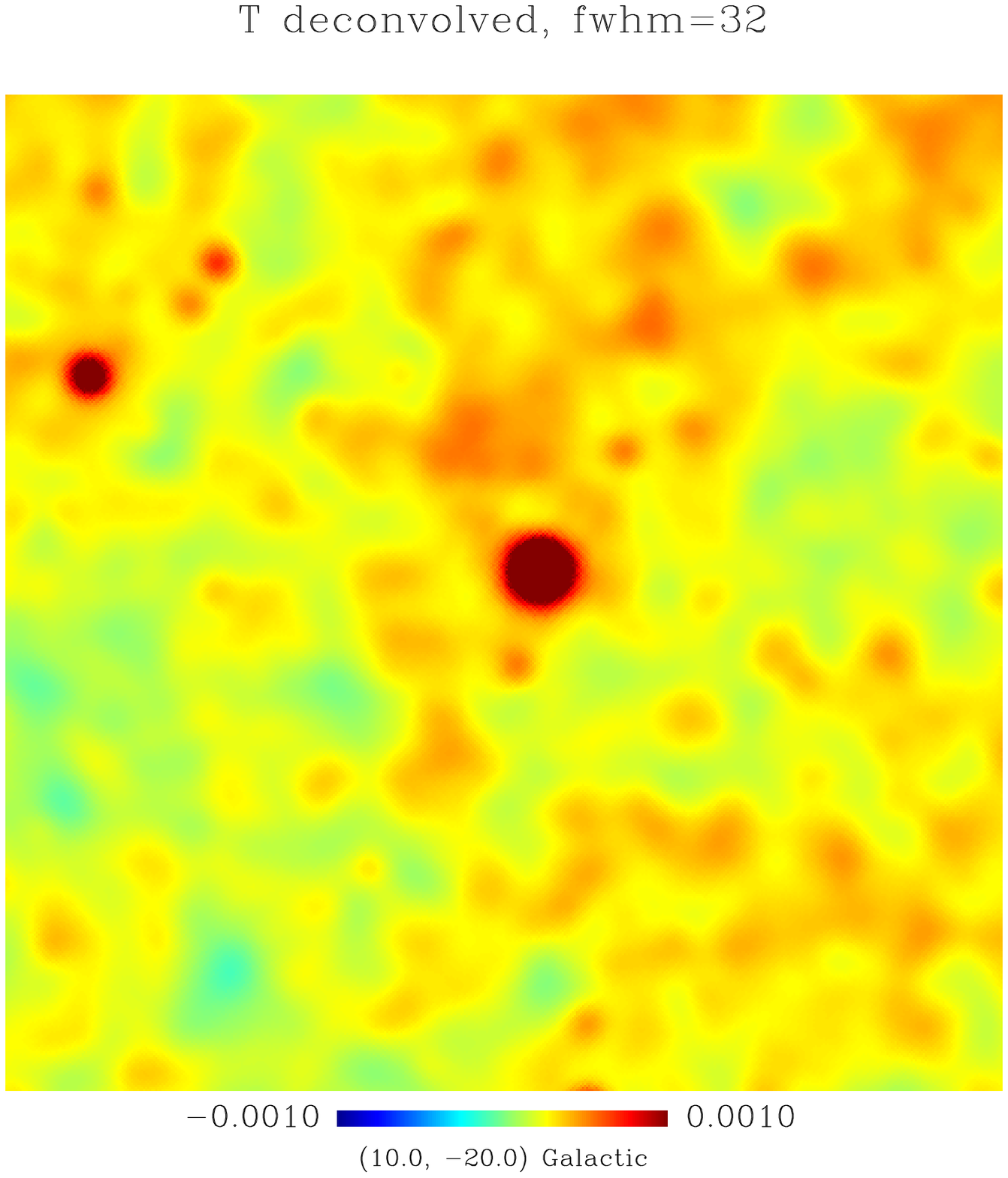}
\caption{Zoom into the 30GHz temperature map without noise: binned (\fleft) and deconvolved  (\fright).
Shown is a 1000\arcmin x1000\arcmin\ patch of the sky.
The deconvolved map can be compared to the map constructed from the input $a_{Tlm}$ (right panel of Fig.~\ref{fig:ringing}).}
\label{fig:gnom_signal_T}
\end{figure}

The difference between the binned and the deconvolved map does not appear dramatic
in the full-sky map image,
but becomes evident when we zoom into a point source.
Figure \ref{fig:gnom_signal_T} shows a randomly selected strong point source
below the galactic plane. While the source in the binned map is clearly
elongated due to beam shape, it appears circular in the deconvolved map.

The deconvolved image can be compared to the right panel of Fig.~\ref{fig:ringing},
where we have shown the same patch of the sky constructed from the input $a_{Xlm}$
coefficients. The maps are nearly identical.

\subsubsection{Polarisation maps}

\begin{figure}
\centering
\includegraphics[height=0.9\columnwidth,angle=90]{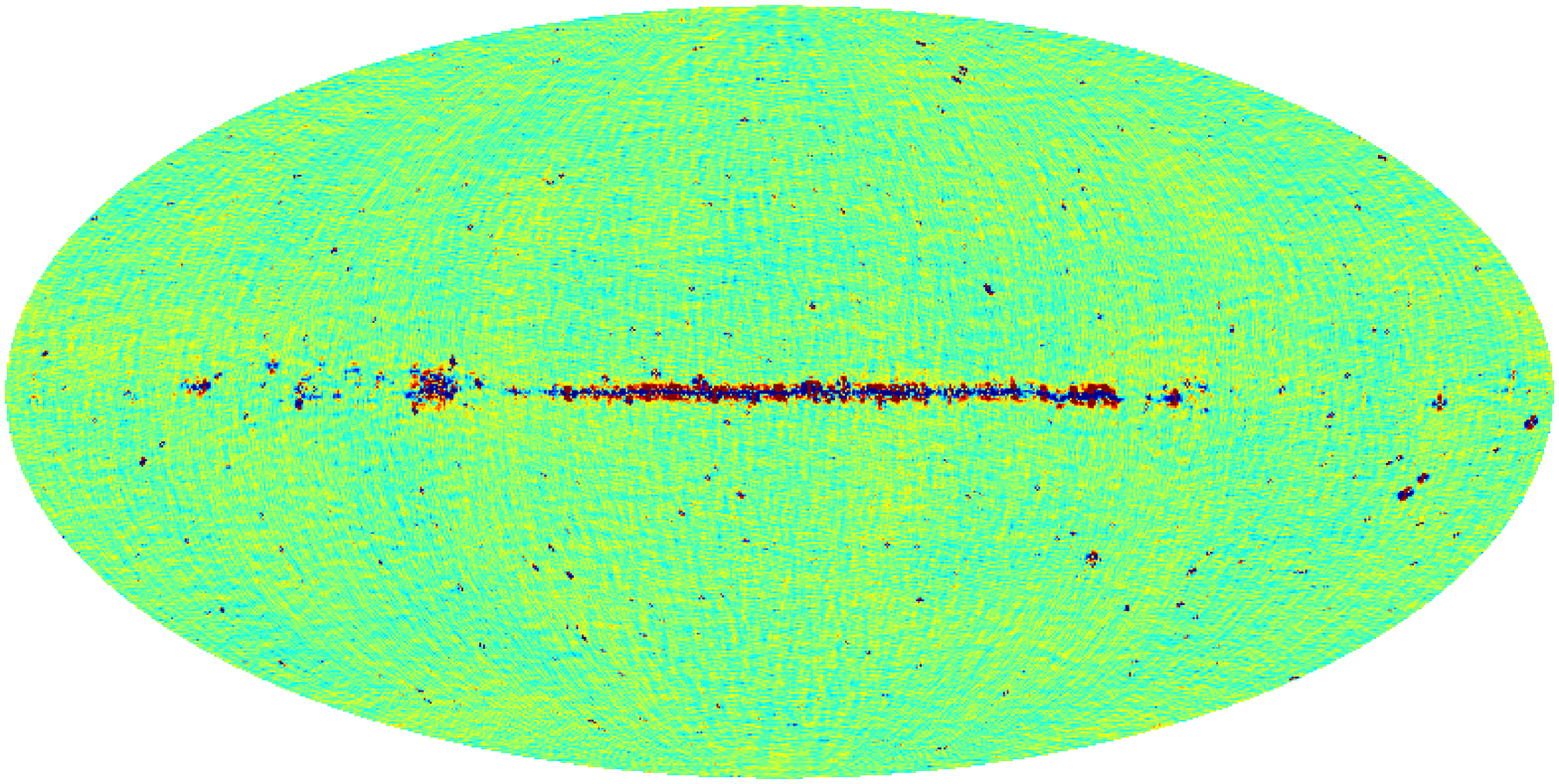}
\includegraphics[height=0.9\columnwidth,angle=90]{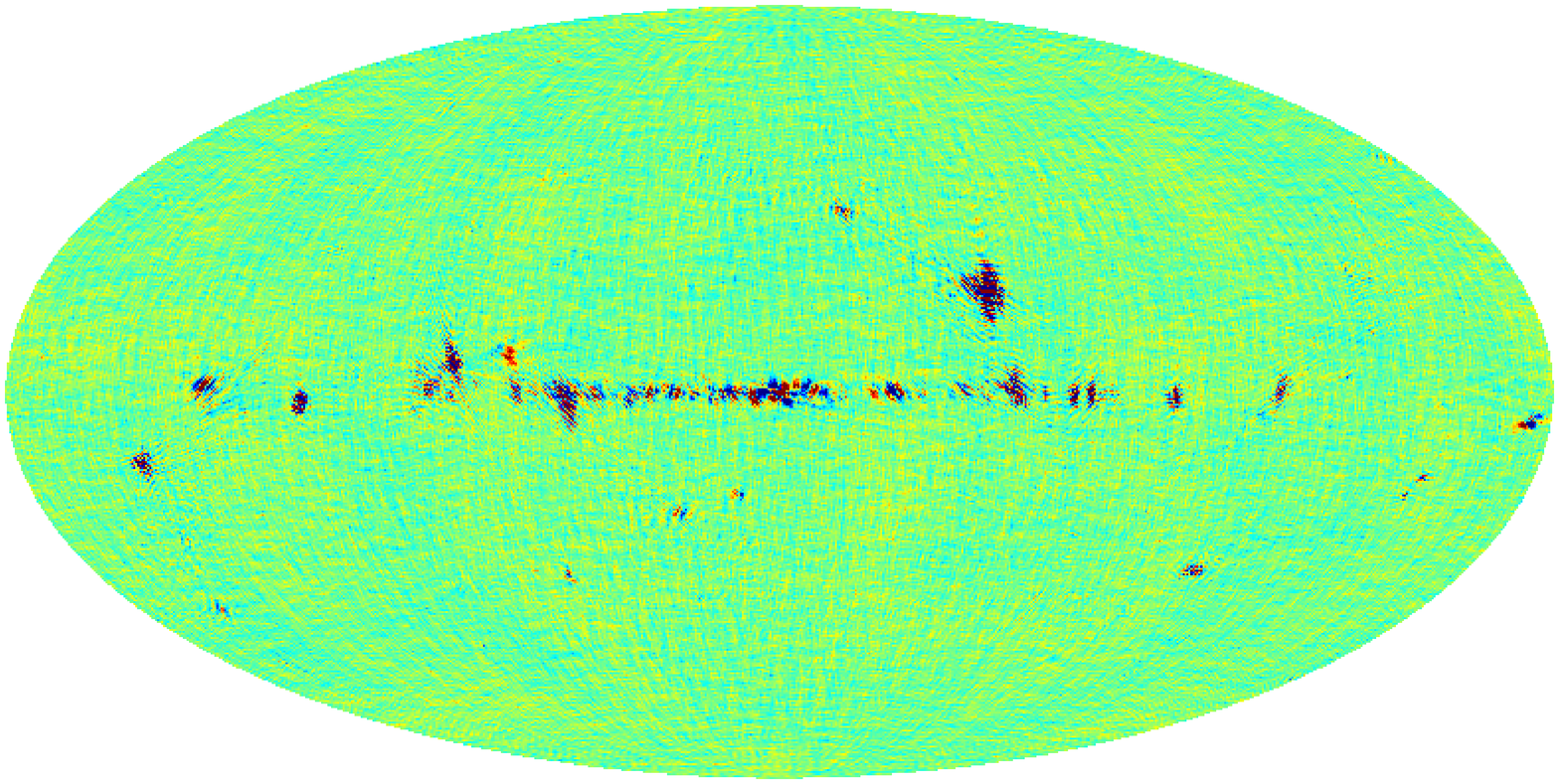}
\includegraphics[height=0.9\columnwidth,angle=90]{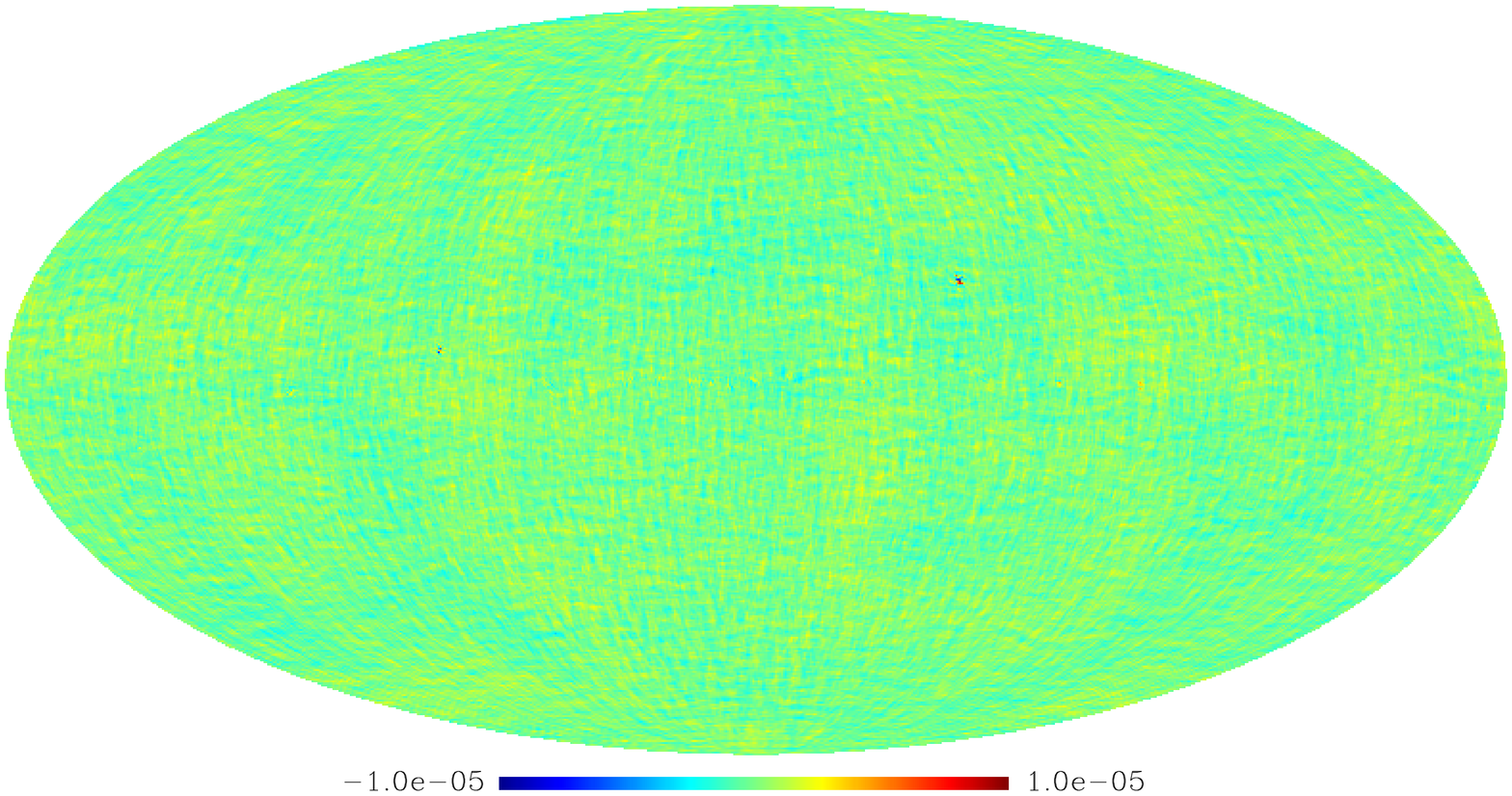}
\caption{30GHz Q polarisation maps, in absence of noise: binned (\ftop), and deconvolved with FWHM=32\arcmin\ (\fmiddle),
or FWHM=45\arcmin\ (\fbottom) smoothing.
The whole galactic structure is a fake polarisation signal arising from beam mismatch.}
\label{fig:moll_signal_Q}
\end{figure}

The beam effects show up more dramatically in the polarisation maps
(Fig.~\ref{fig:moll_signal_Q}).
The temperature signals recorded by two detectors differ due to the different beam shapes.
The binning procedure interprets this erroneously as polarisation signal.
This leads to a leakage of temperature signal into the polarisation maps.
Since our simulation did not include polarised foregrounds, the whole galactic structure
seen in the binned map is temperature leakage through beam mismatch.

We show deconvolved maps with two levels of smoothing below the binned map.
Smoothing with a FWHM=32\arcmin\ beam is not sufficient to fully remove the galactic
residual, but if we smooth the map further to FWHM=45\arcmin, the residual disappears
almost completely.

\begin{figure}
\centering
\includegraphics[width=0.49\columnwidth]{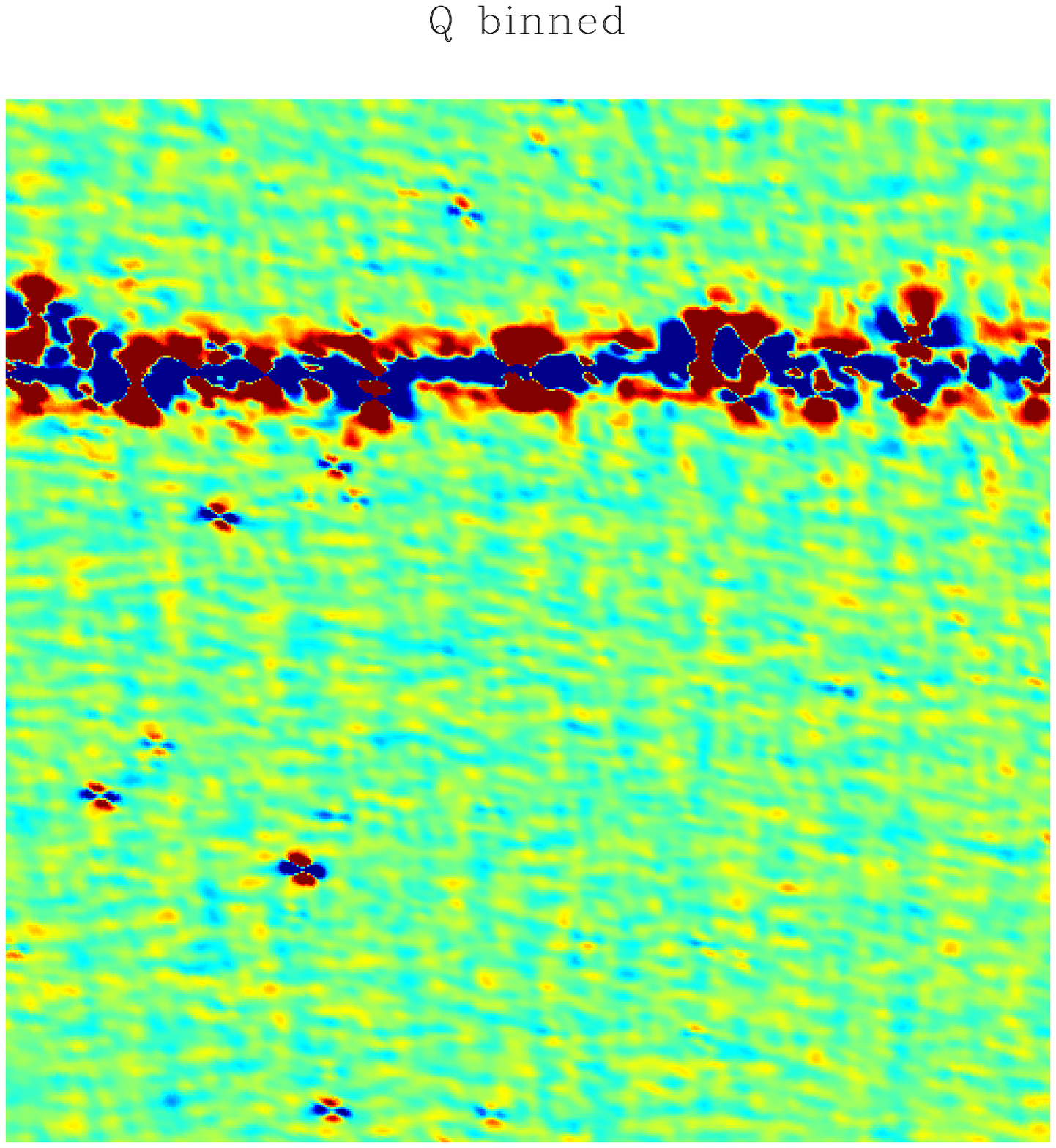}
\includegraphics[width=0.49\columnwidth]{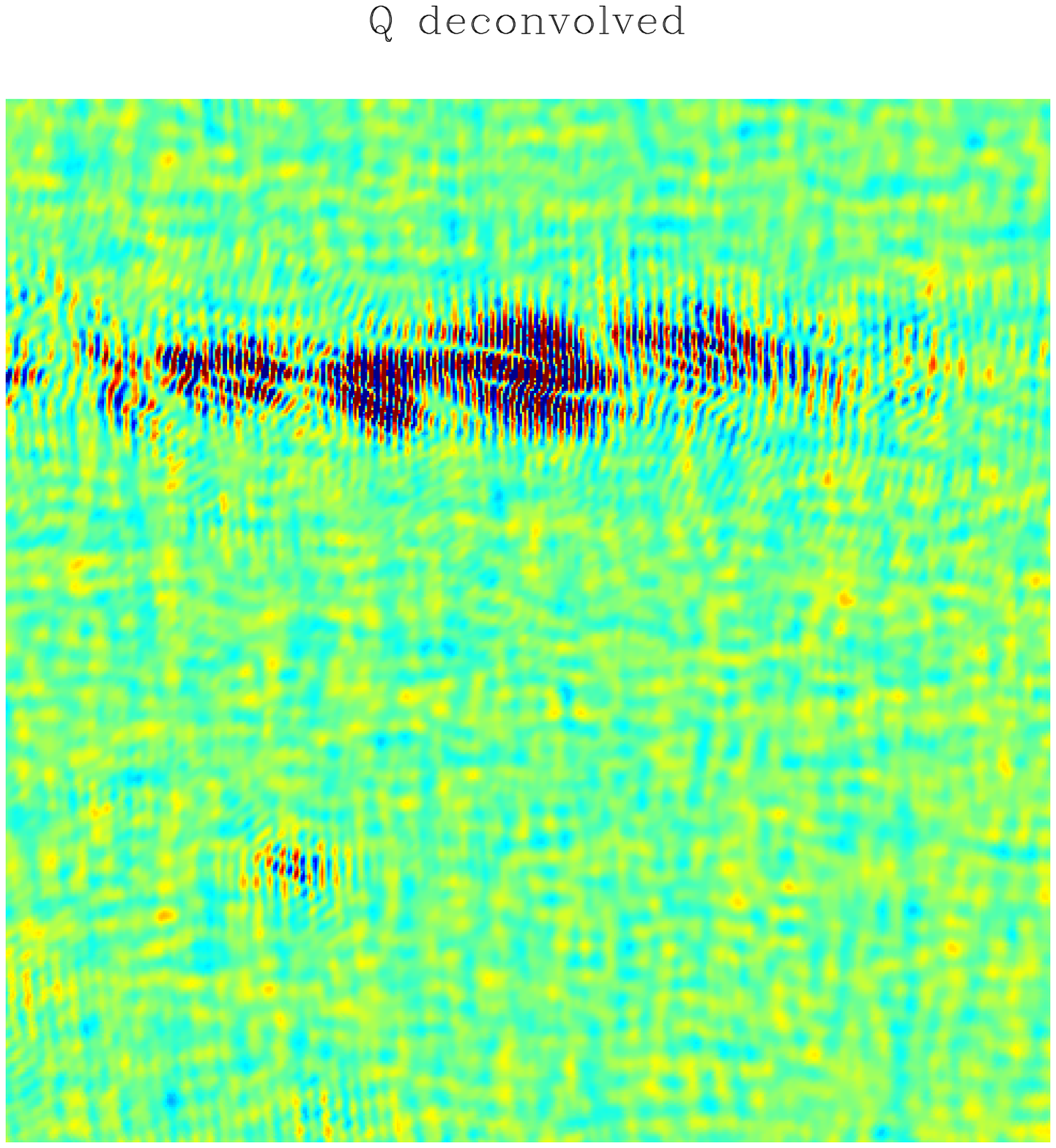}
\includegraphics[width=0.49\columnwidth]{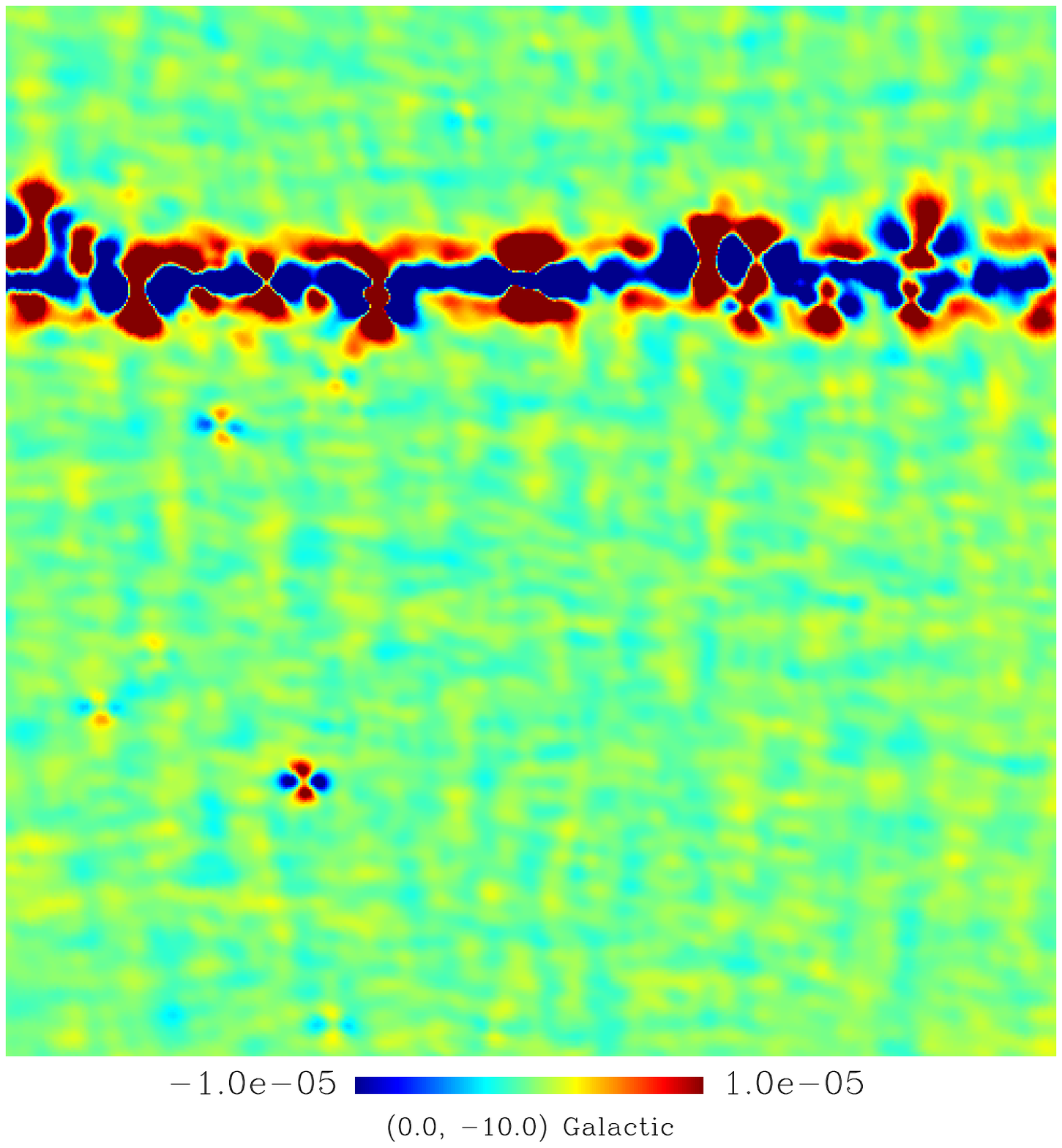}
\includegraphics[width=0.49\columnwidth]{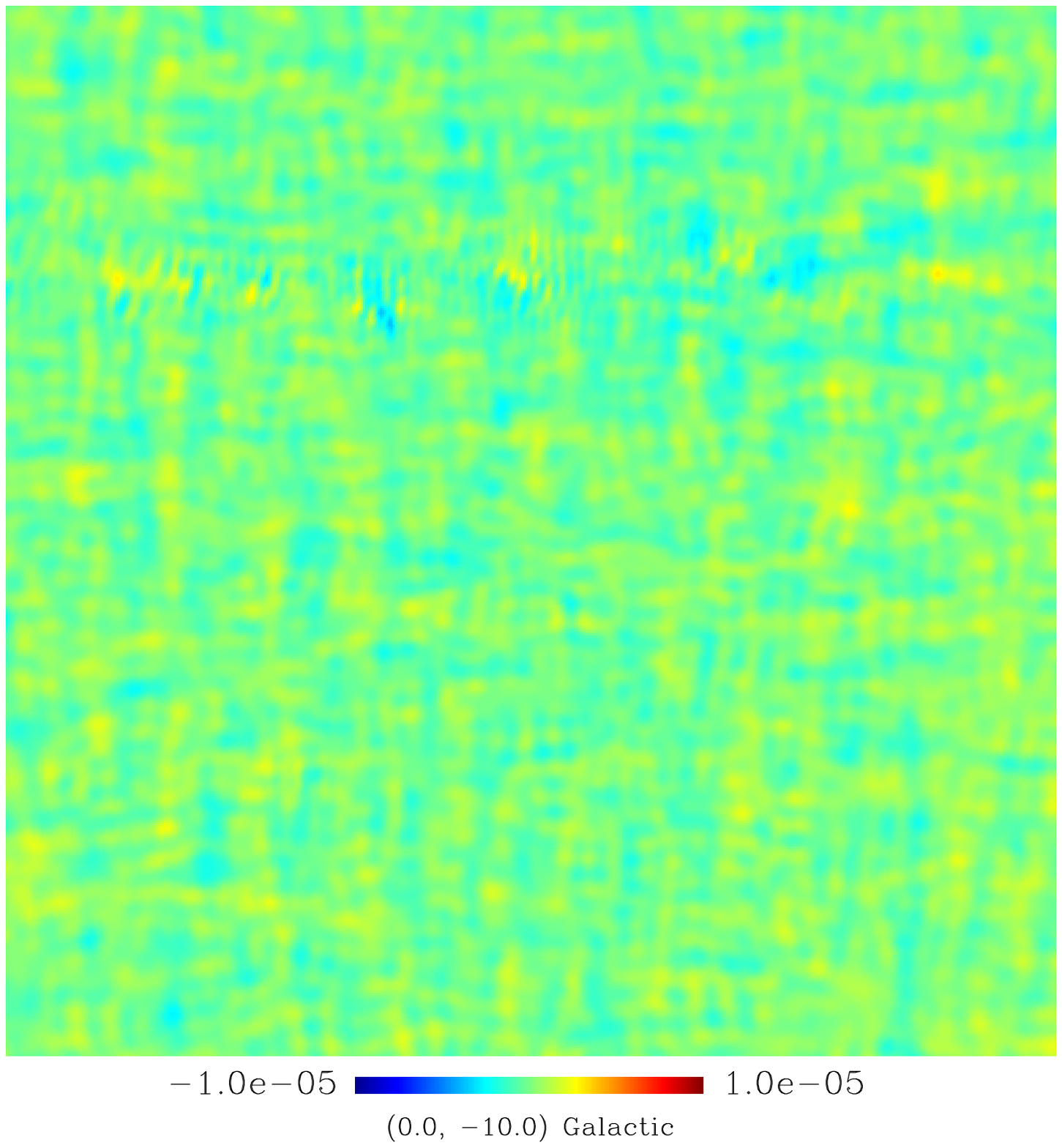}
\caption{Zoom into the 30GHz Q polarisation map without noise: binned (\fleft) and deconvolved (\fright).
The maps were smoothed to resolution FWHM=32\arcmin\ (\ftop), or to 45\arcmin\ (\fbottom).
We show a 2500\arcmin x2500\arcmin\ patch of the sky.
The horizontal structure is spurious signal due to temperature leakage from the galaxy.
}
\label{fig:gnom_signal_Q}
\end{figure}

In Fig.~\ref{fig:gnom_signal_Q}
we show a zoom into the Q polarisation map, just below the galactic plane.
In the binned map we see a strong galactic residual,
and typical ``four-leaf clover'' structures that arise from temperature leakage
at the location of a point source.
We show the deconvolved map again for two levels of smoothing: FWHM=32\arcmin\ and
FWHM=45\arcmin.
The additional smoothing removes the clover structure nearly completely. 
To show that this is not only an effect of more aggressive smoothing,
we smoothed the binned map further
with a Gaussian beam of width FWHM=32\arcmin, which, combined with the 32\arcmin\ of the original beam,
gives a total smoothing of 45\arcmin. This map is shown below the unsmoothed binned map.
The additional smoothing has little effect on the galactic residual.

\subsubsection{Harmonic space}

\begin{figure}
\centering
\includegraphics[width=0.9\columnwidth]{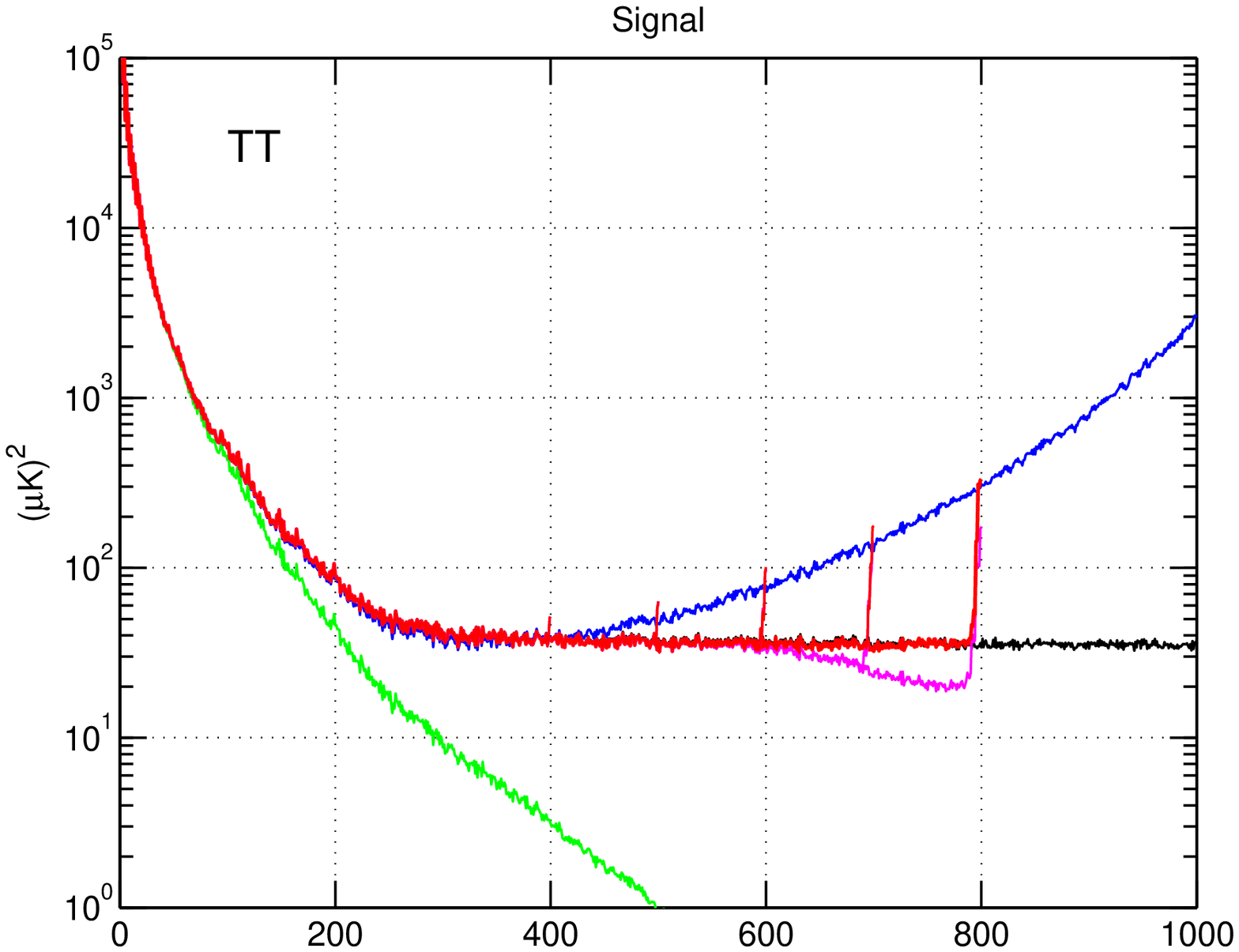}
\includegraphics[width=0.9\columnwidth]{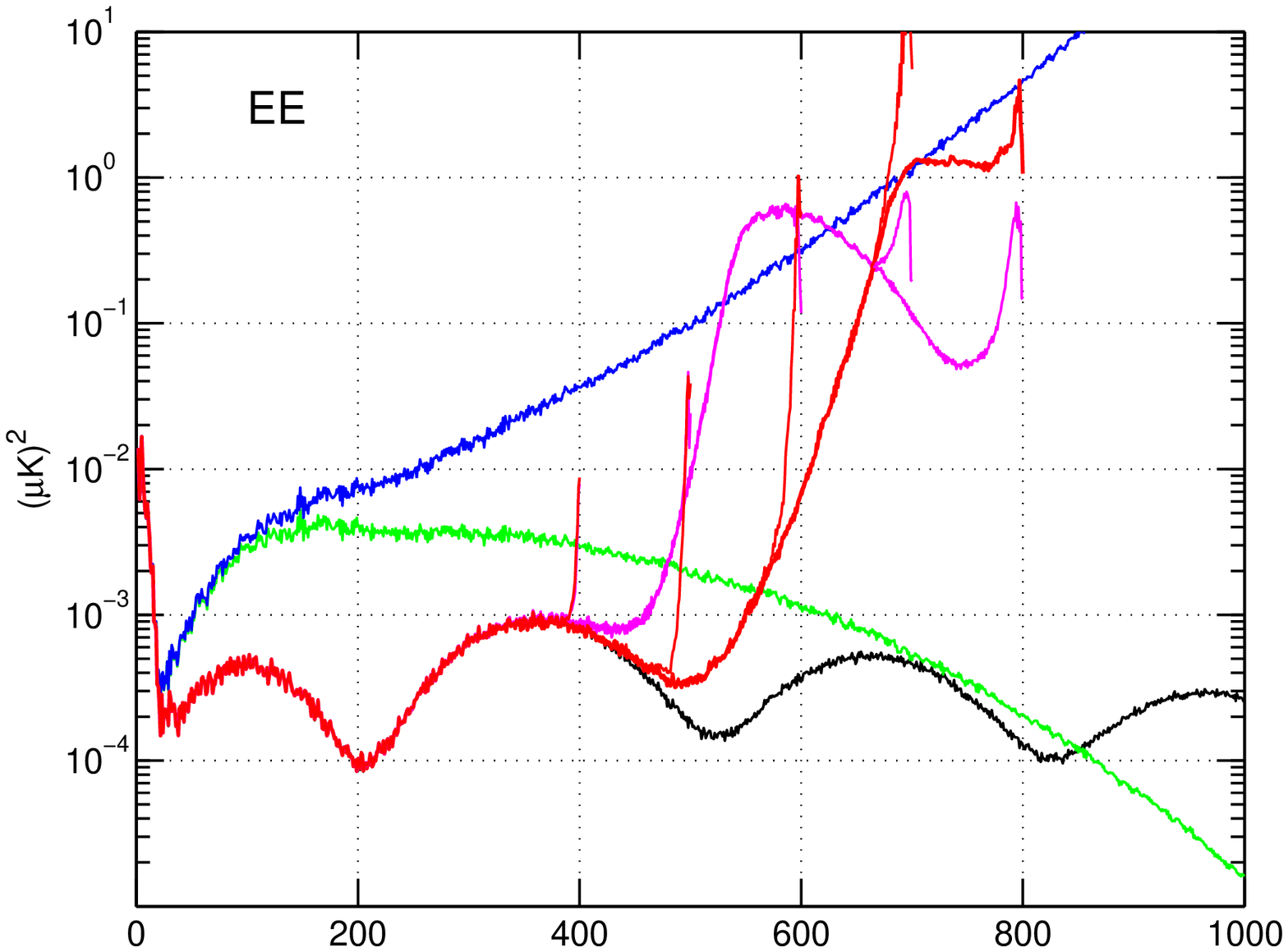}
\includegraphics[width=0.9\columnwidth]{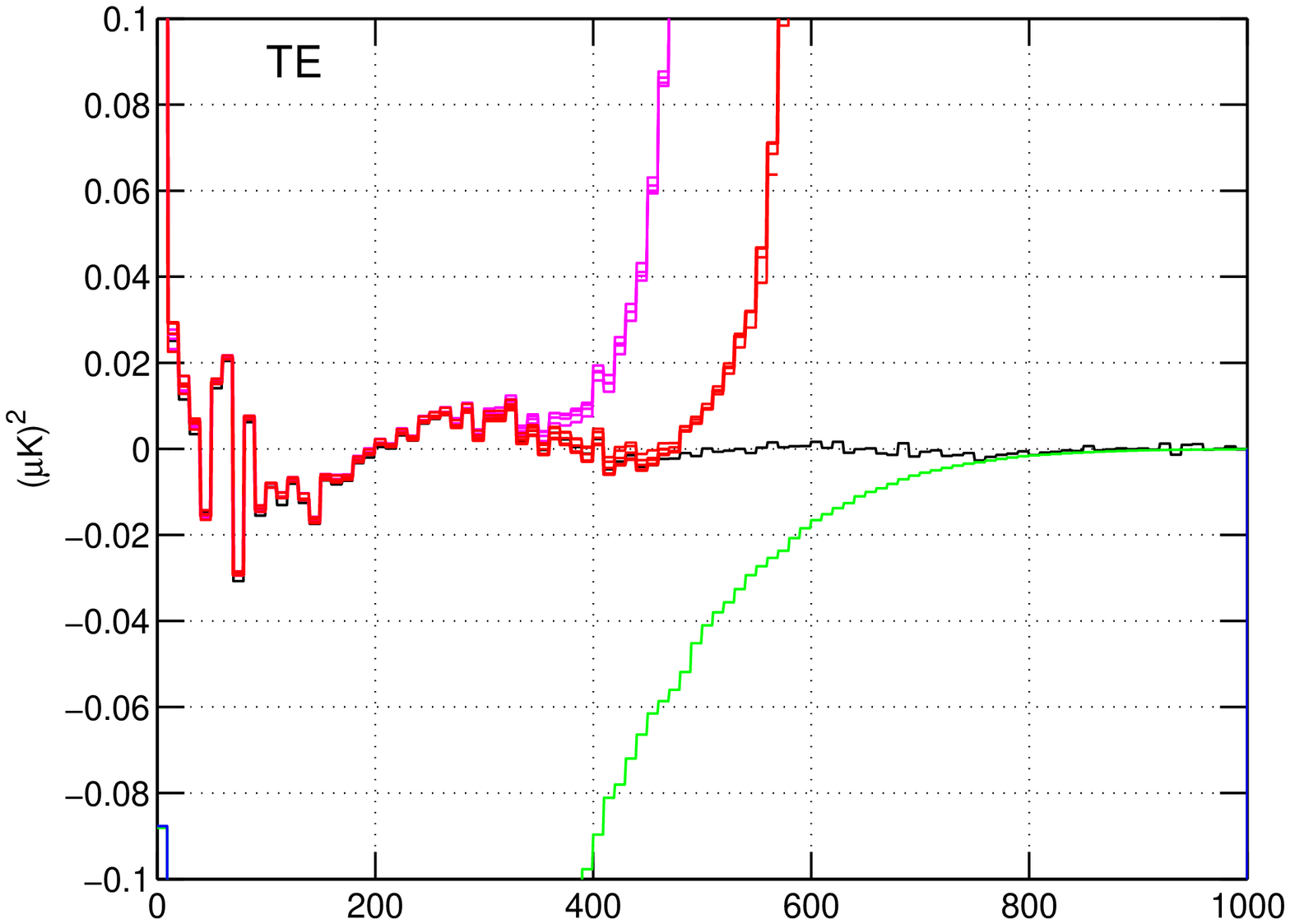}
\caption{Noise-free 30GHz TT (\ftop), EE (\fmiddle), and TE (\fbottom) spectra.
Shown are the input (\black)
and the spectra constructed from the $a_{Xlm}$
coefficients from deconvolution.
Deconvolution results are shown for \kmax=4 (\magenta) and \kmax=6 (\red),
and for \lmax=400, 500, 600, 700, 800.
The best-case results (\lmax=800, \kmax=6) are plotted with a thicker linetype.
For comparison we show also the spectrum obtained from harmonic expansion of
a naive binned map (\green).
The spectrum of the binned map decreases rapidly with $l$ because the binning procedure
inherently produces a beam-smoothed map.
The \blue\ line (TT and EE) is obtained by dividing the latter by the window function of
a symmetric Gaussian beam with FWHM=32\arcmin.
The TE spectra have been binned over ten multipoles to reduce scatter and to
make the plot more readable.}
\label{fig:fig_cl_signal}
\end{figure}

We proceed to analyse the deconvolution results in harmonic space.
We computed the TT, EE, and TE spectra of the deconvolved $a_{Xlm}$ coefficients
and compared them to the corresponding input spectra. The spectrum is constructed as
$$
   C^{XY}_{l}=\frac{1}{2l+1}\sum_{m=-l}^la_{Xlm}^\ast a_{Ylm}\text{,}
$$
where $X$ and $Y$ stand for $T$ and $E$.

We ran the code for all combinations of \kmax=4,6 and \lmax=400,500,600,700,800.
Results for different combinations are shown in
Fig.~\ref{fig:fig_cl_signal}.
We applied no smoothing to the coefficients.

We observe that the optimal value for \kmax\ is coupled to \lmax.
Results for \kmax=4 and \kmax=6 begin to diverge only above $l=400$.
The results for different \lmax\ with \kmax=6 are practically on top of each other,
except for the highest multipoles near \lmax, which are clearly distorted.
When constructing a map, we cut out the last 20 multipoles before making
the harmonic expansion.

For comparison we also show the spectrum obtained from a harmonic expansion
of the binned map, which we computed using the \texttt{anafast} tool of the HEALPix package.
This decreases rapidly as a function of $l$ as a result of beam smoothing.
To correct for the smoothing, we divided the spectrum by the window
function of a symmetric Gaussian beam
with FWHM=32\arcmin. This corrects for the mean smoothing and allows one to recover
the correct spectrum up to \lmax=300.
Above that, the result begins to deviate from the input spectrum due to beam ellipticity,
which cannot be corrected for by application of a window function only.

Deconvolution recovers the TT input spectrum well up to $l$=800.
At higher multipoles there is little information left in the signal,
and the convergence properties of the code become poor.
The weaker EE and TE spectra are recovered up to $l$=400.
Owing to temperature leakage, the binned map
only gives the correct EE spectrum for the ten first multipoles in EE,
and the TE spectrum is useless.

We did one run with parameters \lmax=600, \kmax=8 (not shown), to see if it would help to recover
the EE spectrum to even higher multipoles. The results, however, did not differ significantly
from the \lmax=600, \kmax=6 case.

\subsubsection{Reconstruction error of harmonic coefficients}

A comparison between the reconstructed spectrum and the input spectrum
is not a fully reliable test of the correctness of the deconvolution result.
It is possible to construct a set of harmonic coefficients that provides the correct spectrum,
although the individual coefficients are incorrect.

To assess directly the accuracy of the $a_{Xlm}$ coefficients,
we constructed a quantity we call the error spectrum. 
It is computed as follows.
We take the difference between the $a_{Xlm}$ coefficients we obtain from deconvolution and
the corresponding input coefficients $a^{0}_{Xlm}$. As before, $X$ stands for T or E.
We compute the spectrum of this difference as
$$
   C^{err}_{Xl}=\frac{1}{2l+1}\sum_{m=-l}^l(a_{Xlm}-a^{0}_{Xlm})^\ast(a_{Xlm}-a^{0}_{Xlm}) \text{.}
$$
The resulting spectrum is a measure of the error in the reconstruction of the $a_{Xlm}$ coefficients,
as a function of scale.

\begin{figure}
\centering
\includegraphics[width=0.9\columnwidth]{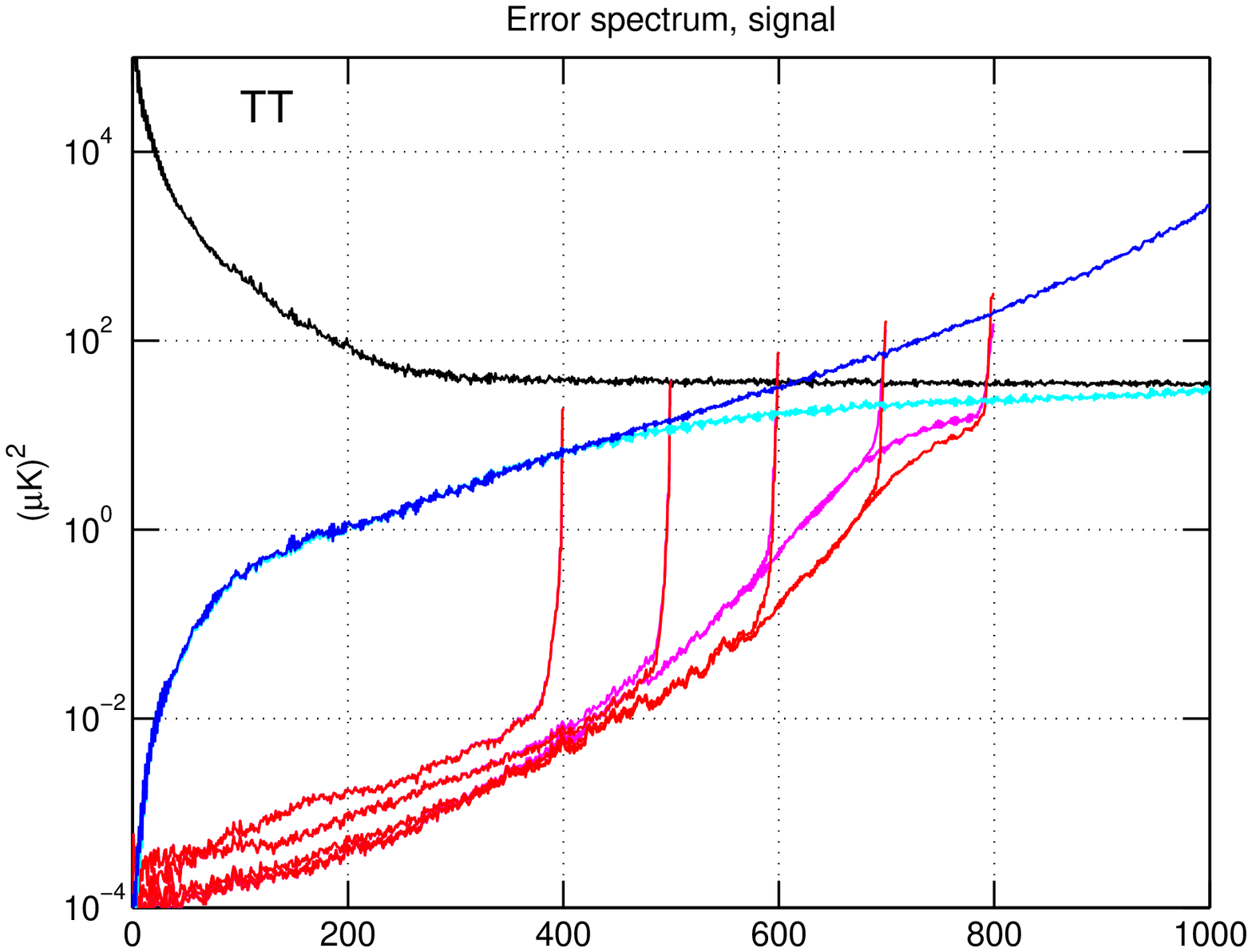}
\includegraphics[width=0.9\columnwidth]{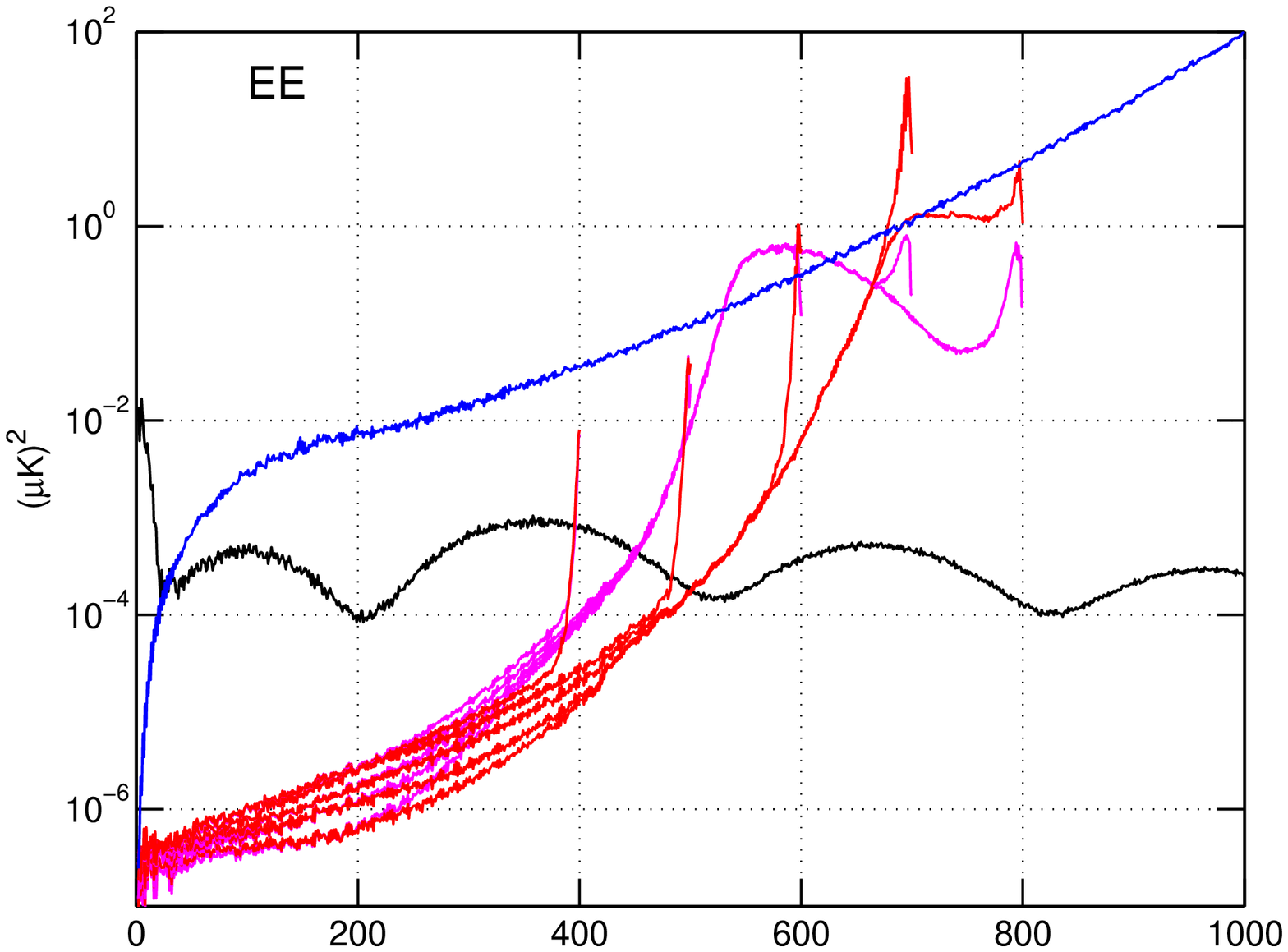}
\caption{30GHz error spectrum, for T (\ftop) and E (\fbottom), without noise. See main text for definition.
The deconvolution results are shown for the same combinations of parameters as in Fig.~\ref{fig:fig_cl_signal}.
The input spectrum is shown in \black.
Also shown is the error spectrum for a harmonic expansion of the binned map,
corrected with a window function of a symmetric Gaussian beam (\blue),
or (T only) with an ideal window function (\emph{light blue}).
The ideal window function was defined as the ratio of the input spectrum and the spectrum 
of the uncorrected binned map.
}
\label{fig:errorspec_signal}
\end{figure}

We show the error spectra of the 30GHz runs for T and E coefficients in Fig.~\ref{fig:errorspec_signal}.
For comparison, we show the input TT or EE spectrum in the same figure.
The $a_{Xlm}$ coefficients are recovered with good accuracy when the error spectrum
is well below the input spectrum.
The meaning of the blue line is the same as in Fig.~\ref{fig:fig_cl_signal}, that is,
we compute the harmonic expansion of the binned 
map and divide them by the window function of a symmetric Gaussian beam (FWHM=32\arcmin).

We made another comparison with an ideal window function,
which we computed as the ratio of the uncorrected spectrum of the binned map, and the input spectrum 
(green and black lines in the top panel of Fig.~\ref{fig:fig_cl_signal}).
The error spectrum for this is shown in light blue in Fig.~\ref{fig:errorspec_signal} (TT spectrum only).
By construction, applying the ideal window function to the spectrum of the binned map
returns exactly the correct input spectrum. The error spectrum, however, is not improved
with respect to the result obtained with Gaussian window function,
which shows that the failure of the binning procedure to recover the correct spectrum
 above $l=300$ in Fig.~\ref{fig:fig_cl_signal} was not just a result
 of a poorly chosen window function.
 
 The error spectrum reveals that the deconvolved $a_{Xlm}$ coefficients are more
 accurate than those obtained by harmonic expansion of the binned map
 already at the lowest multipoles.

\subsection{Simulations with noise}

The first simulation was quite idealised, since the data were noise-free.
We made another simulation, where we added white noise to the TOD.
In other aspects the simulation was identical to the first one.
No correlated noise was added. We are implicitly assuming that correlated noise components, if present,
were removed
to sufficient accuracy prior to deconvolution.

As expected, noise made the deconvolution process more difficult, and the code did not
converge for the highest values of \lmax. We reached full convergence for
the combinations \lmax=600, \kmax=6, and \lmax=700, \kmax=4.
Case \lmax=700, \kmax=6 did not converge anymore.

\begin{figure}
\centering
\includegraphics[width=0.49\columnwidth]{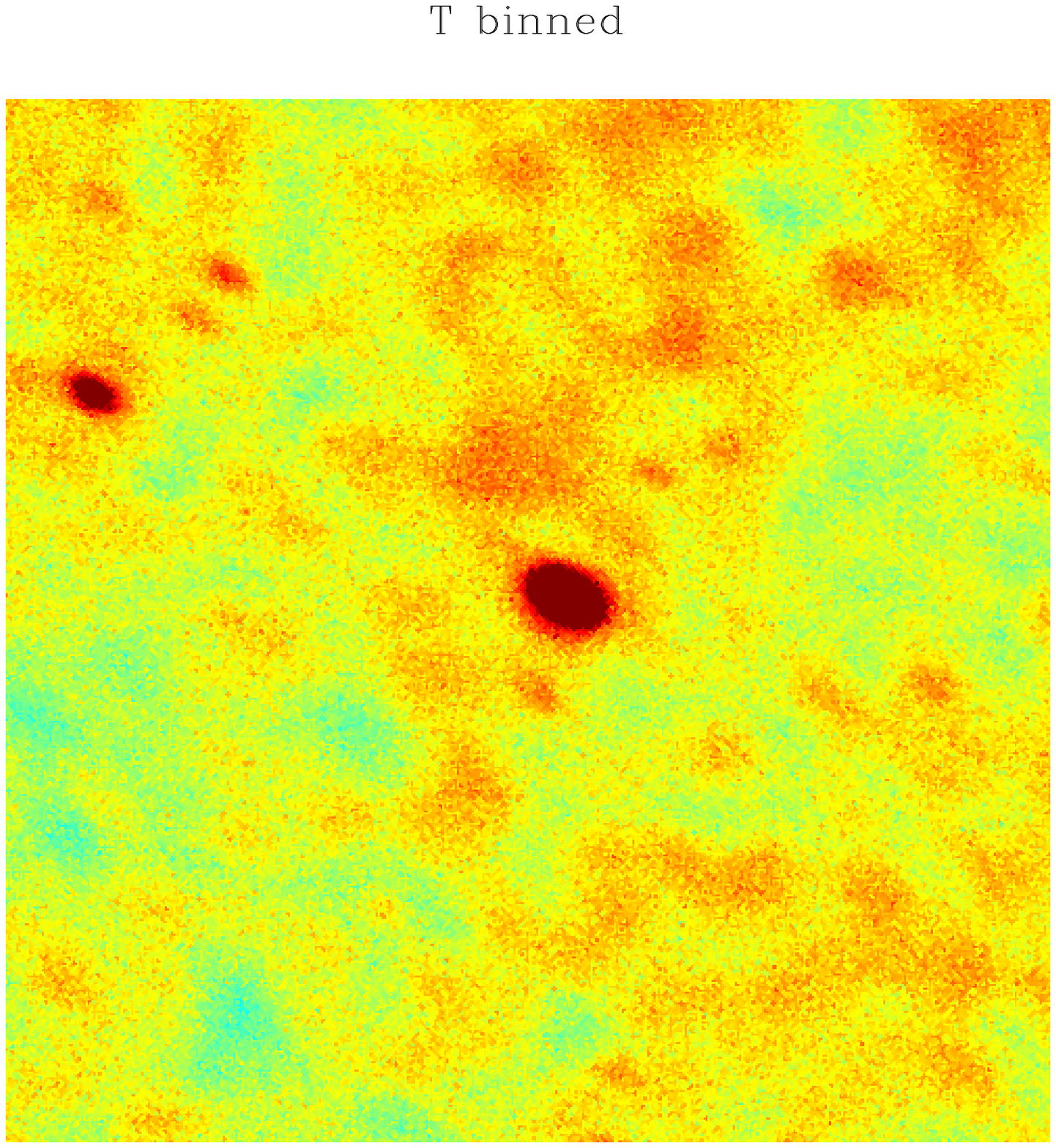}
\includegraphics[width=0.49\columnwidth]{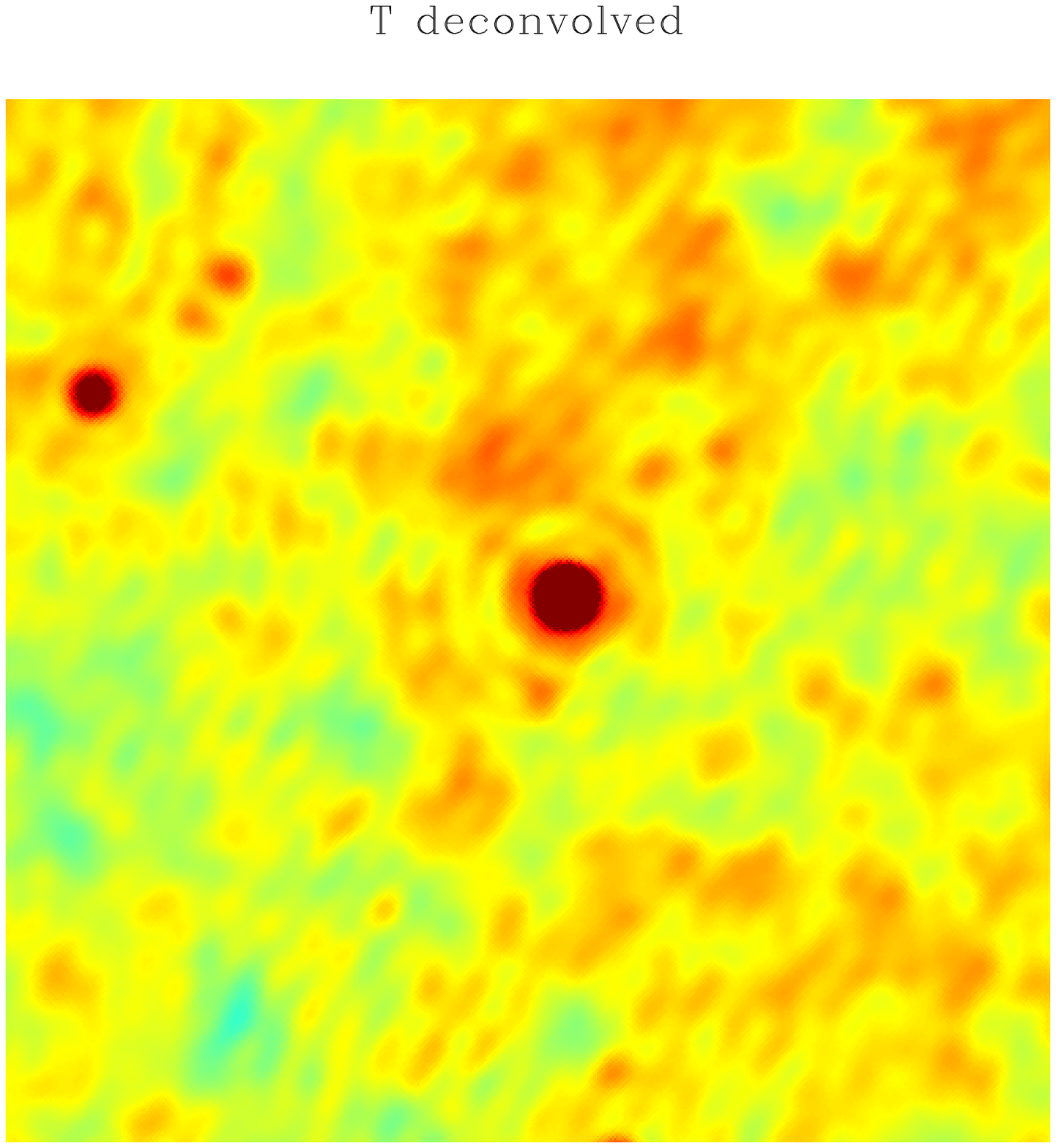}
\includegraphics[width=0.49\columnwidth]{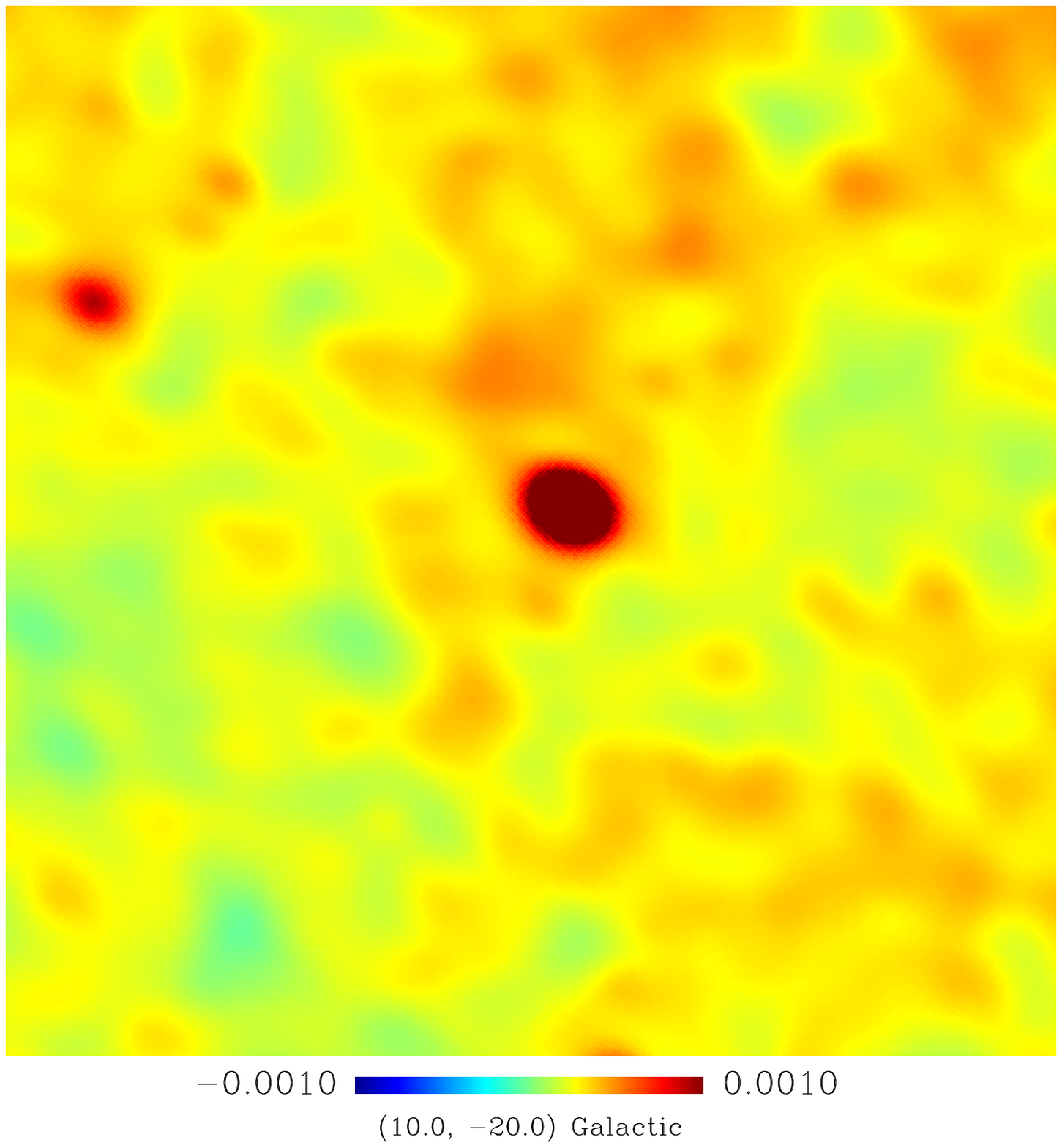}
\includegraphics[width=0.49\columnwidth]{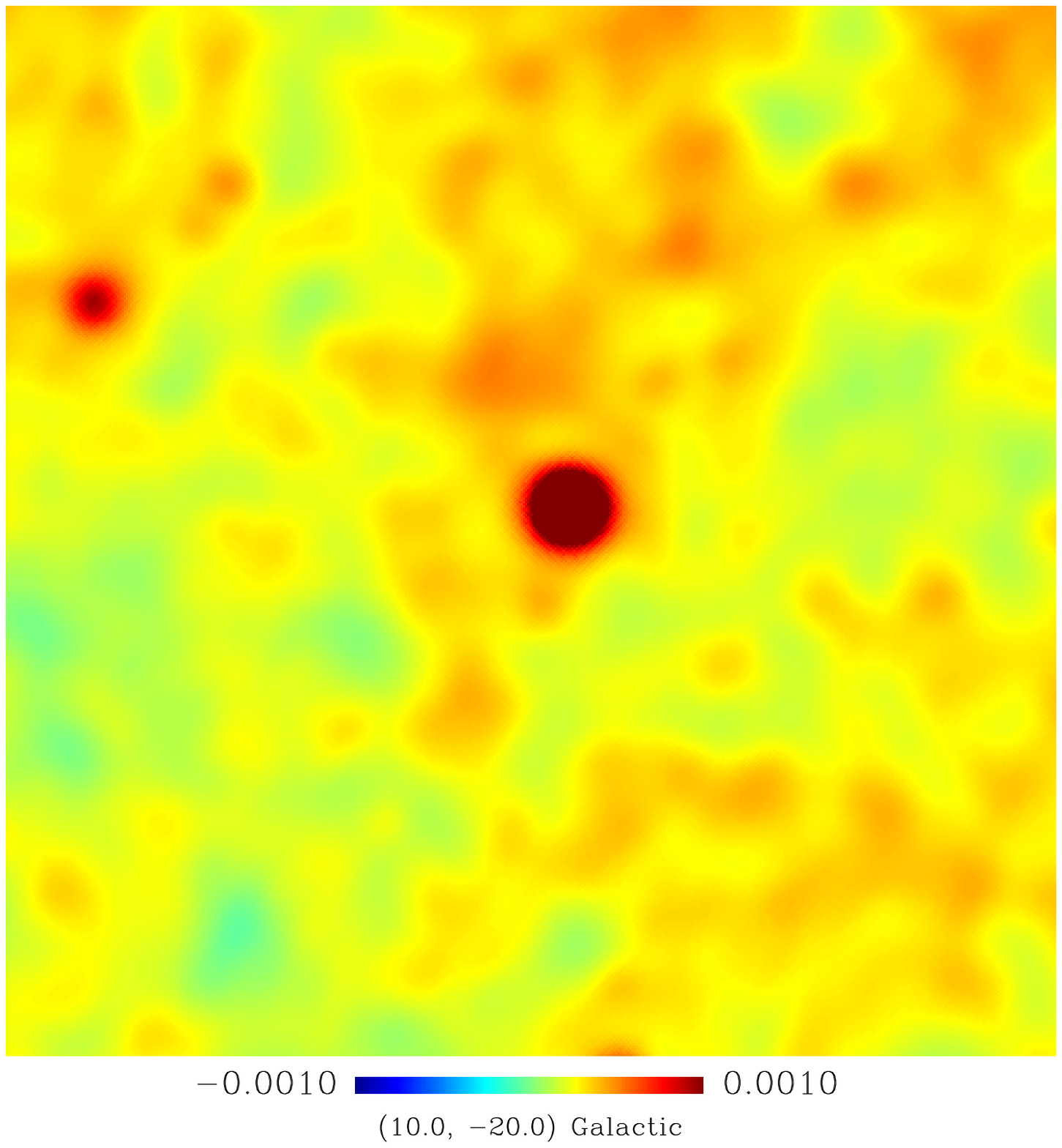}
\caption{Zoom into the 30GHz temperature map in presence of noise: binned (\fleft), and deconvolved (\fright).
In the \ftop\ row we show the unsmoothed binned map and the deconvolved map with FWHM=32\arcmin\ smoothing.
The deconvolved map is distorted by ringing and amplification of noise.
In the \fbottom\ row we show the same maps smoothed further to FWHM=45\arcmin.
The additional smoothing removes the distortion.}
\label{fig:gnom_sigwn_T}
\end{figure}

Fig.~\ref{fig:gnom_sigwn_T} shows a zoom into a point source, the same as
in Fig.~\ref{fig:gnom_signal_T}, in presence of white noise.
The deconvolution parameters were \lmax=600, \kmax=6.
The deconvolved map smoothed with the fiducial FWHM=32\arcmin\ is contaminated by distorted noise.
Also, we see some ringing around the point source due to the moderately low \lmax.
We therefore smoothed the map further to FWHM=45\arcmin, which was enough to remove
the distortion. For comparison we also smoothed the binned map to comparable resolution
by applying a smoothing by another FWHM=32\arcmin\ to it. This, combined with the original beam,
gives a total resolution of roughly FWHM=45\arcmin.
These two maps are shown in the bottom row. Even after the additional smoothing,
the point source in the binned map is elongated.
Deconvolution restores the circular shape of the source.

\begin{figure}
\centering
\includegraphics[width=0.49\columnwidth]{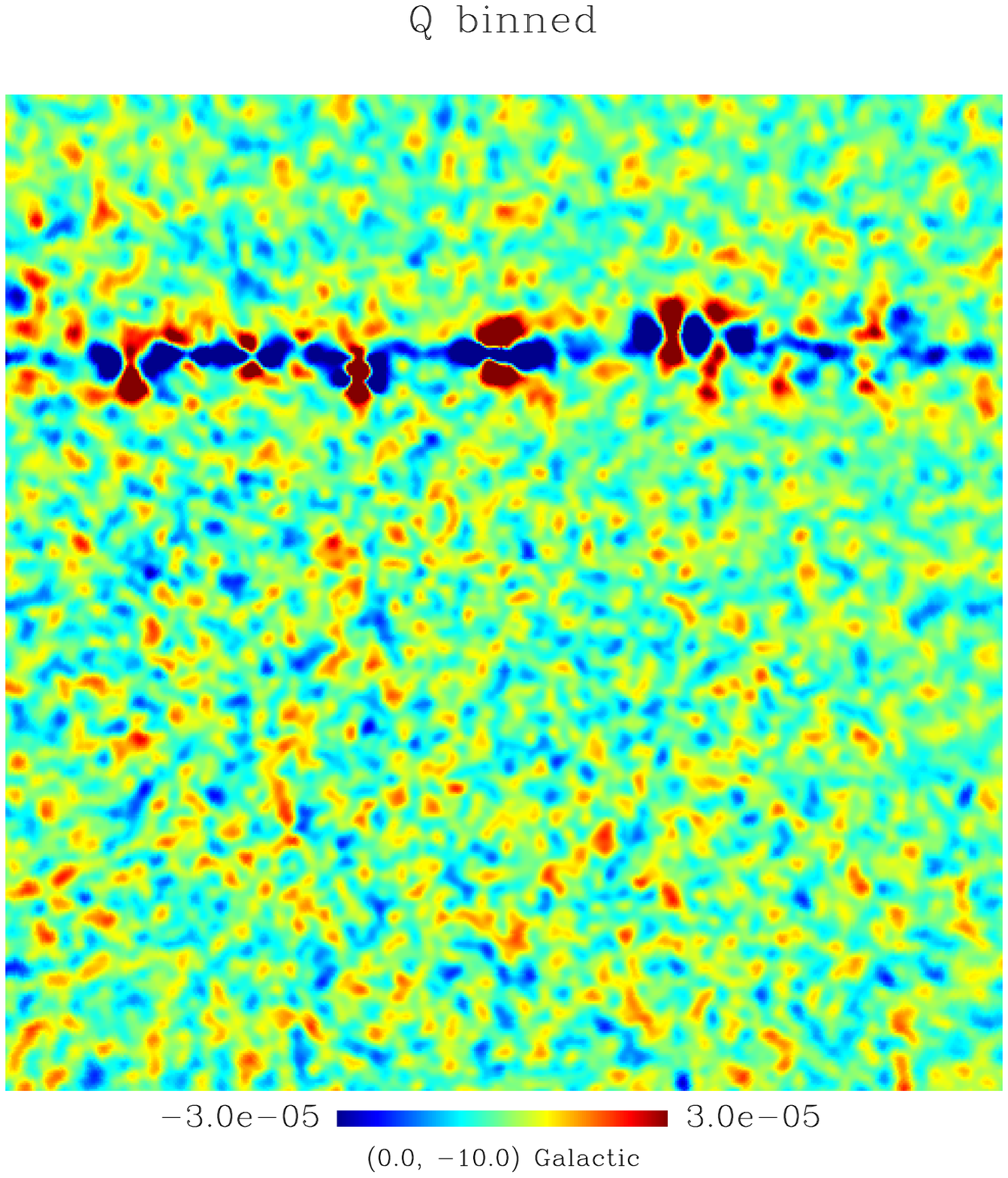}
\includegraphics[width=0.49\columnwidth]{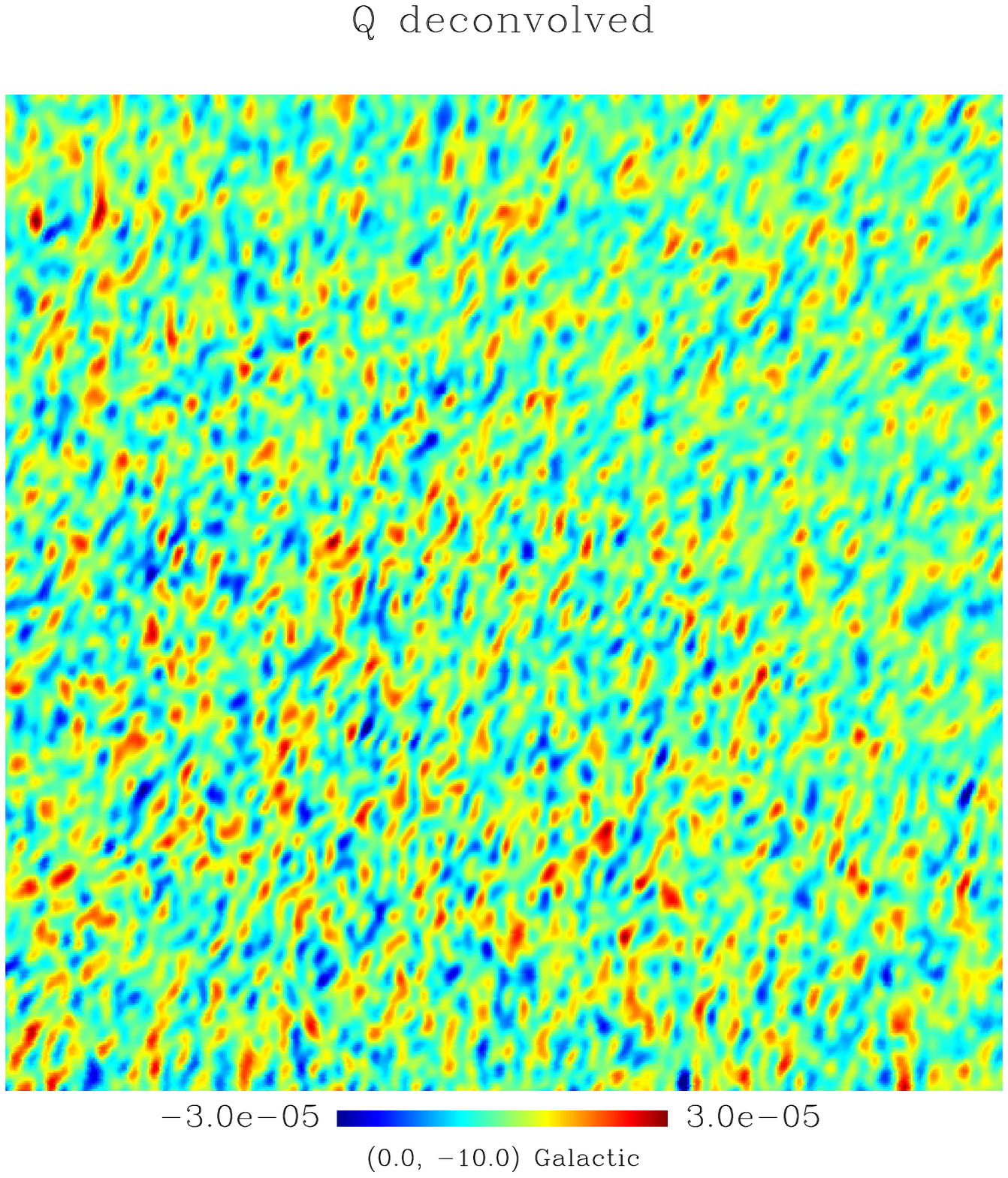}
\caption{Zoom into the 30GHz Q polarisation map: binned (\fleft) and deconvolved (\fright)
in presence of noise.
The maps were smoothed to FWHM=45\arcmin. As in the noiseless case, the smoothed
deconvolved map does not exhibit the spurious leakage signal in the galactic plane.
}
\label{fig:gnom_sigwn_Q}
\end{figure}

Figure \ref{fig:gnom_sigwn_Q} shows a zoom into the Q polarisation map.
We show the same patch of the sky as in Fig.~\ref{fig:gnom_signal_Q}.
The raw binned map (not shown) is fully noise-dominated.
We show maps smoothed to FWHM=45\arcmin.
Again, deconvolution removes the galactic leakage that is apparent in the binned map.

\begin{figure}
\centering
\includegraphics[width=0.9\columnwidth]{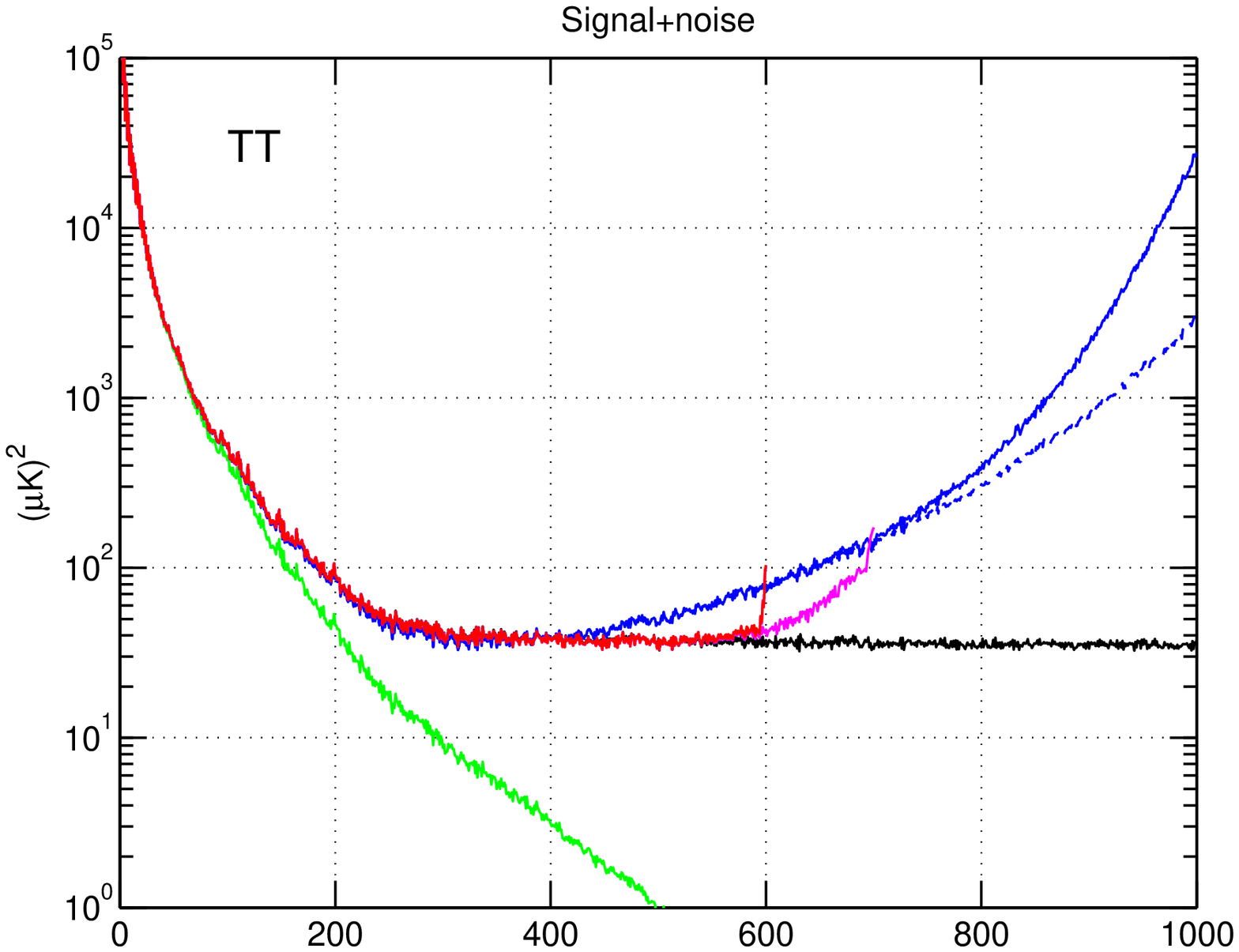}
\includegraphics[width=0.9\columnwidth]{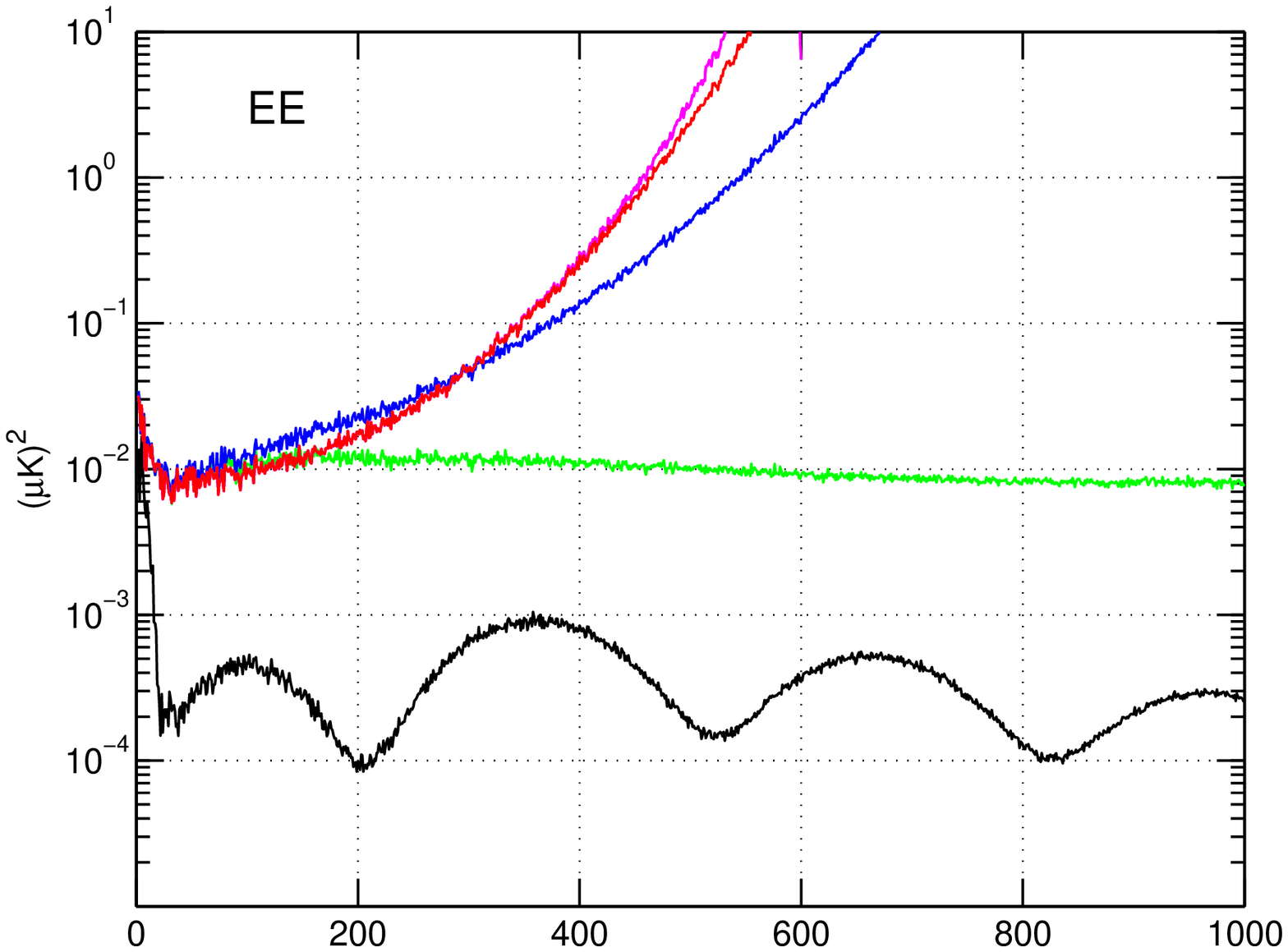}
\includegraphics[width=0.9\columnwidth]{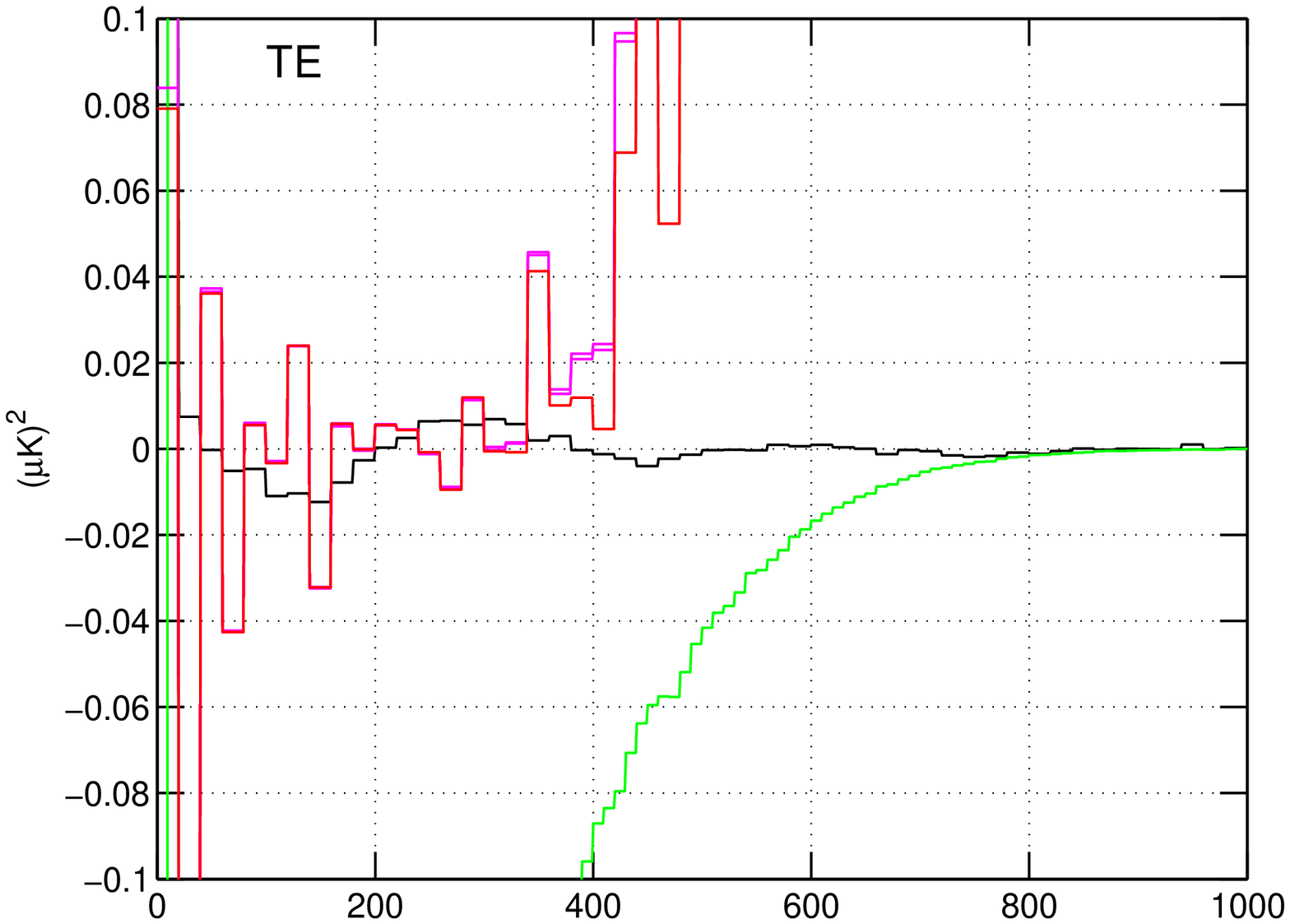}
\caption{30GHz TT, EE, and TE spectra with white noise included.
The TE spectra have been binned over 20 multipoles.
The deconvolution results are shown for three combinations of parameters: 
\kmax=4 and \lmax=600,700 (\magenta), and \kmax=6,\lmax=6 (\red).
The spectrum of a naive binned map is shown in \green.
The \blue\ line shows the spectrum of the binned map corrected for a symmetric beam.
The corresponding noise-free result is shown with \dashed\ linetype in the TT plot,
 to indicate the region where noise begins to dominate over signal.
}
\label{fig:fig_cl_sigwn}
\end{figure}

Figure \ref{fig:fig_cl_sigwn} shows the spectra constructed from the deconvolved $a_{Xlm}$ coefficients
for the parameter combinations for which the code converged fully.
Again we also show the spectrum of the binned map, and the same corrected for a symmetric
Gaussian beam with FWHM=32\arcmin.
Deconvolution recovers the true TT spectrum up to nearly $l=600$, after which noise begins to dominate.
The spectrum of the binned map begins to deviate already below $l=400$.
In the case of the EE spectrum,
both deconvolution and binning are unable to recover the true spectra except for the very lowest multipoles.
In the case of the TE spectrum, deconvolution performs considerably better than binning,
and the result follows the input spectrum up to $l=300$, although it is noisy.

\begin{figure}
\centering
\includegraphics[width=0.9\columnwidth]{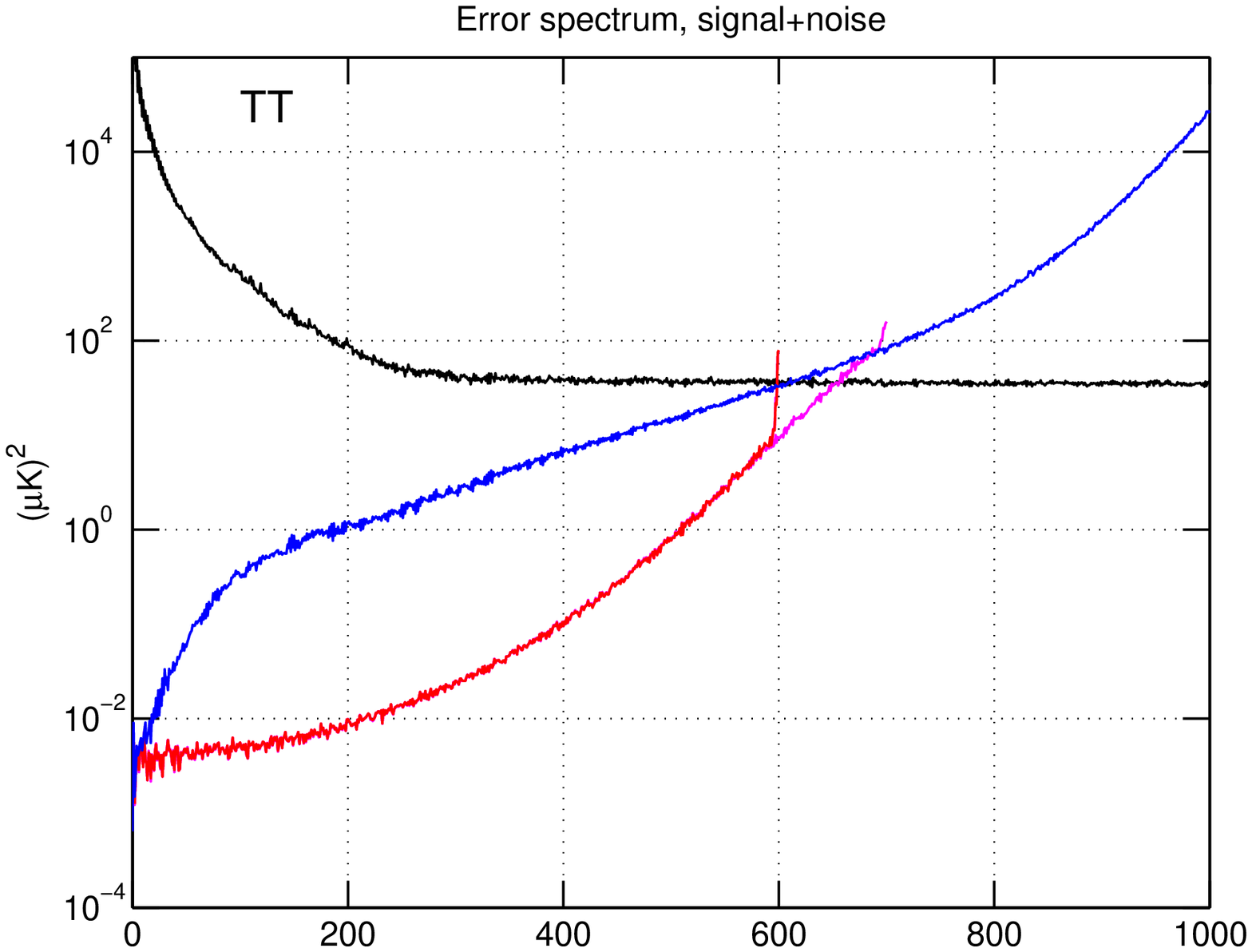}
\includegraphics[width=0.9\columnwidth]{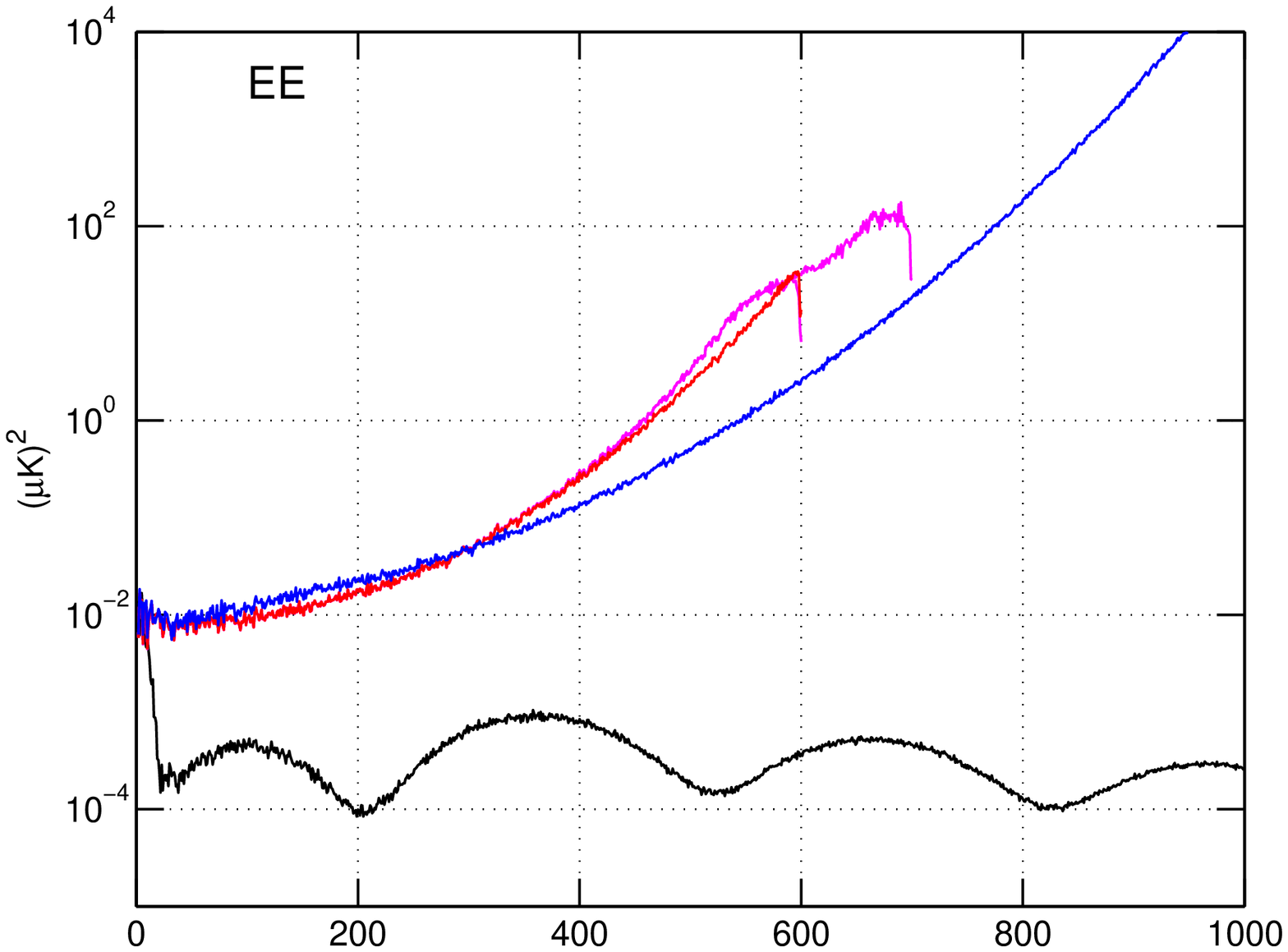}
\caption{30GHz error spectra in presence of white noise.
The \magenta\ and \red\ lines correspond to the same combinations of deconvolution parameters
as in \ref{fig:fig_cl_sigwn}. The \blue\ line shows the error spectrum for the harmonic
expansion of a binned map, corrected for a symmetric beam with FWHM=32\arcmin.}
\label{fig:errorspec_sigwn}
\end{figure}

Finally, in Fig.~\ref{fig:errorspec_sigwn} we show the error spectra in presence of noise.
Deconvolution recovers the $a_{Tlm}$ coefficients well up to $l=600$.
In the case of $a_{Elm}$ coefficients, the error is well above the input spectrum for all but the lowest multipoles.

It is a well-known and often quoted fact that deconvolution tends to amplify noise at high multipoles.
We see this clearly in the upper deconvolved map in Fig.~\ref{fig:gnom_sigwn_T}.
The effects can, however, be suppressed by applying a sufficiently strong smoothing to the map,
as we see from the lower panels of the same figure.

We can look at the amplification of noise also in harmonic space (Fig.~\ref{fig:fig_cl_sigwn}).
While the TT spectrum of the binned map (green line) continues to fall until it levels out at $l=900$
(not seen in the plot), the deconvolution spectrum begins to rise above $l=600$.
Note, however, that for the whole multipole range where we have a deconvolution result,
the extra power in the deconvolved map is still below that in the binned map corrected for
a symmetric beam (blue line).

In the EE spectrum, the power of the deconvolved map exceeds that in the binned map corrected
for a symmetric beam, around $l=300$.
This is in the domain where we have full noise domination in any case.

\subsection{70GHz simulation}
\label{run70}

To test the performance of the code at higher multipoles,
we generated one noise-free simulation mimicking the data voulme of the Planck 70GHz channel.
The used elliptic beams have an average FWHM of 13\arcmin; for more
parameters see Table \ref{table:beamparam}. We ran the deconvolution code with parameters \lmax=1700, \kmax=6.

\begin{figure}
\centering
\includegraphics[width=0.49\columnwidth]{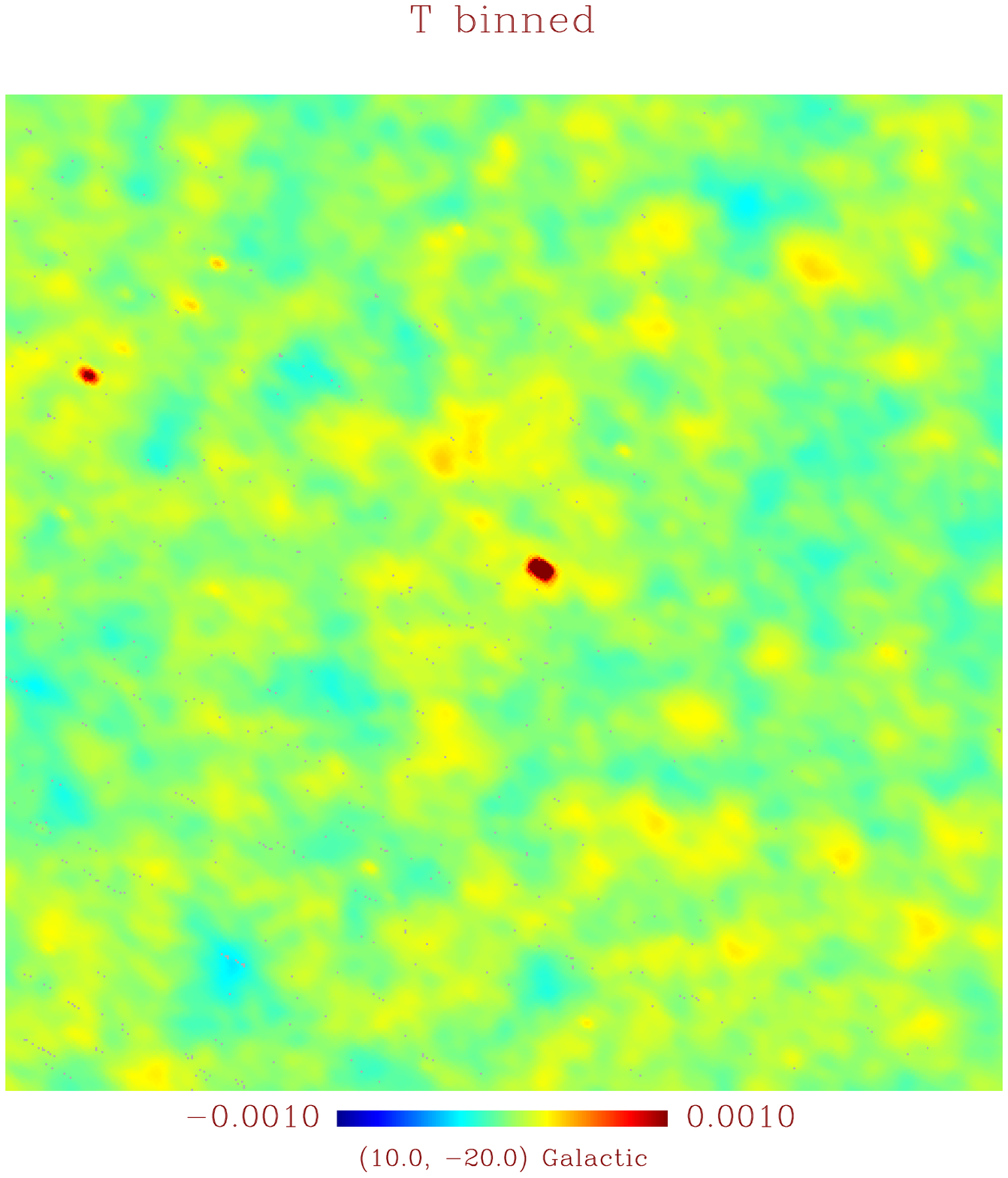}
\includegraphics[width=0.49\columnwidth]{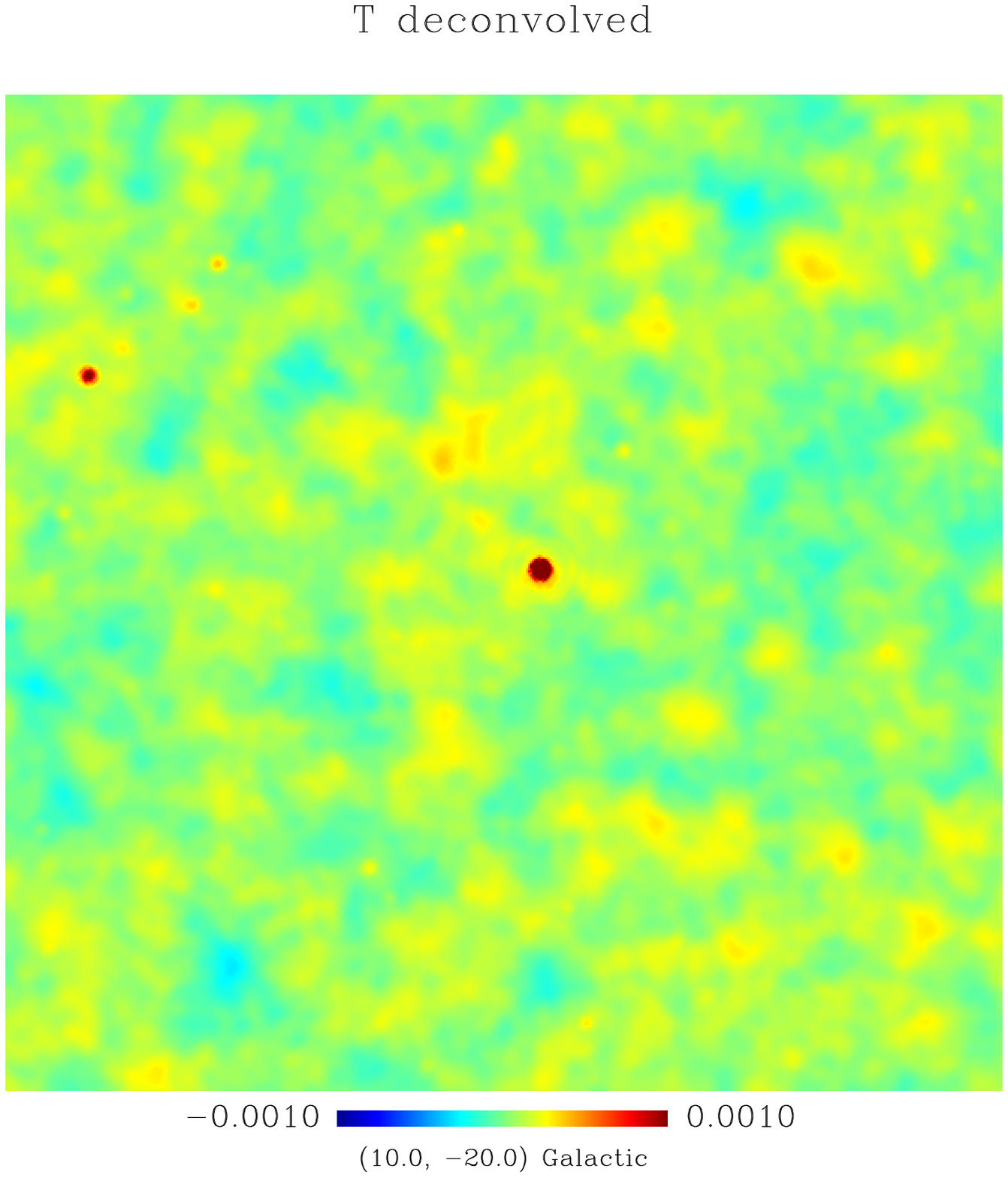}
\caption{70GHz temperature map without noise: binned (\fleft) and deconvolved (\fright).
The deconvolved map was smoothed to resolution FWHM=13\arcmin, corresponding to the mean width of the actual beam.
Shown is the same patch of sky as in Fig.~\ref{fig:gnom_signal_T}.}
\label{fig:gnom_signal_T_70}
\end{figure}

\begin{figure}
\centering
\includegraphics[width=0.49\columnwidth]{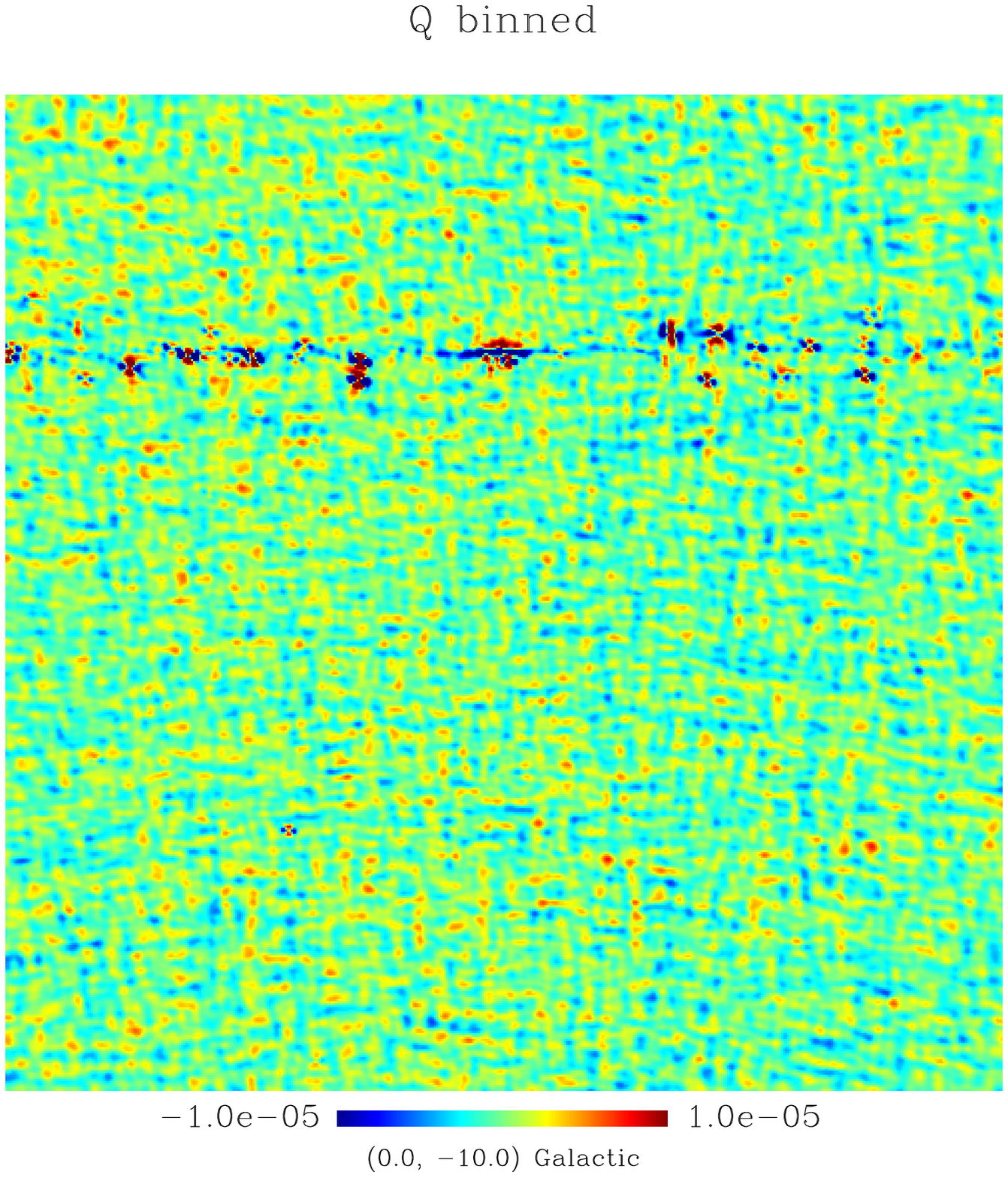}
\includegraphics[width=0.49\columnwidth]{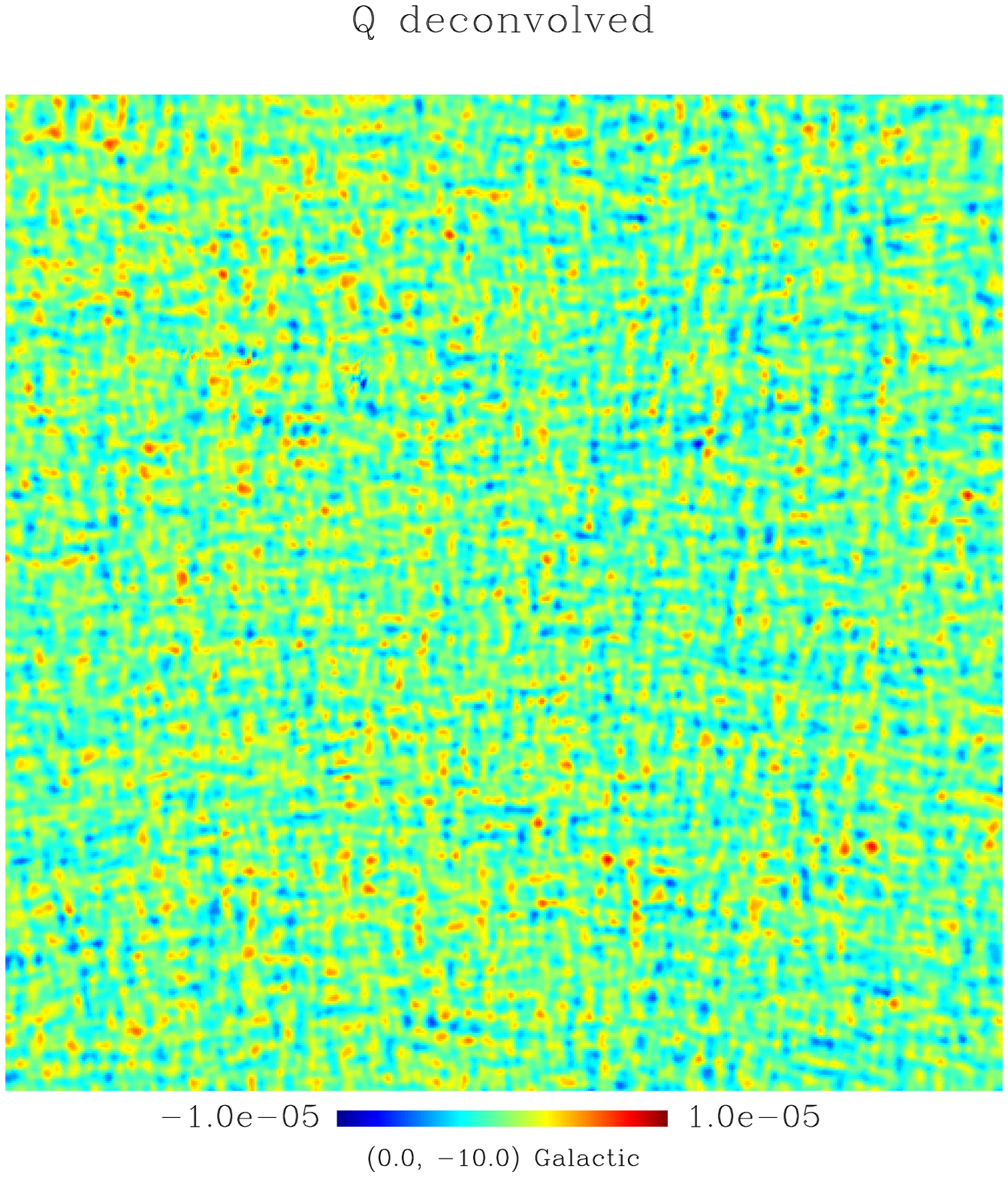}
\caption{70GHz Q polarisation map without noise: binned (\fleft) and deconvolved (\fright).
Both maps were smoothed to final resolution FWHM=18\arcmin.
Shown is the same patch of sky as in Fig.~\ref{fig:gnom_signal_Q}.
Deconvolution removes the spurious galactic signal.}
\label{fig:gnom_signal_Q_70}
\end{figure}

Temperature and polarisation maps constructed from the $a_{Xlm}$ with resolution FWHM=13\arcmin\ (T), or FWHM=18\arcmin\ (Q)
are shown in Figs.~\ref{fig:gnom_signal_T_70} and \ref{fig:gnom_signal_Q_70} together with the corresponding binned maps.
We show the same region of the sky as with 30GHz simulations. Again, deconvolution has worked well,
returning the circular shape of the point source, and removing the galactic leakage in the Q polarisation map.

\begin{figure}
\centering
\includegraphics[width=0.9\columnwidth]{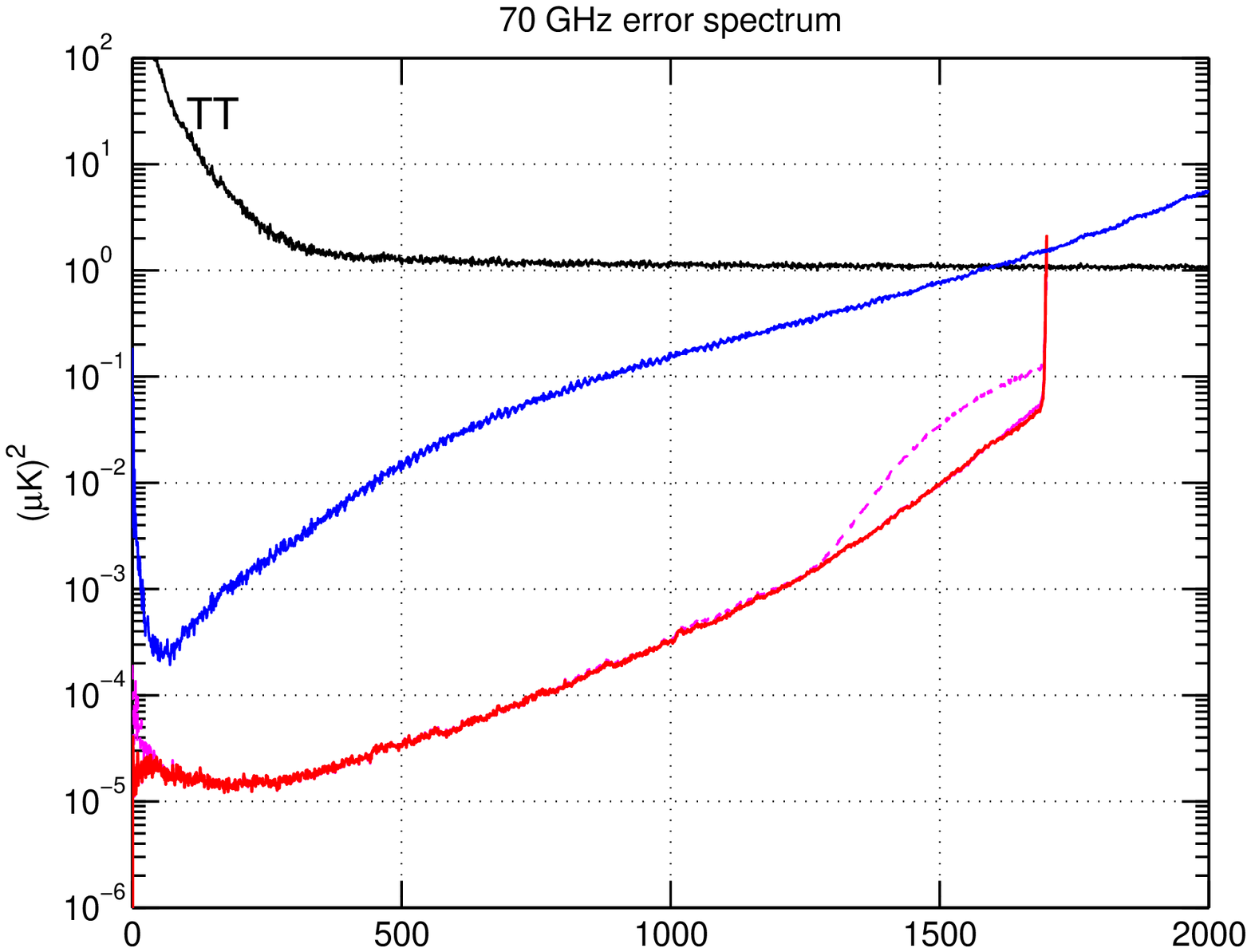}
\includegraphics[width=0.9\columnwidth]{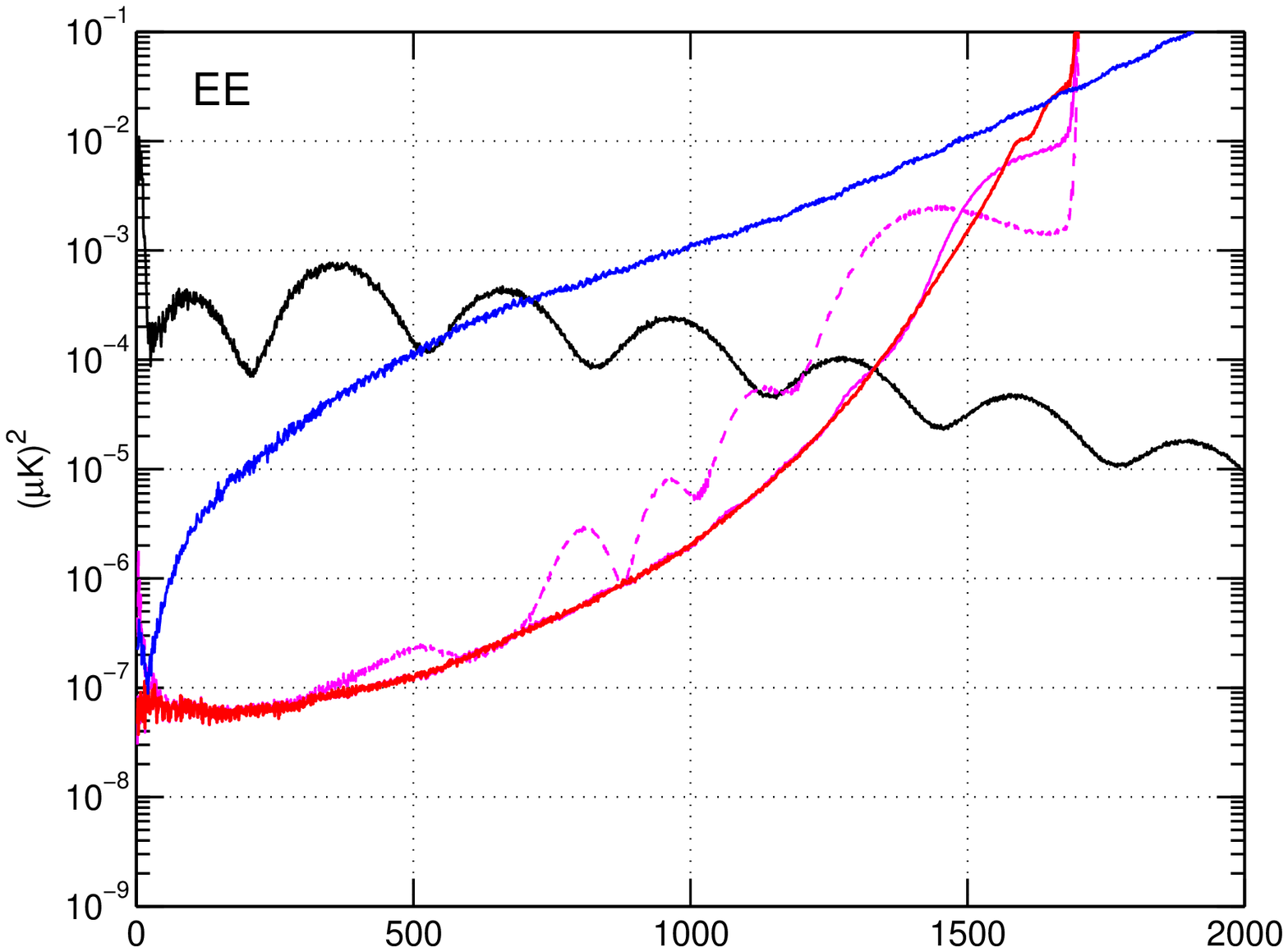}
\caption{70GHz TT (\ftop) and EE (\fbottom) error spectra, without noise.
The \blue\ line represents harmonic expansion of the binned map, corrected for a symmetric beam with FWHM=13\arcmin.
The fully converged deconvolution result (\lmax=1700, \kmax=6), is shown by a \red\ line.
We also show the error after 200 (\magenta\ \dashed) and 500 (\magenta\ \solid) iteration steps.}
\label{fig:convergence}
\end{figure}

In Fig.~\ref{fig:convergence} we show the TT and EE error spectra for the 70GHz simulation.
The $a_{Tlm}$ are recovered well nearly up to $l=1700$, and the $a_{Elm}$ coefficients well above $l=1000$.
Thus we have shown that, with sufficient computational resources, our deconvolution method can be applied to 
realistic data sets of volume and resolution of the Planck 70GHz channel.

\subsection{Convergence}

The convergence criterion we had set in our code is rather strict. We require that the square of the residual
vector has fallen below a fraction of $10^{-12}$ of its initial value.

To see if a less strict criterion would help in reducing the computation time, we studied the convergence of the solution.
We dumped the $a_{Xlm}$ coefficients after every 50 steps,
and computed the error spectrum with respect to the input coefficients. 

We show the TT and EE spectra after 200 and 500
iteration steps in Fig.~\ref{fig:convergence}, along with the final results.
We observe that the solution changes only very little after the first few hundred iterations,
and 200 steps already give a significant improvement over the harmonic expansion of the binned map.
In fact, after the first 100 iteration steps we can see no visible improvement in the reconstructed sky map.
We could thus reduce the computation time by a large factor if we used a more relaxed convergence criterion and would not significantly lose accuracy.

\subsection{Complicated beam shape}
\label{dbeam}

\begin{figure}
\centering
\includegraphics[width=0.4\columnwidth]{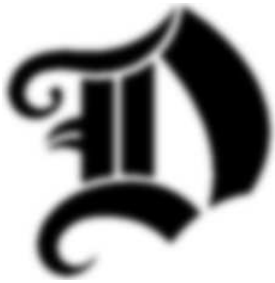}
\caption{Unconventional beam pattern}
\label{fig:Dbeam}
\end{figure}

To demonstrate the power of the deconvolution method,
we made yet another simulation with an unconventional D-shaped beam
(D for ``deconvolution''),
which is shown in Fig.~\ref{fig:Dbeam}. The linear dimension of the beam was approximately $12\degr$ squared.
For demonstration purposes we included only point sources and considered temperature only.
For the binning process into the 3D map, we chose $N_\text{side}=1024$ and $n_\psi=256$.
Deconvolution parameters were $\lmax=500$, $\kmax=30$, and only a single full-sky survey
was simulated.
The comparatively high value of \kmax\ was required to properly take into account the complicated azimuthal structure of the beam.

\begin{figure}
\centering
\includegraphics[width=0.49\columnwidth]{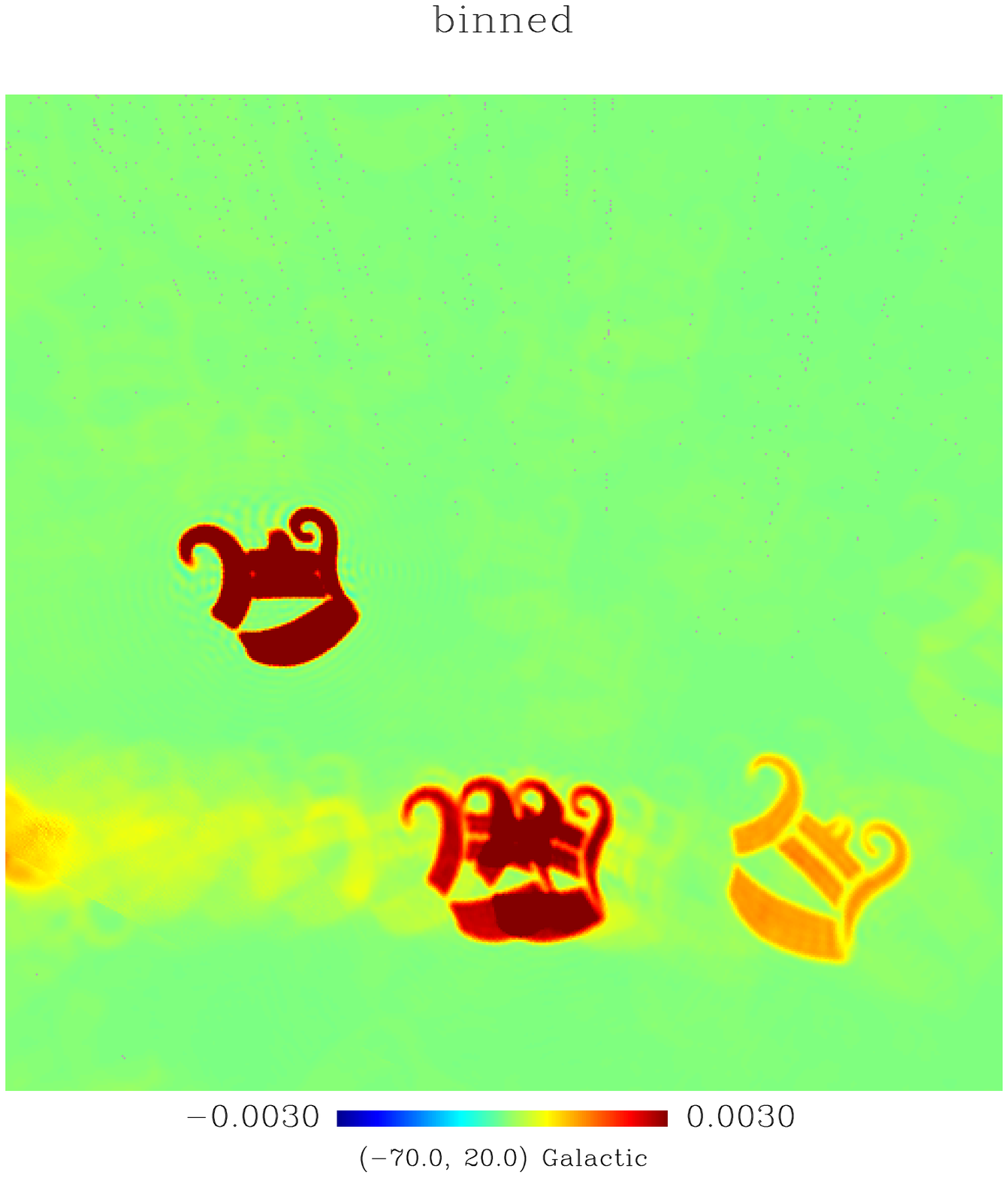}
\includegraphics[width=0.49\columnwidth]{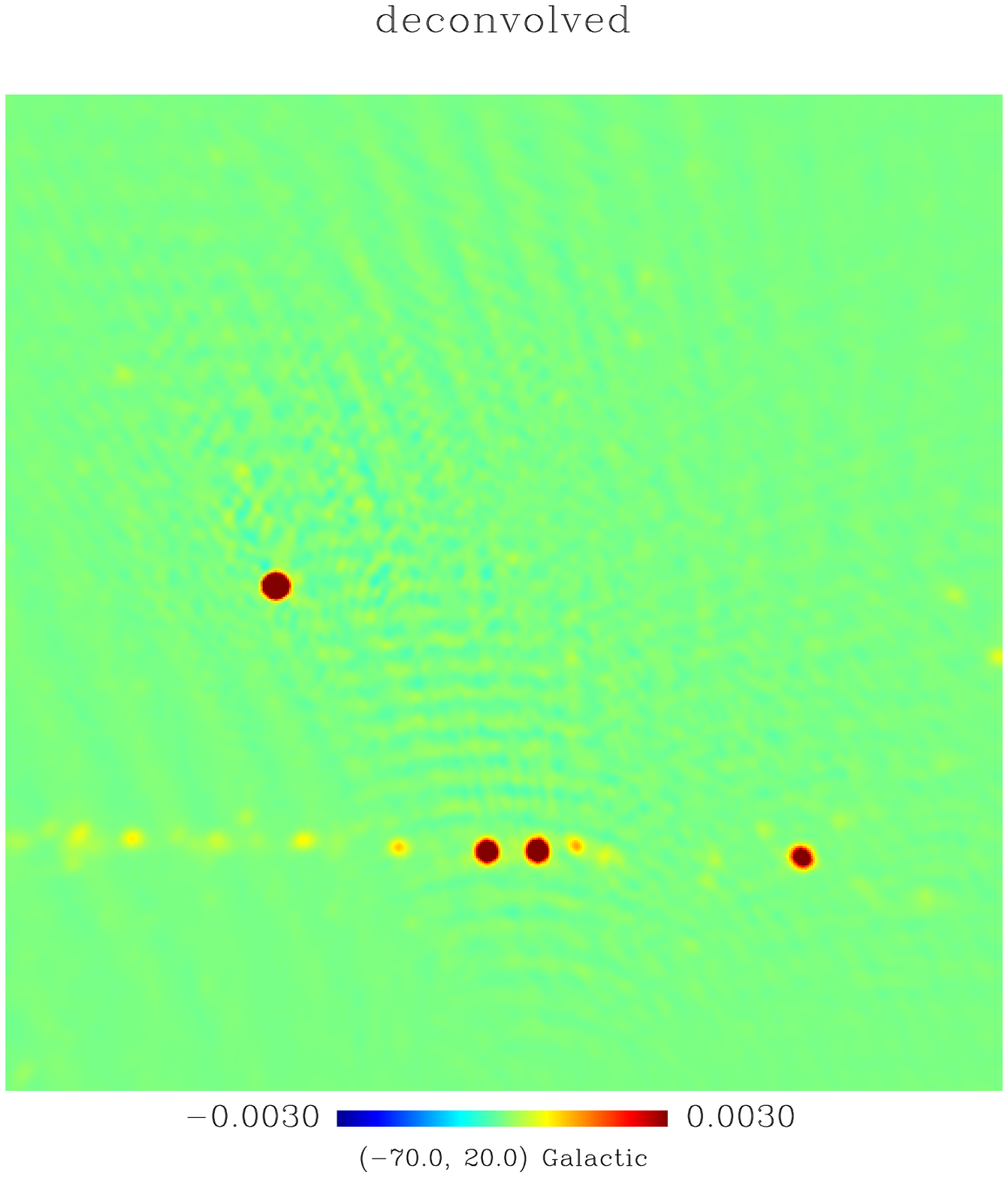}
\caption{Sky seen through a D-beam: binned (\fleft) and deconvolved (\fright) temperature map.
The beam transforms the image of a point source into an image of the beam. 
Deconvolution recovers the original shapes of the sources.
The size of the shown patch is $83.3\degr$ squared. The horizontal
structure near the bottom of the plots is the accumulation of weak point
sources in the galactic plane.}
\label{fig:gnom_Dbeam}
\end{figure}

A map binned directly from the TOD is shown in the left panel of Fig.~\ref{fig:gnom_Dbeam}.
The beam transforms the image of a point source into a D-shaped pattern, resembling the beam.
The deconvolved map is shown on the right; it was smoothed by a symmetric Gaussian
beam with FWHM=$1\degr$ to reduce ringing.
Even with this complicated beam shape we are able to recover the underlying point sources.

It is interesting that in this case it was sufficient to smooth the map with a beam of
FWHM=$1\degr$ to obtain a very satisfactory map, although the full dimension of the beam is much
larger.

\section{Conclusions}

We have presented an efficient beam deconvolution code, designed for absolute CMB experiments,
and tested it on simulated data.
The code is released under the terms of the GNU General Public License and
can be obtained from \url{http://sourceforge.net/projects/art-deco/}.

We looked at maps constructed from the recovered $a_{Tlm}$, $a_{Elm}$, and $a_{Blm}$ coefficients,
and examined the coefficients directly in harmonic space.
These reflect different aspects of the solution.
We compared the results to a harmonic expansion of a binned map.
We have also shown that with sufficient computational resources, we can extend the method to \lmax=1700,
which would be sufficient for the Planck 70GHz channel.

In absence of noise we could recover the $a_{Tlm}$ coefficients to a high
accuracy, and remove the effects of beam asymmetry.
When white noise was added, we were able to reach a lower value of \lmax, but the advantage of
deconvolution over binning was still clear.

In the case of polarisation, deconvolution worked well in absence of noise.
We were able to almost completely remove the temperature leakage due to beam mismatch.
When noise was added, results were less clear. Deconvolution removed the visible galactic residual
that arises from beam mismatch, but in harmonic space deconvolution did not seem to bring a clear benefit over binning.
The recovered $a_{Elm}$ coefficients were dominated by noise at nearly all multipoles,
but this was true for both deconvolved and binned maps.

Our simulation used beams that were highly asymmetrical and
also had a strong mismatch between them, which leads to a significant polarisation leakage.
We did this to show the beam effects more clearly,
but at the same time we also made the deconvolution problem very challenging.
The accuracy we can expect when the code is applied to a real experiment
strongly depends on the beam shapes of the particular experiment.

Some topics for future study can be mentioned.
We have set a very strict convergence criterion for the CG iteration.
Our convergence study indicates that a much more relaxed criterion could be
sufficient for practical purposes. A future improvement would be to define a more suitable
convergence criterion. This could reduce the computation time by a significant factor.

We applied the method to a simulation data set that provides full sky coverage.
Our code is built on a formalism that makes no assumptions on the sky coverage.
In practice, good convergence requires nearly full coverage.
This is intuitively evident, since we are solving the harmonic coefficients, which necessarily represent the complete celestial
sphere. If the input data only cover a part of the sky, the coefficients cannot be well constrained.
A straighforward extension of the method would be to insert a prior that constrains the solution in the region which
is not covered by the data, but this is beyond the scope of this paper and is a subject for future study.

Yet another topic for further study is determining the noise bias present in the TT and EE spectra at high multipoles.

Though the simulations used in this work mimic some aspects of the Planck mission,
the parameters used do not reflect the properties of the actual instrument.
The results presented in this paper are thus not representative of the sensitivity
of the Planck experiment, and the authors do not represent the Planck collaboration in this context.

\appendix

\section{Definitions}
\label{appendix:definitions}

In this appendix we present definitions and conventions used 
in the paper.
We follow the conventions of \citet{varshalovich-etal-1988}.

\subsection{Harmonic expansion of the CMB radiation field}

The CMB temperature and polarisation fields can be presented by
spin-weighted harmonic functions with spin $s=0,\pm2$.
Spin $s=0$ represents the temperature field, 
while components $s=\pm2$ represent polarisation.

The temperature field on the celestial sphere is expanded as
\begin{equation}
   T(\theta,\phi) = \sum_{l=0}^\infty \sum_{m=-l}^l 
                     a_{0lm} {\,}_0 Y_{lm}(\theta,\phi).
        \label{temp_field}
\end{equation}
The Q and U Stokes fields, which describe linear polarisation,
can be written as
\begin{equation}
    (Q\pm iU)(\theta,\phi) = \sum_{l=0}^\infty \sum_{m=-l}^l
                                a_{\pm2lm} {\,}_{\pm2}Y_{lm}\text{.}
      \label{pol_field}
\end{equation}
Here ${}_{s}Y_{lm}$ are spin-weighted harmonic functions,
and $a_{slm}$ are the corresponding (complex) coefficients.
The harmonic functions and coefficients obey the symmetry relation
\begin{align}
   {}_sY_{lm}(\theta,\phi) =& (-1)^{m-s}{}_{-s}Y^{\ast}_{l,-m}(\theta,\phi) \\
    a_{slm} =& (-1)^{m-s} a^{\ast}_{-s,l,-m}\text{.} \nonumber
\end{align}
The harmonic functions are related 
to the \emph{Wigner functions} $D^{l}_{sm}(\theta,\phi,\psi)$ through
\begin{equation}
   {}_sY_{lm}(\theta,\phi) = 
   \sqbeam
   D^{l\ast}_{m,-s}(\phi,\theta,0),   \label{harmonics_def}
\end{equation}
where
\begin{equation}
\sqbeam=\left (\frac{2l+1}{4\pi}\right)^{1/2}.
\end{equation}

It is convenient to define further
\begin{align}
    a_{Tlm} =& \,a_{0lm} \nonumber\\
    a_{Elm} =& -\frac12 (a_{2lm}+a_{-2lm}) \\
    a_{Blm} =& -\frac{1}{2i} (a_{2lm}-a_{-2lm})\text{.} \nonumber
\end{align}

The CMB radiation field is expected to be a
statistically isotropic Gaussian spin 0,2 field.
The statistical properties of the harmonic expansion coefficients
are then fully characterised by four spectra
$C^{T}_l,C^{E}_l,C^{B}_l,C^{C}_l$,
such that
\begin{align}
  \langle a^\ast_{Tlm}a_{Tl'm'} \rangle
 =&  C^T_l\delta_{ll'}\delta_{mm'}  \nonumber \\
  \langle a^\ast_{Elm}a_{El'm'} \rangle 
 =& C^E_l\delta_{ll'}\delta_{mm'}  \nonumber \\
   \langle a^\ast_{Blm}a_{Bl'm'} \rangle 
 =& C^B_l \delta_{ll'}\delta_{mm'}            \\
  \langle a^\ast_{Tlm}a_{El'm'} \rangle
 =& C^C_l \delta_{ll'}\delta_{mm'}\text{.}            \nonumber
\end{align}

\subsection{Wigner functions and rotations on a sphere}
\label{wignerrot}

The harmonic coefficients transform under rotations as follows.

Assume we are given the coefficients $a_{slm}$ in one coordinate system.
We then rotate the coordinate system, as described by Euler angles $\phi,\theta,\psi$:
\begin{enumerate}
\item Rotate by angle $\psi$ around the $z$-axis (in positive direction).
\item Rotate by angle $\theta$ around the \emph{initial} $y$-axis.
\item Rotate by angle $\phi$ around the \emph{initial} $z$-axis.
\end{enumerate}

\noindent The rotation is equivalent to rotating first by angle $\phi$ around the $z$-axis,
then by angle $\theta$ around the \emph{new} $y$-axis, and finally
by angle $\psi$ around the \emph{new} $z$-axis.
The harmonic coefficients in the rotated coordinate system are given by
\begin{equation}
   a'_{slm} = \sum_{m'} a_{slm'}
                   D^{l\ast}_{m'm}(\phi,\theta,\psi)\text{,} \label{rotation}
\end{equation}
where $D^{l}_{m'm}$ are Wigner functions.

A Wigner function may be split into a product of three terms, 
each of which only depends on one of the angles:
\begin{equation}
  D^l_{mk}(\theta,\phi,\psi) = 
    e^{-im\phi} d^l_{mk}(\theta) e^{-ik\psi}.  \label{reduced_def}
\end{equation}
Functions $d^l_{mk}(\theta)$ are the \emph{reduced Wigner functions}.
They have the property
\begin{equation}
   d^l_{mk}(0) = \delta_{mk}.  \label{zero_theta}
\end{equation}
They are real-valued and obey the symmetry relation
\begin{equation}
   d^l_{mk}(\theta) = (-1)^{m+k}d^l_{-m,-k}(\theta).
\end{equation}
The symmetry relation for the full Wigner functions is
\begin{equation}
   D^l_{mk}(\phi,\theta,\psi) = (-1)^{m+k}D^{l\ast}_{-m,-k}(\phi,\theta,\psi).
\end{equation}

\subsection{Wigner expansion}
\label{wigner_expansion}

Wigner functions $D^l_{mk}(\phi,\theta,\psi)$ 
for integer values of $l,m,k$
are defined in domain $V$:
\begin{equation}
  V: \qquad 0\le\phi\le 2\pi, \quad
            0\le\theta\le \pi, \quad
            0\le\psi\le 2\pi\text{.}
\end{equation}
Domain $V$ can be interpreted as the space of Euler angles.
The volume element in domain $V$ is 
$dV=\sin\theta d\theta d\phi d\psi$,
and $V$'s total volume is $8\pi^2$.

Any finite function $f(\phi,\theta,\psi)$ defined in domain $V$
may be expanded in a series of Wigner functions as
\begin{equation}
   f(\phi,\theta,\psi)
     = \sum^\infty_{l=0}\sum^l_{m=-l}\sum^l_{k=-l}
               c^l_{mk} D^{l}_{mk}(\phi,\theta,\psi)\text{,}
      \label{wigner_series}
\end{equation}
where the coefficients $c^l_{mk}$ are given by
\begin{equation}
   c^l_{mk} =
     \frac{2l+1}{8\pi^2}
     \int_0^{2\pi}d\phi
     \int_0^\pi\sin\theta d\theta
     \int_0^{2\pi} d\psi
      f(\phi,\theta,\psi) D^{l\ast}_{mk}(\phi,\theta,\psi)\text{.}
        \label{wigner_coeff}
\end{equation}
Expansion \eqref{wigner_series} can be understood as the three-dimensional
equivalent of the spherical harmonics expansion.

In particular, a product of two Wigner functions 
$D^{l\ast}_{mk}(\omega) D^{l'}_{m'k'}(\omega)$
can be expanded as a series of Wigner functions as
\begin{align}
   & D^{l\ast}_{mk}(\omega) D^{l'}_{m'k'}(\omega)  \label{dproduct_integral}\\
  =& \sum_{l_2m_2k_2} D^{l_2}_{m_2k_2}(\omega) \frac{2l_2+1}{8\pi^2}
    \int  D^{l\ast}_{mk}(\omega') D^{l'}_{m'k'}(\omega')
          D^{l_2\ast}_{m_2k_2}(\omega')d\omega' . 
            \nonumber
\end{align}
An integral over a product of three Wigner functions is given by
Wigner $3j$ symbols as
\begin{align}
    & \int D^{l\ast}_{mk}(\omega') D^{l'}_{m'k'}(\omega')
            D^{l_2\ast}_{m_2k_2}(\omega') d\omega'  \\
    =& 8\pi^2 (-1)^{m'+k'}
        \left( \begin{array}{ccc}
            l & l' & l_2 \\ m & -m' & m_2
         \end{array} \right)
          \left( \begin{array}{ccc}
             l & l' & l_2 \\ k & -k' & k_2
         \end{array} \right)    \nonumber
\end{align}
\citep{varshalovich-etal-1988}.

The following properties of the $3j$ symbols help
in restricting the range of summation.
A $3j$ symbol is non-vanishing
\begin{equation}
          \left( \begin{array}{ccc}
             l_1 & l_2 & l_3 \\ k_1 & k_2 & k_3
                 \end{array} \right) \ne 0 
\end{equation}
only if the two conditions
\begin{equation}
           k_1+k_2+k_3 = 0 
\end{equation}
and
\begin{equation}
|l_1-l_2| \le l_3 \le l_1+l_2
\end{equation}
are fulfilled.
We can thus eliminate the summation over $k_2$ and $m_2$.
Product \eqref{dproduct_integral} becomes
\begin{align}
   D^{l\ast}_{mk}(\omega) D^{l'}_{m'k'}(\omega) =&
    (-1)^{m'+k'}\sum_{l_2} (2l_2+1) D^{l_2}_{m'-m,k'-k}(\omega) \\
    & \times
        \left( \begin{array}{ccc}
            l & l' & l_2 \\ m & -m' & (m'\!-\!m)
         \end{array} \right)
          \left( \begin{array}{ccc}
             l & l' & l_2 \\ k & -k' & (k'\!-\!k)
         \end{array} \right).       \nonumber
\end{align}


\subsection{Beam}

\subsubsection{Detector signal and definition of beam coefficients}

Consider a detector measuring CMB temperature and linear polarisation.

We assume now, very generally, that the signal recorded by a detector, apart from noise,
is proportional to the $a_{slm}$ coefficients.
We define the beam coefficients $b_{slk}$ as follows.
When the detector is at some (predefined) fiducial position
and orientation (for instance at the north pole), the signal detected 
in the absence of noise is
\begin{equation}
   t_\text{fid} = \sum a_{slm}b^\ast_{slm}.  \label{beam_def}
\end{equation}
It is clear from the definition that
the beam coefficients must contain all information not only of 
the beam shape, but also of the possible polarisation
sensitivity and orientation.

We may then rotate the beam to another position and orientation,
as described in Section \ref{wignerrot}.
The rotation brings the beam onto location 
$\theta,\phi$ in spherical coordinates, 
in an orientation defined by angle $\psi$.
The harmonic coefficients transform under rotations 
according to \eqref{rotation}.
The signal seen by the rotated detector is then
\begin{equation}
 t = \sum_{slmk} a_{slm}b^\ast_{slk}
     D^{l\ast}_{mk}(\phi,\theta,\psi).
\end{equation}

In the following we derive expressions for the beam 
coefficients in some special cases.

\subsubsection{Non-polarised beam}

Consider first a detector with an ideal beam, beam width 
equal to zero, and no sensitivity to polarisation.
The signal detected by the beam is directly given by 
\eqref{temp_field}.
We see readily, by comparing \eqref{temp_field} and \eqref{beam_def},
that if the chosen fiducial beam position 
is at the north pole ($\theta=0,\phi=0$), the beam coefficients must be
\begin{equation}
   b_{slm} = {}_sY^\ast_{lm}(0,0) \delta_{s0} = 
     \sqbeam  \delta_{s0}\delta_{m0}.
\end{equation}

A general non-polarised beam with sensitivity $g(\theta,\phi)$
gives for the $s=0$ components
\begin{align}
   b_{0lm} =&\iint g(\theta,\phi)
            {}_0Y^\ast_{lm}(\theta,\phi)\sin\theta d\theta d\phi    \label{temp_blm} \\
           =&\sqbeam\iint g(\theta,\phi)
            D^l_{m0}(\phi,\theta,0)\sin\theta d\theta d\phi  \nonumber
\end{align}
and $b_{slm}=0$ for $s=\pm2$.


\subsubsection{Polarisation measurements}

A polarisation-sensitive detector requires careful treatment,
because the Stokes parameters Q and U are not defined at the pole.
The beam coefficients can still be defined in a meaningful way, 
as we show in the following.

Consider again a beam at a fiducial position and orientation 
at the north pole. A general beam may have different polarisation
sensitivity at different locations.
We therefore define the beam shape $g(\theta,\phi)$ separately 
for temperature and polarisation.

The signal detected may be written in a general form as
\begin{align}
   t =& \iint  \sin\theta d\theta d\phi
            \Big\{ g_T(\theta,\phi) T(\theta,\phi) +
            \label{detected} \\
     &
        +g_P(\theta,\phi) \left[
             Q(\theta,\phi)\cos(2\psi(\theta,\phi))
            +U(\theta,\phi)\sin(2\psi(\theta,\phi)) \right] \Big\}\text{.}
      \nonumber
\end{align}
Here $\psi(\theta,\phi)$ is the \emph{local} direction
of polarisation sensitivity,
measured as an angle from the local meridian.
Functions $g_{T}(\theta,\phi)$, $g_{P}(\theta,\phi)$, and $\psi(\theta,\phi)$
together define a general beam.

We proceed to calculate the beam coefficients.
We write the trigonometric 
functions in exponential form, to obtain
\begin{align}
  t =& \iint  \sin\theta d\theta d\phi
           \Big\{ g_T(\theta,\phi) T(\theta,\phi)+
            \label{detected_exponential}  \\
    & +\frac{g_P(\theta,\phi)}{2}
               \left[ (Q-iU)(\theta,\phi) e^{i2\psi(\theta,\phi)}
                    + (Q+iU)(\theta,\phi) e^{-i2\psi(\theta,\phi)} 
               \right] \Big\}\text{.}
           \nonumber
\end{align}
We can now insert the harmonic expansions \eqref{temp_field} and
\eqref{pol_field} into \eqref{detected_exponential}
and read the beam coefficients
\begin{align}
    b_{ 0lk} =& \iint g_T(\theta,\phi)
         {\,}_0Y^\ast_{lk}(\theta,\phi)
       \sin\theta d\theta d\phi    \nonumber  \\
    b_{\pm 2lk} =& \frac12 \iint g_P(\theta,\phi)
         {\,}_{\pm2}Y^\ast_{lk}(\theta,\phi)
       e^{\pm i2\psi(\theta,\phi)}
       \sin\theta d\theta d\phi\text{.}
\end{align}
We then use definition \eqref{harmonics_def} and write
the harmonic functions in terms of Wigner functions.
In terms of the reduced Wigner functions we have
\begin{equation}
   {}_sY^\ast_{lk}(\theta,\phi) = 
   \sqbeam
   e^{-ik\phi}d^{l}_{k,-s}(\theta)\text{.}
\end{equation}
The $\psi$-term can be transferred inside the Wigner function.
We may thus write for a general polarised beam
\begin{align}
   b_{ 0lk} =& \sqbeam
                 \iint g_T(\theta,\phi)
         D^l_{k,0}(\phi,\theta,0)
       \sin\theta d\theta d\phi                  \nonumber \\
    b_{\pm 2lk} =& \frac{\sqbeam}{2}
                 \iint g_P(\theta,\phi)
         D^l_{k,\mp 2}(\phi,\theta,\psi(\theta,\phi))
        \sin\theta d\theta d\phi\text{.} \label{beam_general}
\end{align}
The above formulas for the beam coefficients are completely general
as long as the beam response is linear.

We now make the simplifying assumption that the polarisation
sensitivity is ``in the same direction''
everywhere on the beam.
By ``same direction'' we mean that the direction of polarisation sensitivity
at an arbitrary location is obtained by rotating the polarisation direction
vector at the north pole to the desired location along a meridian.
The local polarisation orientation angle then becomes
\begin{equation}
   \psi(\theta,\phi) = \psipol -\phi\text{,}
\end{equation}
where \psipol\ is the deviation of polarisation direction
from $\phi=0$ at the north pole.

At this point it is convenient to write the Wigner functions 
in terms of the reduced Wigner functions 
\eqref{reduced_def}.
The coefficients for a beam with unique polarisation direction become 
\begin{align}
   b_{ 0lk} =& \sqbeam
        \int \sin\theta d\theta \,
         d^l_{k,0}(\theta)
       \int g_T(\theta,\phi) e^{-ik\phi} d\phi  \label{beam_unique} \\
    b_{\pm2lk} =& \frac{\sqbeam}{2}
           e^{\pm i2\psipol}
          \int \sin\theta d\theta \,
         d^l_{k,\mp 2} (\theta)
            \int g_P(\theta,\phi) e^{-i(k\pm 2)\phi} d\phi\text{.} \nonumber
\end{align}

We readily see that for a symmetric beam the only non-vanishing coefficients are those with $k+s=0$.
Especially, for a perfect beam with zero width and perfect polarisation
sensitivity ($g_P=g_T$)
we obtain
\begin{align}
   &b_{ 0lk}&   &=& &\sqbeam d^l_{k,0}(0)&
              &=& &\sqbeam \delta_{k,0}&          \\
    &b_{\pm 2lk}&  &=& &\frac{\sqbeam}{2}
                 d^l_{k,\mp 2}(0) e^{\pm i2\psipol}&
               &=& &\frac{\sqbeam}{2}
                 e^{\pm i2\psipol} \delta_{k,\mp 2}\text{.}& \nonumber
\end{align}
We have assumed normalisation
$\int g_T(\Omega)d\Omega=1$.

In this work we have constructed the beam coefficients according to (\ref{beam_unique}),
but the assumption of unique polarization direction is not crucial for the deconvolution method.
A more general beam presentation is given by (\ref{beam_general}).


\section{Methods used to accelerate computation}
\label{technical_tweaks}

\subsection{Load balancing}
In an MPI-parallel code, good load balancing among the individual tasks
is crucial. As was mentioned in Section \ref{algorithm}, some quantities
are distributed across tasks depending on their colatitude $\theta$
or the index $m$. Since the computational
cost of the algorithm is not independent of $\theta$ and $m$, but rather
an unknown, smoothly varying function of these two numbers, it is advisable
not to assign sequential ranges of those indices to individual MPI tasks,
because choosing index ranges that distribute the load evenly is very hard in
that scenario.

To avoid this complication, we adopted a ``round robin'' strategy for data
distribution. Assuming $n$ MPI tasks numbered 0 to $n-1$, we assign to
a particular task $i$ the $\theta$ indices $\theta_i$, $\theta_{i+n}$,
$\theta_{i+2n}$,~\dots, and deal with the index $m$ in analogous fashion.
This results in near-perfect load balancing.

\subsection{Convolution optimisation}
A significant part of total CPU time is spent in the FFTs
necessary for the convolution with the quantity $N$ (see section
\ref{algorithm}). Fortunately, several measures can be taken to decrease the
resource requirements. First of all, we are making use of the highly optimised
FFTW\fnurl{http://fftw.org} library \citep{frigo-johnson-1998}, which provides
close to optimal performance for FFTs of specific size.

Furthermore, it is important to notice that the execution time for an FFT of
length $n$ depends sensitively on the prime factorisation of $n$; in general,
an $n$ composed of only small prime factors is much preferable.

Since we are performing a convolution operation, we have some freedom in the
choice of array dimensions; they can be increased and zero-padded if desired.
In our case, the minimal array dimensions are $4\lmax+1$ and
$4\kmax+1$, respectively; if necessary, we increase both of them to the next
larger number, which is a product of the prime factors 2, 3, and 5 only.
Depending on the prime factorisation of the original array dimensions, this
can decrease the run-time for this part of the code by integral factors.

Finally, the quantities that are to be convolved exhibit the symmetry
\[ V_{i,k} = V_{n_i-i,n_k-k}^* \qquad \text{for } i,k> 0\text{,} \]
which allows us to employ the specialised ``real-valued'' FFT, resulting
in another substantial speedup and a size reduction of the precomputed
Fourier transform of $N$ (see Section \ref{n_precomp}).

\subsection{Efficient computation of the reduced Wigner functions}
During code execution, the reduced Wigner functions $d^l_{mk}(\theta)$
are required twice in every conjugate gradient iteration step. Storing them
in memory is prohibitively expensive, so it is important to regenerate them
efficiently whenever needed. For this purpose we used code originally presented
by \citet{prezeau-reinecke-2010}, which was extended to make use of the SSE2
instruction set present in all modern Intel-compatible CPUs. We also used
the symmetry properties of the $d$ matrix to avoid redundant computations.

\subsection{Shortcuts for symmetrical beams}
If a beam used for deconvolution is symmetric with respect to a $180^{\circ}$
rotation around its axis (which is true for elliptical beams, for example),
all its $b_{slm}$ coefficients for odd $m$ vanish
by definition. This fact is used to skip unnecessary computations as well
as save space by not storing any array elements that are known to be zero.
It also reduces the minimum second dimension of the convolution arrays from
$4\kmax+1$ to $2\kmax+1$, saving even more time on the FFT operations and
reducing memory consumption.

\subsection{Optional precomputation}
\label{n_precomp}
The convolution step mentioned in section \ref{algorithm} requires the Fourier
transform of array $N$, which is constant throughout a run.
The code permits one to precompute and store this quantity, which reduces total CPU
time (because only two FFT operations are then needed for each convolution step,
instead of three). However, especially for runs with high \lmax\ and
\kmax\ this increases the total memory consumption considerably.
So for situations where the available main memory is insufficient to store
this precomputed array, we provide the option of recomputing the
Fourier-transformed $N$ whenever necessary, approximately halving the
memory consumption of the code, but also slowing it down by 5-15 percent.
For the 70GHz runs described in this paper, we had to make use of this
space-saving feature.

\subsection{Supplying an initial guess}
While not strictly a performance optimisation, the code allows the user
to supply an initial guess from which to start the CG iteration, which is then
used instead of a vector of zeroes. This is
particularly helpful if one tries to compute series of deconvolutions
on the same input data set, but with different parameters. For example,
if a solution for a deconvolution problem with a given \lmax\ and
\kmax\ has already been obtained, it can be used as a good starting
point for other calculations with higher \lmax\ and/or \kmax.
In any case, the convergence criterion will still be evaluated with respect
to an all-zero $a_{slm}$, not the supplied guess.

\section{Resource usage}
\label{benchmarks}

Most computations discussed in this paper were performed on a computer
with 64GB of RAM and 24 Intel Xeon E7450 cores. Since this computer is used
interactively, we only ran our calculations on 16 of these cores and ensured
that interactive usage did not exceed the capability of the remaining 8 cores.
Nevertheless, the measured timings contain small uncertainties and should be
interpreted as upper limits.

\begin{table}
\caption{Resource usage for various deconvolution runs}
\begin{center}
\begin{tabular}{crrrrrrr}
\hline\hline
Name & \lmax\vphantom{\Large I} & \kmax & $T_\text{wall}$ & $T_\text{setup}$ & $T_\text{iter}$  & $N_\text{iter}$ & Mem\\
\hline
\hline
30a & 400\vphantom{\Large I}  &  6 &     10:07 &  1:13 &   3.9 &  140 &  6.6 \\
30b & 500\vphantom{\Large I}  &  6 &     25:49 &  1:20 &   5.5 &  270 &  8.1 \\
30c & 600\vphantom{\Large I}  &  4 &   1:24:57 &  1:07 &   5.1 &  995 &  6.4 \\
30d & 600\vphantom{\Large I}  &  6 &   1:08:53 &  1:35 &   6.9 &  591 &  9.6 \\
30e & 600\vphantom{\Large I}  &  8 &   1:40:40 &  2:24 &  10.3 &  573 & 11.9 \\
30f & 700\vphantom{\Large I}  &  6 &   3:41:13 &  1:48 &   8.5 & 1567 & 11.3 \\
30g & 800\vphantom{\Large I}  &  6 &   5:10:41 &  1:56 &  10.6 & 1759 & 12.8 \\
 D & 500\vphantom{\Large I}  & 30 &   10:31:38 &  8:34 &  20.3 & 1842 & 17.6 \\
70a & 1700\vphantom{\Large I} &  6 &  63:51:56 & 15:27 &  78.3 & 2923 & 28.8 \\
70x & 1700\vphantom{\Large I} & 6 &    9:43:05 &  5:58 &  11.8 & 2955 & --- \\
70y & 1700\vphantom{\Large I} & 6 &    5:25:32 &  7:20 &   6.6 & 2922 & --- \\
\end{tabular}
\end{center}
\tablefoot{$T_\text{wall}$ denotes the total wall clock time of
the run, $T_\text{setup}$ is the time required for the code startup
(i.e.\ data input, computing the RHS, Wigner transform of the hit count and
preconditioner), $T_\text{iter}$ gives the average wall clock time for a single
CG step in seconds, $N_\text{iter}$ is the required number of iterations until
convergence, and ``Mem'' is the total memory usage in
GB. Please note the additional comments in the text.}
\label{table:bench}
\end{table}

Table \ref{table:bench} gives an overview of the parameters of the various
runs and their impact on resource usage. Runs 30a-g are the 30GHz calculations
discussed in Sect.~\ref{run30}; runs 70a, 70x, and 70y are related to the 70GHz
high-resolution study (Sect.~\ref{run70}), and run ``D'' is the deconvolution
with the D-shaped beam presented in Sect.~\ref{dbeam}.
For all runs except the D-beam, the
beam was assumed to be symmetrical with respect to a $180\degr$ rotation
around its axis. Also, for all runs except the 70GHz ones the FFT of array $N$
was cached (see Section \ref{n_precomp}).

For the 70GHz run on the test machine described above (run 70a), each CG iteration
step needed significantly more time than for the calculations using the 30GHz data set;
in combination with the high number of iterations, this leads to an inconveniently
long computation time.
As a consequence, deconvolution runs of this magnitude are better suited for execution
on massively parallel computers. We therefore repeated the computation on the Louhi
supercomputer of CSC (Cray XT4/XT5) on 128 and 512 processor cores, respectively.
The corresponding performance figures are listed as runs 70x and 70y in Table
\ref{table:bench}; memory consumption was not measured for these runs.

The computational power of a single CPU core is roughly the same
on both computers used for our tests. Looking at the total wall clock time
for the 70GHz runs, it becomes obvious that the scaling from 16 to 128 cores
is close to ideal, but becomes much worse when going from 128 to 512 cores.
In this range, MPI communication and redundant computations begin to
dominate. For deconvolution problems with larger \lmax, we expect that this transition
will occur at a higher number of MPI tasks.

The numbers of iterations required for the 70GHz runs are not identical, as
one might intuitively expect. This is because in an MPI-parallel
code some computation steps -- most notably summations and other reductions of
quantities residing on all tasks -- are not carried out in a strictly defined
order, but depend on the relative timing of the tasks during such an operation,
which is nondeterministic. Since floating-point algebra with finite precision
is non-associative, the results of such operations can differ slightly from run
to run, causing the conjugate gradient solver to take different paths towards
the solution, but of course ultimately solving the same numerical problem.

Compared to earlier publications on deconvolution map-making, it is evident
that both memory and CPU requirements of the presented algorithm are quite
modest and allow deconvolution up to values of \lmax\ that were prohibitively
expensive up to now. Making use of massively parallel computing clusters,
we expect that the algorithm can be applied to data sets with even higher
resolution than those presented here.

\begin{acknowledgements}
Some of the results in this paper have been derived using the HEALPix
\citep{gorski-etal-2005} package. We acknowledge the use of the Planck Sky Model, 
developed by the Component Separation Working Group (WG2) of the Planck Collaboration.
MR is supported by the German Aeronautics Center and Space Agency (DLR), under
program 50-OP-0901, funded by the Federal Ministry of Economics and
Technology. EK is supported by Academy of Finland grant 253204.
This work was granted access to the HPC resources of CSC made
available within the  Distributed European Computing Initiative by
the PRACE-2IP, receiving funding from the European Community's Seventh
Framework Programme (FP7/2007-2013) under grant agreement RI-283493. 
We thank CSC -- the IT Center for Science Ltd.\ (Finland) -- for computational resources.
We thank Torsten En\ss{}lin for constructive discussions and feedback,
Valtteri Lindholm for help in using the Planck Sky Model,
and Ronnie James Dio for the D-beam.
\end{acknowledgements}

\bibliographystyle{aa}
\bibliography{planck}

\end{document}